\documentclass[journal,onecolumn,10pt]{IEEEtran}

\usepackage{setspace}
\usepackage{bm}
\usepackage{graphics,
           psfrag,
           epsfig,
           amsthm,
           cite,
           amssymb,
           url,
           dsfont,
           subfigure,
           algorithm,
           algorithmic,
           balance,
           enumerate,
           color,
           setspace
}
\usepackage{amsmath}
\usepackage{epstopdf}
\newtheorem{example}{Example}

\newtheorem{lemma}{Lemma}

\newtheorem{theorem}{Theorem}
\newtheorem{corollary}{Corollary}

\newtheorem{remark}{Remark}
\newcommand{\cororef}[1]{Corollary \ref{#1}}
\newcommand{\eref}[1]{(\ref{#1})}
\newcommand{\sref}[1]{Section~\ref{#1}}
\newcommand{\appref}[1]{Appendix~\ref{#1}}
\newcommand{\fref}[1]{Figure~\ref{#1}}

\newcommand{\cref}[1]{Constraint~\ref{#1}}
\newcommand{\thref}[1]{Theorem~\ref{#1}}
\newcommand{\lref}[1]{Lemma~\ref{#1}}

\usepackage{lipsum}
\usepackage{graphicx}
\allowdisplaybreaks
\usepackage{blkarray, bigstrut} 
\usepackage{mathtools}
\usepackage{bbm}
\usepackage{multirow}
\usepackage{caption}
\usepackage{diagbox}
\usepackage{makecell}
\allowdisplaybreaks
\usepackage{url}

\usepackage{mathtools}
\DeclarePairedDelimiter\ceil{\lceil}{\rceil}
\DeclarePairedDelimiter\floor{\lfloor}{\rfloor} 
 
 \usepackage{amsmath}
\usepackage{pifont}

\newcommand{\xmark}{\text{\ding{55}}}
\title{
Centralized Multi-Node Repair Regenerating Codes}
\author{Marwen Zorgui,~\IEEEmembership{Student Member,~IEEE}, 
Zhiying Wang,~\IEEEmembership{Member,~IEEE}
\thanks{
M. Zorgui and Z. Wang are with the Center for Pervasive Communications and Computing,
University of California at Irvine, Irvine, CA 92697 USA (e-mail: mzorgui@uci.edu,
zhiying@uci.edu). Part of this work has been presented in the 54th Annual Allerton Conference on Communication, Control, and Computing (Allerton), 2016 \cite{allerton_multi_node} and IEEE International Symposium on Information Theory, 2017 \cite{online_version}.
}
}

\begin{document}
\maketitle
\thispagestyle{empty}
\begin{abstract}
In a distributed storage system, recovering from multiple failures is a critical and frequent task that is crucial for maintaining the system's reliability and fault-tolerance. In this work, we focus on the problem of repairing multiple failures in a centralized way, which can be desirable in many data storage configurations, and we show that a significant repair traffic reduction is possible. First, the fundamental tradeoff between the repair bandwidth and the storage size for functional repair is established. Using a graph-theoretic formulation, the optimal tradeoff is identified as the solution to an integer optimization problem, for which a closed-form expression is derived. Expressions of the extreme points, namely the minimum storage multi-node repair (MSMR) and minimum bandwidth multi-node repair (MBMR) points, are obtained. Second, we describe a general framework for converting single erasure minimum storage regenerating codes to MSMR codes. The repair strategy for $e$ failures is similar to that for single failure, however certain extra requirements need to be satisfied by the repairing functions for single failure. For illustration, the framework is applied to product-matrix codes and interference alignment codes. Furthermore, we prove that the functional MBMR point is not achievable for linear exact repair codes. We also show that exact-repair minimum bandwidth cooperative repair (MBCR) codes achieve an interior point, that lies near the MBMR point, when $k \equiv 1  \mod e$, $k$ being the minimum number of nodes needed to reconstruct the entire data. Finally, for $k> 2e, e\mid k$ and $e \mid d$, where $d$ is the number of helper nodes during repair, we show that the functional repair tradeoff is not achievable under exact repair, except for maybe a small portion near the MSMR point, which parallels the results for single erasure repair by Shah et al.
\end{abstract}
%%%%%%%%%%%%%%%%%%%%%%%%%%%%%%%%%%%%%%%%%%%%%%%%%%%%%%%%%%%%%%%%%%%%%%%%%%%%%%%%%%
%%%%%%%%%%%%%%%%%%%%%%%%%%%%%%%%%%%%%%%%%%%%%%%%%%%%%%%%%%%%%%%%%%%%%%%%%%%%%%%%%%
%%%%%%%%%%%%%%%%%%%%%%%%%%%%%%%%%%%%%%%%%%%%%%%%%%%%%%%%%%%%%%%%%%%%%%%%%%%%%%%%%%
%%%%%%%%%%%%%%%%%%%%%%%%%%%%%%%%%%%%%%%%%%%%%%%%%%%%%%%%%%%%%%%%%%%%%%%%%%%%%%%%%%
\section{Introduction}\label{introduction}
Ensuring data reliability is of paramount importance in modern storage systems. Reliability is typically achieved through the introduction of redundancy. Traditionally, simple replication of data has been adopted in many systems. For instance, Google file systems opted for a triple replication policy \cite{ghemawat2003google}. However, for the same redundancy factor, replication systems fall short on providing the highest level of reliability. 
On the other hand, erasure codes can be optimal in terms of the redundancy-reliability tradeoff.
In erasure codes, a file of size $\mathcal{M}$ is divided into $k$ fragments, each of size $\frac{\mathcal{M}}{k}$. The $k$ fragments are then encoded into $n$ fragments using an $(n,k)$ maximum distance separable (MDS) code and then stored at $n$ different nodes. Using such a scheme, the data is guaranteed to be recovered from any $n-k$ node erasures, providing the highest level of worst-case data reliability for the given redundancy. However, traditional erasure codes suffer from high repair bandwidth. In the case of a single node erasure, they require downloading the entire data of size $\mathcal{M}$ to repair a single node storing a fragment of size $\frac{\mathcal{M}}{k}$. This expansion factor made erasure codes impractical in some applications using distributed storage systems. In the last decade, the repair problem has gained increasing interest and motivated the research for a new class of erasure codes with better repair capabilities. The seminal work in \cite{dimakis2010network} proposed regenerating codes that optimally solve the repair bandwidth problem. Interestingly, the authors in \cite{dimakis2010network} proved that one can significantly reduce the amount of bandwidth required for repair and the bandwidth decreases as each node stores more information. Formally, suppose any $k$ out of $n$ nodes are sufficient to recover the entire file of size $\mathcal{M}$. Assuming that $d$ nodes, termed helpers, participate in the repair process, denoting the storage capacity of each node by $\alpha$ and the amount of information downloaded from each helper by $\beta$, then, an optimal $(\mathcal{M},n,k,d,\alpha,\beta)$ regenerating code satisfies
\begin{align}
\label{tradeoff}
\mathcal{M}= \sum_{i=0}^{k-1} \min \{\alpha, (d-i) \beta \}.
\end{align}
Equation \eref{tradeoff} describes the fundamental tradeoff between the storage capacity $\alpha$ and the bandwidth $\beta$. Two extreme points can be obtained from the tradeoff. Minimum storage regenerating (MSR) codes correspond to the best storage efficiency with $\alpha=\frac{\mathcal{M}}{k}$, while minimum bandwidth regenerating (MBR) codes achieve the lowest possible bandwidth at the expense of extra storage per node. 

If we recover the exact same information as the failed node, we call it \emph{exact repair}, otherwise we call it \emph{functional repair}. Using network coding \cite{ahlswede2000network, ho2006random}, it is possible to construct functional regenerating codes satisfying \eqref{tradeoff} \cite{dimakis2010network}.
Following the seminal work in \cite{dimakis2010network}, there has been a flurry of interest in designing exact-repair regenerating codes that achieve the optimal tradeoff, focusing mainly on the extreme MSR and MBR points, e.g., 
\cite{shah2012interference,suh2011exact,wu2009reducing,papailiopoulos2013repair,wang2014explicit
,rawat2016progress,goparaju2017minimum,cadambe2013asymptotic,Zigzag_Codes_IT, ye2017nearly}.
For interior points that are between the MBR and MSR points in the tradeoff of \eqref{tradeoff}, \cite{non_achievability} showed that most points are not achievable for exact repair.
Moreover, there has been a growing literature focused on understanding the fundamental limits of exact-repair regenerating codes. Other outer bounds for exact repair include \cite{duursma2014outer,duursma2015shortened,mohajer2015new} for general parameters, and \cite{sasidharan2014improved} for linear codes. 
The aforementioned references, as most of the studies on regenerating codes in the literature, focus on the single erasure repair problem. However, in many practical scenarios, such as in large scale storage systems, multiple failures are more frequent than a single failure. Moreover, many systems (e.g., \cite{bhagwan2004total}) apply a lazy repair strategy, which seeks to limit the repair cost of erasure codes.  Instead of immediately repairing every single failure, a a lazy repair strategy waits until $e$ erasures occur, $e \le n-k$, then, the repair is done by downloading the equivalent of the total information in the system to regenerate the erased nodes. However, a natural question of interest is, whether one can reduce the amount of download in such scenarios.

In this work, we consider centralized repair. 
%The centralized repair framework is applicable to many practical situations.
Indeed, there are situations in which, due to architectural constraints, it is more desirable to regenerate the lost nodes at a central server before dispatching the regenerated content to the replacement nodes \cite{bhagwan2004total}. For instance, one can think of a rack-based node placement architecture \cite{rawat2016centralized} in which failures frequently occur to nodes corresponding to a particular rack. In this scenario, a centralized repair of the entire rack is favorable as opposed to repairing the rack on a per-node basis. Furthermore, \cite{rawat2016centralized} showed that a centralized repair framework can have interesting applications in communication-efficient secret sharing. Finally, centralized repair can be used in a broadcast network, where the repair information is transmitted to all replacement nodes (e.g. \cite{hu2015broadcast}). %For the above reasons, characterizing the repair-bandwidth tradeoff under the centralized repair framework is important from both an information-theoretic and also a practical perspective. 

Our centralized repair framework requires the content of any $k$ out of $n$ nodes in the system to be sufficient to reconstruct the entire data. Upon the failure of $e$ nodes in the system, the repair is carried out by contacting any $d$ helpers out of the $n-e$ available nodes, $d \le n-e$, and downloading $\beta$ amount of information from each of the $d$ helpers. Our objective is to characterize the functional repair tradeoff between the storage per node $\alpha$ and the repair bandwidth $\beta$ under the centralized multiple failure repair framework. 
%\bl{???is this part repeated in the single erasure case? Remove??? Under functional repair, the repaired nodes are not necessarily the same as the failed nodes. Exact repair however requires that the replacement nodes recover exactly the content of the failed nodes.} 
We also seek to investigate the achievability of the functional tradeoff under exact repair.

\subsection{Related work}
Cooperative regenerating codes (also known as coordinated regenerating codes) have been studied to address the repair of multiple erasures \cite{kermarrec2011repairing,closed_form_cooperative_regene} in a distributed manner. In this framework, each replacement node downloads information from $d$ helpers in the first stage. Then, the replacement nodes exchange information between themselves before regenerating the lost nodes. Cooperative regenerating codes that achieve the extreme points on the cooperative tradeoff have been developed; namely, minimum storage cooperative regenerating (MSCR) codes \cite{li2014cooperative,closed_form_cooperative_regene,ye2018optimal} 
%\bl{?? Add new references by Ye??} 
and minimum bandwidth cooperative regeneration (MBCR) codes\cite{wang2013exact}. 
%In \cite{li2014cooperative}, the authors showed that, given an instance of linear exact MSR codes, it is possible to construct an instance of exact linear MSCR codes for 2 erasures. 

The number of nodes involved in the repair of a single node, known as locality, is another important measure of node repair efficiency \cite{locality}. Various bounds and code constructions have been proposed in the literature \cite{locality,family_locality}. Recent works have investigated the problem of multiple node repair under locality constraints \cite{locality_two_erasures,song2015locally}.

The problem of centralized repair has been considered in \cite{cadambe2013asymptotic}, in which the authors restricted themselves to MDS codes, corresponding to the point of minimum storage per node. \cite{cadambe2013asymptotic} showed the existence of MDS codes with optimal repair bandwidth in the asymptotic regime where the storage per node (as well as the entire information) tends to infinity. In \cite{wang2016optimal}, the authors proved that Zigzag codes, which are MDS codes designed initially for repairing optimally single erasures \cite{Zigzag_Codes_IT}, can also be used to optimally repair multiple erasures in a centralized manner. In \cite{rawat2016centralized}, the authors independently proved that multiple failures can be repaired in Zigzag codes with optimal bandwidth. Moreover, \cite{rawat2016centralized} defines the minimum bandwidth multi-node repair codes as codes satisfying the property of having the downloaded information $d \beta$ matching the entropy of $e$ nodes\footnote{The definition of minimum bandwidth multi-node repair codes in our paper is simply the minimum bandwidth point on the functional tradeoff, which is different from \cite{rawat2016centralized} for $e \nmid k$.}. Based on that, the authors derived a lower bound on $\beta$ for systems having a certain entropy accumulation property and then showed achievability of the minimum bandwidth codes using MBCR codes. However, the optimal storage  size per node  $\alpha$ is not known under these conditions. In \cite{ye2017explicit}, the authors presented an explicit MDS code construction that provides optimal repair for all $e\le n-k$ and $k \le d \le n-e$ simultaneously. The authors in \cite{hu2015broadcast} studied the problem of broadcast repair for wireless distributed storage which is equivalent to the model we study in this paper.  
It is worth pointing out that the previous constructions are for high-rate codes, with large subpacketization $\alpha$. 
%\bl{?? why need this??? For scalar MDS codes, i.e., $\beta=1$, it is shown that exact repair cannot be achieved when $\frac{k}{n}> \frac{1}{2}+ \frac{2}{n}$.}
%\bl{Add other recent results on Reed Solomon codes. Also explain here that most previous MSMR constructions are for high-rate codes. When the rate is less than 1/2, the sub-packetization size needs to be $k^2$ (? please verify the parameters of the references).} 
In \cite{li2015enabling}, the authors presented an approach that enables single erasure MSR codes to recover from multiple failures simultaneously with near-optimal bandwidth. Based on simulations, \cite{li2015enabling} showed that their approach can provide efficient recovery of most of the failure patterns, but not all of them. 
The repair problem of Reed Solomon codes has been recently investigated in \cite{guruswami2017repairing} for single erasure and in
\cite{dau2016repairing,dau2017repairing,bartan2017repairing,ye2017repairing} for
multiple erasures.
%. Repairing multiples failures in Reed Solomon codes has been investigated in. 
In \cite{IA_code}, the authors proved that the interference alignment MSR construction of \cite{suh2011exact}, originally
designed for repairing any single node failure, can recover from multiple failures in a cooperative way. 
%\bl{??If we think the result of this paper is not correct, we may omit the details?? 
Specifically, it is shown that any set of systematic nodes, set of parity-check nodes, or pair of nodes can be repaired cooperatively with optimal bandwidth. 
%}
\subsection{Contributions of the paper}
The main contributions of this paper are the characterization of functional tradeoff, and the examination of its achievability under exact repair for the extreme points and the interior points.
They are summarized as follows.
\begin{itemize}
\item	We first establish the explicit functional tradeoff between the repair bandwidth and the storage size for functional repair (Theorems \ref{min_cut_theorem}, \ref{optimal_cut_result},
\ref{tradeoff_expression}). We obtain the tradeoff using information flow graphs. From the functional tradeoff, we characterize the minimum storage multi-node repair (MSMR) point, and the minimum bandwidth multi-node repair (MBMR) point.
\item	When the number of erasures $e$ satisfies $e \ge k$, $k$ being the minimum number of nodes needed to reconstruct the entire data, the tradeoff reduces to a single point, for which we provide an explicit code construction. 

\item	We formalize a construction for exact-repair MSMR codes. Given an instance of an exact linear MSR code, we present a framework to construct an instance of an exact linear MSMR regenerating code. We note here that \cite{li2014cooperative} and \cite{li2015enabling} used a similar approach for MSCR codes and their numerical results, respectively. Based on this framework, we study the product-matrix (PM) MSR codes \cite{Rashmi_Product_Matrix} and the interference alignment (IA) construction in \cite{suh2011exact}. 
We prove the existence of PM and IA MSMR codes for any number of failures $e$, $e \le n-k$ (Theorems \ref{thm_PM_2_erasures}, \ref{thm_PM_general}, \ref{thm_IA_general}). Moreover, for the IA code, we prove that the code can always efficiently recover from any set of $e \le n-k$ node failures as long as the failed nodes are either all systematic nodes or all parity nodes (Theorem \ref{thm_IA_1_group}); for failures including both systematic and parity nodes, we derive explicit design conditions under which exact recovery is ensured, for some particular system parameters (Theorems \ref{thm_IA_2_erasures}, \ref{thm_IA_3_erasures}). We note here that unlike previous constructions, our codes are applicable when the code rate is low and they use a small subpacketization size of $\alpha=k-1$ or $k$.
%We prove the existence of product-matrix MSMR codes for any number of failures $e$, $e \le n-k$. For the IA code, we prove that one can always efficiently recover from any set of $e \le n-k$ node failures as long as the failed nodes are either all systematic nodes or all parity nodes. For failures including both systematic and parity nodes, we derive explicit design conditions under which exact recovery is ensured, for some particular system parameters. We provide an existence proof of interference alignment MSMR codes for any $e \le n-k$. 
\item	We prove that, to our surprise, functional MBMR point is not achievable for linear exact repair codes for $1 < e < k$ (Theorems \ref{non_existence_MBMR_1}, \ref{non_existence_MBMR_2}), while linear codes achieve such point for single erasure \cite{Rashmi_Product_Matrix}.

\item	We show that exact-repair MBCR codes achieve an interior point, that lies near the MBMR point, when $k\equiv 1  \mod e $ (Theorem \ref{MBCR_achievability}). 

\item	We show that the functional repair tradeoff is not achievable under exact repair for interior points between MBMR and MSMR points, except for maybe a small portion near the MSMR point, for
$k,d$ being multiples of $e$ and $ k >2 e$
% $e \mid k,  e \mid d, k >2 e$ 
 (Theorems \ref{non_achivability_1}, \ref{non_achivability_2}), which parallels the results for single erasure repair \cite{non_achievability}. The achievability of the functional tradeoff under exact repair is summarized in Table \ref{tab:title}.

\item	Finally, we study the adaptive repair problem of multiple erasures in MBR codes and present an MBR construction with optimal repair, simultaneously for varying numbers of helpers and varying numbers of erasures (Theorem \ref{MBR_costruction}).
\end{itemize}

\begin{figure*}
\begin{center}
 
\centering

\begin{tabular}{ | m{1.5cm} | m{3.5cm}| m{3.5cm} |m{7.5cm}| } 

\hline
 & \center{MSMR point} & \center{MBMR point }& \qquad\qquad\qquad \qquad Interior points  \\ 
\hline
\center{$e=1$} &  \center{$\checkmark$ \cite{Zigzag_Codes_IT,suh2011exact,Rashmi_Product_Matrix}}   & \center{$\checkmark$ \cite{Rashmi_Product_Matrix}} &  $\xmark$, except maybe for a small portion near the MSMR point \cite{non_achievability}.
\\ 
\hline
\center{$1 < e <k$} &  \center{$\checkmark$ \cite{ye2017explicit,cadambe2013asymptotic}, [Sections \ref{approach_s1ection}, \ref{PM_codes}, \ref{IA_codes}] } & \center{$\xmark$  (for linear codes)} [\sref{MBMR_codes}]& 
$\bullet$ if $k \equiv 1 \mod e$: an interior point near the MBMR point is achievable [\sref{MBCR_codes}].

$\bullet$ if $e\mid k, e\mid d, k > 2e: $ $\xmark$, except maybe for a small portion near the MSMR point [\sref{interior_points}].
\\ 
\hline
\center{$e\geq k$} & \center{$\checkmark$ \sref{construction}} &    \center{$\checkmark$\sref{construction}} &  \qquad\qquad\qquad\qquad $\checkmark $ \sref{construction} \vspace{-.17cm} \\ 
\hline

\end{tabular}
\captionof{table}{Summary of achievability results of functional repair tradeoff under exact repair for an $(n,k,d,e,\alpha, \beta)$ distributed storage system. MSMR and MBMR points are defined to be the minimum storage point and the minimum bandwidth point on the functional tradeoff, respectively. Here $e \mid k, e \mid d$ means that $k,d$ are multiples of $e$. The symbol $\checkmark$ denotes achievability while $\xmark$ denotes non-achievability, both of which are under exact repair. \label{tab:title} }
\end{center}
\end{figure*}
%%%%%%%%%%%%%%%%%%%%%%%%%%%%%%%%%%%%%%%%%%%%%%%%%%%%%%%%%%%%%%%%%%%%%%%%%%%%%%%%%%%%%%%%%%%%%%%%%%%%%%%%%%%%%%%%%%%%%%%%%%%%%%%%%%%%%%%%%%%%%%%%%%%%%%%%%%%%%%%%%%%%%%%%%%%%%%%%%%%%%%%%%%%%%%%%%%%%%%%%%%%%%%%%%%%%%%%%%%%%%%%%%%%%%%%%%%%%%%%%%%%%%%%%%%%%%%%%%%%%%%%%%%%%%%%%%%%%%%%%%%%%%%%%%%%%%%%%%%%%%%%%%%%%%%%%%%%%%%%%%%%%%%%%%%%%%%%%%%%%%%%%%%%%%%%%%%%%%%%%%%%%%%%%
\subsection{Organization of the paper}
The remainder of the paper is organized as follows. In \sref{S2}, we first describe the system model before analyzing the fundamental functional repair tradeoff between the storage size and the repair bandwidth. \sref{S3} describes our code construction for the case $e \geq k$, as well as the MSMR codes framework and its application to the product-matrix and the interference alignment codes. We prove the non-achievability of MBMR point under linear exact repair in \sref{MBMR_codes}. The non-achievability of the interior points under exact repair is investigated in \sref{interior_points}. The adaptive repair of multiple erasures for an MBR code is presented in \sref{adaptive_MBR} and \sref{conclusion} draws conclusions. %Finally, some of the proofs are relegated to \sref{Appendices}.

{\bf Notation.} 
$[n]$ denotes the set of elements $\{ 1, \ldots,n\}$. 
$\ceil{\cdot}$ and $\floor{\cdot}$ represent the ceiling and the floor functions.
For a set $\mathcal{A}$,  
 $\mathcal{A}\backslash \{ i\}$ denotes the resultant set after removing item $i$, while $|\mathcal{A}|$ denotes the size of $\mathcal{A}$. The symbol $\mathbbm{1}_{\{ E \}}$ denotes the indicator function of an event $E$, which is 1 if $E$ is true, and 0 otherwise.
The notations $e \mid k$ and $e \nmid k$ are used to denote whether $k$ is a multiple of $e$, or not, respectively. 
The superscript $t$ is used to denote the transpose of a matrix. For a matrix $A$, $|A|$ denotes its determinant and $A_{i,j}$ refers to its entry at position $(i,j)$. $I_n$ denotes the identity matrix of size $n$ and $\text{diag}\{\lambda_1,\ldots, \lambda_n \}$ denotes the $(n \times n)$ diagonal matrix with the corresponding elements. 
Vectors are denoted with lower-case
bold letters. 
$\mathbf{u}=[u_1,\ldots,u_m]$ denotes a vector of length $m$. Note that the notation $[k]$ may refer to a vector of size 1, or the set $\{ 1, \ldots,k\}$, however the meaning is clear from the context. 
$\mathbf{e}_i$ denotes the standard basis vector whose dimension is clear from the context. 
%The notation $A \subseteq B$ means that the set $A$ is included or equal to $B$. 
\section{Functional storage-bandwidth tradeoff}
\label{S2}
\subsection{System model}\label{system_nodel}
The centralized mutli-node repair problem is characterized by parameters $(\mathcal{M},n,k,d,e,
\alpha,\beta)$.
We consider a distributed storage system with $n$ nodes storing $\mathcal{M}$ amount of information. The data elements are distributed across the $n$ storage nodes such that each node can store up to $\alpha$ amount of information. Every node corresponds to a codeword symbol. The system should satisfy the following two properties:
\begin{itemize}
\item	Reconstruction property: a data collector (DC) connecting to any $k \le n$ nodes should be able to reconstruct the entire data.
\item	Regeneration property: upon failure of $e$ nodes, a central node is assumed to contact $d$ helpers, $k \le d \le n-e$, and download $\beta$ amount of information from each of them. New replacement nodes join the system and the content of each is determined by the central node. %The repair bandwidth is given by $ \beta$. 
$\beta$ is called the repair bandwidth.
The total bandwidth is denoted $\gamma= d\beta$.
\end{itemize}
We consider functional repair and exact repair. In the former case, the replacement nodes are not required to be exact copies of the failed nodes, but the repaired code should again satisfy the above two properties. Our objective is to characterize the tradeoff between the storage per node $\alpha$ and the repair bandwidth $ \beta$ under the centralized multiple failure repair framework.
On the optimal functional tradeoff, the minimum bandwidth mutli-node repair point is called \emph{MBMR}, and it has the minimum possible $\beta$, while the minimum storage mutli-node repair point is called \emph{MSMR} and has the minimum possible $\alpha$.
When considering exact repair, the minimum storage and minimum bandwidth points may be different from the above functional extreme points. While it has been shown for single erasure that the extreme points match for functional and exact repair, we will show later that MBMR is not achievable under exact repair.

In the paper, we will use the notation $k= \eta e +r$, such that $\eta= \floor*{\frac{k}{e}}$ and $ 0 \le r \le e-1$. 
%%%%%%%%%%%%%%%%%%%%%%%%%%%%%%%%%%%%%%%%%%%%%%%%%%%%%%%%%%%%%%%%%%%%%%%%%%%%%%%%%%%%%%%%%%%%%%%%%%%%%%%%%%%%%%%%%%%%%%%%%%%%%%%%%%%%%%%%%%%%%%%%%%%%%%%%%%%%%%%%%%%%%%%%%%%%%%%%%%%%%%%%%%%%%%%%%%%%%%%%%%%%%%%%%%%%%%%%%%%%%%%%%%%%%%%%%%%%%%%%%%%%%%%%%%%%%%%%%%%%%%%%%%%%%%%%%%%%
%\section{Functional storage-bandwidth tradeoff}\label{S3}
We now study the fundamental tradeoff between the storage size $\alpha$ and the repair bandwidth $\beta$ for $e$ erasures under functional repair. We use the technique of evaluating the minimum cut of a multi-cast information flow graph similar to the single erasure codes \cite{dimakis2010network} and the cooperative regenerating codes 
\cite{closed_form_cooperative_regene}.
\subsection{Information flow graphs}
\begin{figure*}
\centering
	\includegraphics[width=1\textwidth]{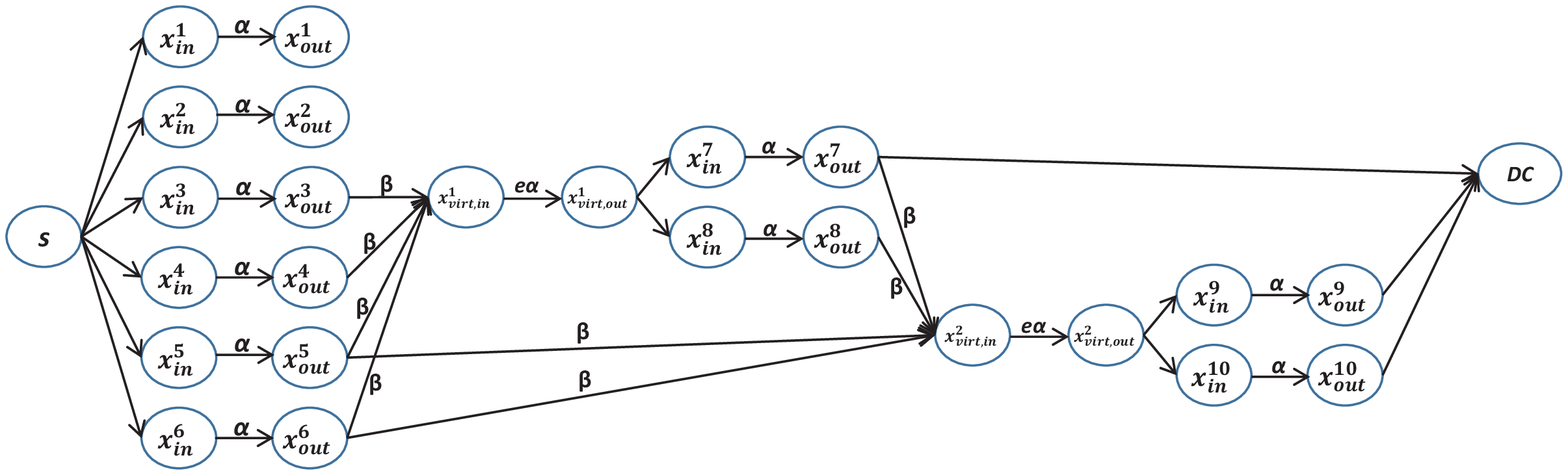}
\caption{Example of an information flow graph: $k=3,d=4, n=6, e=2$. The unlabeled edges have capacity $\infty$. Nodes 1 and 2 are repaired in the first stage and nodes 3 and 4 are repaired in the second stage. A data collector connecting to any 3 nodes should be able to recover the entire information.  
%\bl{In the figure, the source node is $S$, but in the text, the source node is $s$.}
}
\label{IFG}
 \end{figure*}

The performance of a storage system can be characterized by the concept of information flow graphs (IFGs). Our constructed IFG depicts the amount of information transferred, processed and stored during repair. We design our IFG with the following different kinds of nodes (see Figure \ref{IFG}). It contains a single source node $s$ that represents the source of the data object. Each storage node $x^i, i \in [n],$ of the IFG is represented by two distinct nodes: an input storage node $x_{in}^i$ and an output storage node $x_{out}^i$.  Each output node $x_{out}^i$ is connected to its input node $x_{in}^i$ with an edge of capacity $\alpha$, reflecting the storage constraint of each individual node.
The information flow graph is formed with $n$ initial storage nodes, %each with storage size $\alpha$
connected to the source node with edges of capacity $\infty$. 
The IFG evolves with time whereupon failure of $e$ nodes, $e$ new nodes simultaneously join the system. Each of the replacement nodes $x^j, j \ge n,$ is similarly represented by an input node $x_{in}^j$ and an output node $x_{out}^j$, linked with an edge of capacity $\alpha$. To model the centralized repair nature of the system, we add a virtual node $x_{virt}^i, i \ge 1,$ that links the $d$ helpers to the new storage nodes.
Likewise, the virtual node consists of an input node $x_{virt,in}^i$ and an output node $x_{virt,out}^i$. The input node $x_{virt,in}^i$ is connected to the $d$ helpers with edges each of capacity $\beta$. The output node $x_{virt,out}^i$ is connected to the input node $x_{virt,in}^i$ with an edge of capacity $e \alpha$, reflecting the overall size of the data to be stored in the new replacement nodes.
The output node $x_{virt,out}^i$ is then connected to the input nodes $x_{in}^j$ of the replacement nodes, with edges of capacity $\infty$. 
We define a \emph{repair group} to be any set of $e$ nodes that have been repaired simultaneously. In an IFG, a repair group is then associated with the virtual node that performs the repair operation.
%\bl{Define a repair group???}

Each IFG represents one particular history of the failure patterns. The ensemble of IFGs is denoted by $\mathcal{G}(n,k,d,e,\alpha,\beta)$. For convenience, we drop the parameters whenever it is clear from the context. Given an IFG $G \in \mathcal{G}$, there are  $ \binom{n}{k}$ different data collectors connecting to $k$ output storage nodes in $G$ with edges of capacity $\infty$. The set of all data collector nodes in a graph $G$ is denoted by $\text{DC}(G)$. For an IFG $G \in  \mathcal{G}$ and a data collector $t \in \text{DC}(G)$, the  minimum cut (min-cut) value separating the source node $s$ and the data collector $t$ is denoted by $\text{mincut}_G(s,t)$.
% \vspace{-0.5cm}
\subsection{Network coding analysis} 
The key idea behind representing the repair problem by an IFG lies in the observation that the repair problem can be cast as a multicast network coding problem \cite{dimakis2010network}. Celebrated results from network coding \cite{ahlswede2000network,ho2006random} are then invoked to establish the fundamental limits of the repair problem.  

According to the max-flow bound of network coding\cite{ahlswede2000network}, for a data collector to be able to reconstruct the data, the min-cut separating the source to the data collector should be larger or equal to the data object size $\mathcal{M}$. Considering all possible data collectors and all possible failure patterns, and assuming that the number of failures/repairs is bounded, the following condition is necessary and sufficient for the existence of centralized multi-node repair codes %satisfying the reliability constraint
 \cite[Proposition 1]{dimakis2010network}
\begin{align}
\label{min_cut_condition}
\min\limits_{G \in \mathcal{G}} \min\limits_{t \in \text{DC}(G)} \text{mincut}_G(s,t) \geq \mathcal{M}.
\end{align}
%\section{A cut-set bound on the repair bandwidth}
Analyzing the minimum cut of all IFGs result in the following theorem.

\begin{theorem}
\label{min_cut_theorem}
For fixed system parameters $(\mathcal{M}, n,k,d,e,\alpha,\beta)$, assuming that the number of failures/repairs is bounded, regenerating codes satisfying the centralized multi-node repair condition exist if and only if
\begin{align}
\label{cut_general}
&\mathcal{M} \le \min\limits_{\mathbf{u} \in \mathcal{P}} \left(   \sum\limits_{i=1}^{g}  \min(u_i \alpha, (d-\sum\limits_{j=1}^{i-1} u_j) \beta)  \right) \triangleq \min\limits_{\mathbf{u} \in \mathcal{P}} f(\mathbf{u}),
\end{align}
\noindent where
\begin{align}
\label{f_definition}
&f(\mathbf{u})= \sum\limits_{i = 1}^{g}  \min(u_i \alpha, (d-\sum\limits_{j=1}^{i-1} u_j) \beta),\\
& \mathcal{P}= \{ \mathbf{u}=[u_1,\ldots, u_g]: 1 \le u_i \le e, g \in \mathbb{N} \text{ such that } \sum\limits_{i=1}^g u_i =k   \}.
\label{def_P} 
\end{align}
\end{theorem}
Note that $g$ in \eref{def_P} corresponds to the support of $\mathbf{u}$, and it satisfies $\ceil{\frac{k}{e}} \le  g \le k$. We call the vector $\mathbf{u} \in \mathcal{P} $ a recovery \emph{scenario}.
\begin{IEEEproof}
Consider the {scenario} $\mathbf{u} \in \mathcal{P} $ as follows. A data collector DC connects to a subset of $k$ nodes $ \{ x_{out}^j : j \in I\}$, where $I$ is the set of $k$ contacted nodes. The size of the support of $\mathbf{u}$ corresponds to the number of repair groups of size $e$ taking part in the reconstruction process, while $u_i$ corresponds to the number of nodes contacted from repair group $i$. 

As all incoming edges of DC have infinite capacity, we only examine cuts $( U, \bar U)$ with $S \in U$ and  
$ \{ x_{out}^i : i \in I\} \subseteq \bar U$.
Every directed acyclic graph has a topological sorting, which is an ordering ``$<$'' of its vertices such that the existence of an edge $x \to y$ implies $x <y$. We recall that nodes within the same repair group are repaired simultaneously, hence it is possible that all input (or output) nodes in a repair group are adjacent in the the ordering. We thus order the $g$ repair groups connected to DC according to the sorting. Since nodes are sorted, nodes in the $i$-th repair group do not have incoming edges from nodes in the $j$-th repair group, with $j>i, i,j \in [g]$. 

Considering the $i$-th repair group, consider the case $|\left\{x_{\text{in}}^i \in U \right\}|=m$ and the remaining nodes are such that $x_{\text{in}}^i \in \bar U$.
\begin{itemize}
\item
	if $x_{\text{in}}^i \in U$, then the contribution of each node is $\alpha$. The overall contribution of these nodes is $m \alpha$.

\item
	else: $x_{\text{in}}^i \in \bar U$, then if $x_{\text{virt,out}}^i \in U$, the contribution of this node is $\infty$. Thus, we only consider the case $x_{\text{virt,out}}^i \in \bar U$. Then, we discuss two cases
	\begin{itemize}
	\item	if $x_{\text{virt,in}}^i \in U$, the contribution to the cut is $e \alpha$. 
	
	\item	else, since the $i$-th group is the topologically i-th repair group, at most $\sum\limits_{j=1}^{i-1} u_j$ edges come from output nodes in $\bar U$. The contribution is $(d-\sum\limits_{j=1}^{i-1} u_j) \beta$. Thus, the contribution of this node is $\min(e \alpha, (d-\sum\limits_{j=1}^{i-1} u_j) \beta)$. 
	Note that $x_{\text{virt,out}}^i \in \bar U$, we do not need to account for other similar nodes. 
	\end{itemize}
\end{itemize}
Hence, if $m = u_i$, the contribution of the i-th repair group is $u_i \alpha$. If $ m < u_i$, the contribution is $m \alpha+ \min(e \alpha, (d-\sum\limits_{j=1}^{i-1} u_j) \beta)$, which is minimized to be $\min(e \alpha, (d-\sum\limits_{j=1}^{i-1} u_j) \beta)$ when $m=0$. Thus, to lower the cut, either $m=u_i$ in the case of $(d-\sum\limits_{j=1}^{i-1} u_j) \beta > u_i \alpha $ or $m=0$ otherwise. The total contribution of the $i$-th repair group is then
\begin{align*}
\min(u_i \alpha, (d-\sum\limits_{j=1}^{i-1} u_j) \beta).
\end{align*}
Finally, summing all contributions from different repair groups and considering the worst case for $\mathbf{u} \in \mathcal{P}$ implies that
\begin{align*}
&\min\limits_{G \in \mathcal{G}} \min\limits_{t \in \text{DC}(G)} \text{mincut}_G(s,t)= \min\limits_{\mathbf{u} \in \mathcal{P}} \left(   \sum\limits_{i=1}^{g}  \min(u_i \alpha, (d-\sum\limits_{j=1}^{i-1} u_j) \beta)  \right),
\end{align*}
\noindent with $\mathcal{P}$ defined as in \eref{def_P}.
%Therefore, the existence of regenerating codes is guaranteed by \cite{ahlswede2000network} as long as \\
%$\mathcal{M} \le \min\limits_{G \in \mathcal{G}} \min\limits_{t \in \text{DC}(G)} \text{mincut}_G(s,t)$\footnote{Strictly speaking, this is only valid when the number of failures/repairs is bounded. 
The theorem follows according to the necessary and sufficient condition in \eref{min_cut_condition}.
%\bl{??remove??? A rigorous proof is required to drop the boundedness assumption as \cite{wu2007deterministic,closed_form_cooperative_regene}.}
\end{IEEEproof}
%\begin{remark}
Our characterization of Theorem \ref{min_cut_theorem} relies on the boundedness assumption of the total number of failures/repairs. A future direction is to investigate the correctness of Theorem \ref{min_cut_theorem} for arbitrary number of failures/repairs, similar to \cite{closed_form_cooperative_regene,wu2007deterministic}.
%\end{remark}
%In the sequel, we will use the notation $k= a e +r$, such that $a= \floor*{\frac{k}{e}}$ and $r= k \text{ mod } e$. 
%%%%%%%%%%%%%%%%%%%%%%%%%%%%%%%%%%%%%%%%%%%%%%%%%%%%%%%%%%%%%%%%%%%%%%%%%%%%%%%%%%%%%%%%%%%%%%%%%%%%%%%%%%%%%%%%%%%%%%%%%%%%%%%%%%%%%%%%%%%%%%%%%%%%%%%%%%%%%%%%%%%%%%%%%%%%%%%%%%%%%%%%%%%%
\subsection{Solving the minimum cut problem}
In this section, we derive the structure of the optimal scenario $\mathbf{u}$ in \eref{cut_general} for any set of parameters $(\alpha,\beta)$. For instance, we show that for $ m e < k \le (m +1) e$, the number of optimal repair groups $g^*$ (the support of $\mathbf{u}$) is equal to $m+1$. The result is formalized in the following theorem. Recall that we denote $\eta=\lfloor k/e \rfloor, r = k  - \eta e$.
\begin{theorem}
\label{optimal_cut_result}
For fixed system parameters $(\mathcal{M}, n,k,d,e,\alpha,\beta)$, functional regenerating codes satisfying the centralized multi-node repair condition exist if and only if
\begin{align}
&\mathcal{M} \le  f(\mathbf{u^*} )= \sum\limits_{i=1}^{\ceil{\frac{k}{e}}}  \min(u^*_i \alpha, (d-\sum\limits_{j=1}^{i-1} u^*_j) \beta), 
\end{align}
where
\begin{align}
\mathbf{u}^*=
\begin{cases}
  [k],  &\text{if } k \le e,\\
 [\underbrace{e,\ldots,e}_{\eta \text{ times}}], &\text{else if } k=\eta e, \\
[  r,\underbrace{e,\ldots,e}_{\eta  \text{ times}} ], &\text{else if }  k=\eta e + r \text{ and } \alpha \le \frac{d+\eta r - \eta e}{r} \beta, \\
[  \underbrace{e,\ldots,e}_{\eta  \text{ times}}, r ],&\text{otherwise}, \\
\end{cases}
\label{optimal_scenario}
\end{align}
where $0<r<e$.
\end{theorem}
%compare with related work. XXX have independently developed Theorem 1 or equivalent of Theorem 1. XX independently proved Theorem 2, except for the last case.... (and proof techniques are different.)
%We note that \eref{cut_general} was also independently developed in \cite{rawat2016centralized}.
Note that $[k]$ in \eref{optimal_scenario} means a vector with a single entry $k$. We note that \cite{rawat2016centralized,hu2015broadcast} have independently developed Theorem 1 or an equivalent of Theorem 1, without entirely characterizing the optimal solution.  
\cite{zhang2017concurrent} independently proved via a different approach Theorem 2, except for the last case in \eref{optimal_scenario}.  

We denote by $[\mathbf{v},\mathbf{u},\mathbf{w}]$ the vector that is the concatenation of the vectors $\mathbf{v},\mathbf{u},\mathbf{w}$. 
The next lemma shows that the minimum cut can be obtained by optimizing any subsequence of $\mathbf{u}$ first. The proof follows directly from the definition of $f()$ in \eref{f_definition} and is omitted.
\begin{lemma} 
\label{lemma:subseq}
Consider vectors $\mathbf{v},\mathbf{w},\mathbf{u}, \mathbf{u'}$ such that $\sum_i u_i  = \sum_i u'_i$. If
\begin{align}
f(\mathbf{u}) \ge f(\mathbf{u'}),
\end{align}
then,
\begin{align}
f([\mathbf{v},\mathbf{u},\mathbf{w}]) \ge f([\mathbf{v},\mathbf{u}',\mathbf{w}]).
\end{align}
\end{lemma}

In proving the result of \thref{optimal_cut_result}, we first characterize the optimal solution in the case of $k \le e$. Insight and intuition gained from this case are used to  motivate and derive the general optimal solution.
We first state the following lemma, which represents a key step towards proving our result.
\begin{lemma}
Let $\alpha,\beta$ be non-negative reals, $u_1, u_2,d,e,s,l$ be non-negative integers such that $u_1+u_2=s \le e$, then the following inequality holds
\label{jump_inequality}
\begin{align}
f([u_1,\underbrace{e,\ldots,e}_{l \text{ times}},u_2])
\geq \min (f([s,\underbrace{e,\ldots,e}_{l \text{ times}}]),f( [\underbrace{e,\ldots,e}_{l \text{ times}},s])),
\end{align}
where $f(\mathbf{u})$ is defined as in \eref{f_definition}.
%\begin{align}
%\label{f_definition}
%f(\mathbf{u})= \sum\limits_{i \geq 1}  \min(u_i \alpha, (d-\sum\limits_{j=1}^{i-1} u_j) \beta).
%\end{align}
%\begin{align}
%&\min(u_1 \alpha, d \beta)+ \sum\limits_{i=0}^l \min(e \alpha,  (d-i e - u_1) \beta    )\nonumber\\&
%+ 
% \min(u_2 \alpha, (d- (l+1) e - u_1) \beta) \geq \nonumber\\
%& 
%\min\left( 
%\min(s \alpha, d \beta)+ \sum\limits_{i=0}^l \min(e \alpha,  (d-i e - s) \beta    ),
%\right. \nonumber\\& 
%\left.
%  \sum\limits_{i=0}^l \min(e \alpha,  (d-i e ) \beta   )
%+ \min( s\alpha, (d-(l+1)  e ) \beta)
% \right)
%\end{align}
\end{lemma}
\begin{IEEEproof}
To prove the result, we cast it as an optimization problem:
\begin{align}
& \underset{\mathbf{u}=[u_1,u_2]}{\text{minimize}}\qquad
\min(u_1 \alpha, d \beta)+ \sum\limits_{i=0}^{l-1} \min(e \alpha,  (d-i e - u_1) \beta    ) +\min(u_2 \alpha, (d- (l+1) e - u_1) \beta) \nonumber\\
& \text{subject to}  \qquad
   0 \le u_1 \le s, \nonumber\\
& \qquad \qquad \qquad 0 \le u_2 \le e, \nonumber\\
&  \qquad \qquad \qquad u_1 + u_2 =  s. 
\label{jump_optimization}
\end{align} 
%
%\begin{equation}
%\begin{aligned}
%& \underset{\mathbf{u}=[u_1,u_2]}{\text{minimize}}
%& &\min(u_1 \alpha, d \beta)+ \sum\limits_{i=0}^l \min(e \alpha,  (d-i e - u_1) \beta    )
%+
% \min(u_2 \alpha, (d- (l+1) e - u_1) \beta) \\
%& \text{subject to}
%& & 0 \le u_1 \le s \\
%& & &0 \le u_2 \le e \\
%& & & u_1 + u_2 =  s. 
%\end{aligned}
%\label{jump_optimization}
%\end{equation} 
Substituting $u_2$ by $s-u_1$ in \eref{jump_optimization}, using the identity $\min(x,y)= \frac{x+y-|x-y|}{2}$ and after eliminating constant terms, \eref{jump_optimization} becomes equivalent to 
\begin{align}
 & \underset{ u_1 }{\text{minimize}}
 - u_1 l \beta - |u_1 \alpha - d \beta| - \sum \limits_{i=0}^{l-1}| e \alpha - d \beta 
 + i e \beta + u_1 \beta|  -
 | s \alpha - u_1 (\alpha-\beta) - (d-l e)\beta| \nonumber\\
 & \text{subject to} \qquad 0 \le u_1 \le s.
 \label{jump_optimization_2}
 \end{align}
The objective function in \eref{jump_optimization_2}, as a function of $u_1$, is concave over the interval $[0,s]$. The concavity is due to the convexity of $x \to |x|$. Therefore, the minimum is achieved at one of the extreme values. Equivalently, $u_1^*=s$ or $u_1^*=0$.
\end{IEEEproof}
\subsubsection{Case $k \le e$}
In this scenario, connecting to $k$ nodes from the same repair group yields the worst case scenario from an information flow perspective. 
%Intuitively, considering repair from a single set of failed nodes should decrease the amount of overall information as this repair set represents a bottleneck in the reconstruction process. We state the following lemma that will help confirm the intuition.
Given a particular repair scenario characterized by a vector $\mathbf{u}$, for any two adjacent repair groups (i.e., two adjacent entries in $\mathbf{u}$) with $u_1$ and $u_2$ nodes respectively, we have $u_1+ u_2 \le e$. One can combine these two groups into a single repair group to achieve a lower cut value. Indeed, from the cut expression in \eref{cut_general}, the contribution of the initial set $[u_1, u_2]$ to the cut is
$
\min( u_1 \alpha,  l \beta ) + \min( u_2 \alpha,  (l-u_1) \beta ),
$  
for some non-negative integer $l$. After combining the groups into a single repair group, the contribution of the newly formed repair group is 
$
\min( (u_1+u_2) \alpha,  l \beta )
$, which is lower than the initial contribution by virtue of \lref{jump_inequality}, thus achieving a lower cut. This means that starting from an IFG, we construct a new IFG that has one less repair group and lower min-cut value.
This process can be repeated until we end up with a single repair group consisting of $k \le e $ nodes, which corresponds to the minimum cut over all graphs in this case. 

Therefore, the tradeoff in \eref{cut_general} is simply characterized by $\mathcal{M} \le \min( k \alpha, d \beta)$. Moreover, $\alpha_{\text{MSMR}}=\alpha_{\text{MBMR}}= \frac{\mathcal{M}}{k}$ and $\beta_{\text{MSMR}}=\beta_{\text{MBMR}}=\frac{\mathcal{M}}{d}$. Equivalently, the functional storage bandwidth tradeoff reduces to a single point given by $(\alpha_{\text{MSMR}},\beta_{\text{MSMR}})=(\alpha_{\text{MBMR}},\beta_{\text{MBMR}})=(\frac{\mathcal{M}}{k},\frac{\mathcal{M}}{d})$.
\subsubsection{Case $e<k $}
Motivated by the previous case, the intuition is that, given a scenario $\mathbf{u}$, one should form a new scenario which exhibits as many groups of size $e$ as possible. Subsequently, one constructs a scenario $\mathbf{u}$ such that all its entries, except maybe one entry, are equal to $e$. \lref{jump_inequality} addresses the case $u_1+u_2 \le e$. Generalizing it to the case where $e \le u_1+u_2 \le 2e$ follows the same approach. 
\begin{lemma}
\label{move_part}
Let $\alpha,\beta$ be non-negative reals, $u_1, u_2,d,e,s,l$ be non-negative integers such that $u_1+u_2=e+s$ and $0 \le u_1,u_2,s \le e$. Then, the following inequality holds
\begin{align}
f([u_1,\underbrace{e,\ldots,e}_{l \text{ times}},u_2])
\geq \min (f([s,\underbrace{e,\ldots,e}_{l+1 \text{ times}} ]),
f([ \underbrace{e,\ldots,e}_{l+1 \text{ times}},s])),
\end{align}
\noindent where $f(\mathbf{u})$ is defined as in \eref{f_definition}.
%\begin{align}
%&\min(u_1 \alpha, d \beta)+ \sum\limits_{i=0}^l \min(e \alpha,  (d-i e - u_1) \beta    )
%\nonumber\\ &\qquad \qquad+
% \min(u_2 \alpha, (d- (l+1) e - u_1) \beta)  \nonumber\\ 
%&  \geq \min(A,B),
%\end{align}
%where $A=\min(s \alpha, d \beta)+ \sum\limits_{i=0}^l \min(e \alpha,  (d-i e - s) \beta    )
%+\min(e \alpha, (d- (l+1) e - s) \beta)$ and $B=\min(e \alpha, d \beta)+ \sum\limits_{i=0}^l \min(e \alpha,  (d-i e - e) \beta    )+\min(s \alpha, (d- (l+1) e - e) \beta)$.
\end{lemma}
\begin{IEEEproof}
First, we notice that $u_1=e+s-u_2 \geq s$ as $u_2 \le e$. Then, the proof follows along similar lines as that of \lref{jump_inequality} by replacing the constraint in \eref{jump_optimization_2} by $s \le u_1 \le e$. 
\end{IEEEproof}
For a fixed $\beta$, we denote the cut corresponding to
$\mathbf{u}=[\underbrace{e,\ldots,e}_{j \text{times}}, r,\underbrace{e,\ldots,e}_{\eta-j \text{ times}} ]$, as a function of $\alpha$, by $C_j(\alpha), j=0,\ldots,\eta$. 
As will be shown later in the proof of \thref{optimal_cut_result}, a careful analysis of the behavior of the $\eta+1$ different scenarios $C_j(\alpha), 0 \le j \le \eta,$ is needed to determine the overall optimal scenario. We state the result in the following lemma, whose proof is relegated to \appref{General_case}.
\begin{lemma}
\label{alpha_c_lemma}
Assume $e\nmid k$. There exists a real number $\alpha_c(\eta)\in [\frac{d }{e}\beta,\frac{d }{ r}\beta]$ such that, for any $0\le j \le \eta$, 
\begin{align}
\label{c_j_compare}
C_j(\alpha)  \begin{cases}
\geq C_0(\alpha), & \text{if } \alpha \le \alpha_c(\eta),\\
\geq C_\eta(\alpha), & \text{if } \alpha \ge \alpha_c(\eta),
\end{cases}
\end{align}
\noindent with
\begin{align}
\label{alpha_c_a}
\alpha_c(\eta)= \frac{d+\eta r - \eta e}{r} \beta.
\end{align}
\end{lemma}
\begin{IEEEproof}[Proof of \thref{optimal_cut_result}]
Now that we have the necessary machinery, we proceed as follows: given any scenario $\mathbf{u}$, we keep combining and/or changing repair groups by means of successive applications of
\lref{jump_inequality} and \lref{move_part} on subsequences of $\mathbf{u}$ until we can no longer reduce the minimum cut. By Lemma \ref{lemma:subseq} we reduced the overall minimum cut. The algorithm terminates because at each step, either the number of repair groups in $\mathbf{u}$ is reduced by one, or the number of repair groups of full size $e$ is increased by one. As the number of repair groups is lower bounded by $\eta+1$, and as the number of repair groups of full size $e$ is upper bounded by $\eta$, the algorithm must terminate after a finite number of steps. 
It can be seen then that the above reduction procedure has a finite number of outcomes, given by
\begin{itemize}
\item	$\mathbf{u}=[\underbrace{e,\ldots,e}_{\eta \text{ times}}]$ if $k=\eta e$,

\item	$\mathbf{u}=[\underbrace{e,\ldots,e}_{j \text{times}}, r,\underbrace{e,\ldots,e}_{\eta-j \text{ times}} ]$ when $k=\eta e +r$,\\\noindent with $0<r<e$ and $j \in \{ 0, \ldots, \eta\}$.
\end{itemize}
Therefore, if $e \mid k$, then the optimal scenario corresponds to considering exactly $\eta$ repair groups. On the other hand, if $e \nmid k$, then, it is optimal to consider exactly $\eta+1$ repair groups. However, the optimal position of the repair group with $r$ nodes needs to be determined. Then, using \lref{alpha_c_lemma}, the result in \thref{optimal_cut_result} follows. 
\end{IEEEproof}
\begin{example}
Let $\mathbf{u}=[1,3,2,3,2 ]$ with $e=3$. Then, one can start by reducing the first three repair groups $[1,3,2]$. This leads to $\mathbf{u}=[3,3,3,2 ]$. Another approach would be to consider the last three repair groups $[2,3,2]$. Reducing this vector leads to either $\mathbf{u}=[1,3,3,3,1 ]$ or $\mathbf{u}=[1,3,1,3,3 ]$. Reducing further $\mathbf{u}=[1,3,3,3,1 ]$ leads to $\mathbf{u}=[2,3,3,3 ]$ or $\mathbf{u}=[3,3,3,2]$. Reducing $\mathbf{u}=[1,3,1,3,3 ]$ leads to 
$\mathbf{u}=[3,2,3,3 ]$ or $\mathbf{u}=[2,3,3,3 ]$. It remains to compare the cuts given by $\mathbf{u}=[3,3,3,2]$, $\mathbf{u}=[3,3,2,3 ]$, $\mathbf{u}=[3,2,3,3]$ and $\mathbf{u}=[2,3,3,3]$. Following \thref{optimal_cut_result}, either $\mathbf{u}=[2,3,3,3]$ or $\mathbf{u}=[3,3,3,2]$ gives the lowest min-cut.
\end{example}
%%%%%%%%%%%%%%%%%%%%%%%%%%%%%%%%%%%%%%%%%%%%%%%%%%%%%%%%%%%%%%%%%%%%%%%%%%%%%%%%%%%%%%%%%%%%%%%%
\subsection{Explicit expression of the tradeoff}
Having characterized the optimal scenario generating the minimum cut in the last section, we are now ready to state the admissible storage-repair bandwidth region for the centralized multi-node repair problem, the proof of which is in \appref{tradeoff_1}. 
%
%\bl{In the expressions below, is $\alpha^*$  in terms of $\beta$ or $\gamma$? Please use consistent notation.???}
\begin{theorem}
\label{tradeoff_expression} 
For an $(\mathcal{M}, n,k,d,e,\alpha,\beta)$ storage system, there exists a threshold function $\alpha^*(\mathcal{M},n,k,d,e,\gamma)$ such that for any $\alpha \geq \alpha^*(\mathcal{M},n,k,d,e,\gamma)$, regenerating codes exist. For any $\alpha < \alpha^*(\mathcal{M},n,k,d,e,\gamma)$, it is impossible to construct codes achieving the target parameters. The threshold function $\alpha^*(\mathcal{M},n,k,d,e,\gamma)$ is defined as follows:
\\
$\bullet$ if $k \le e$, then: $\alpha^* = \frac{\mathcal{M}}{k}, \ \gamma \in [  \mathcal{M} ,+\infty)$,
\\
$\bullet$ if $k=\eta e, \eta \ge 2$, then:
\begin{align}
      \alpha^*=\begin{cases}
              \frac{\mathcal{M}}{k} ,  
         			& \gamma \in [f_0(\eta-1), +\infty ), \\
              \frac{  \mathcal{M}- \gamma g_0(i)}  { ie}  ,  
              &\gamma \in [  f_0(i-1) ,   f_0(i)],
               i=\eta-1,\ldots 1,
    \end{cases}
\end{align}
 $\bullet$ if $k=\eta e+ r$ with  $\eta\ge 1, 1 \le r \le e-1$, then:
\begin{align}
\label{tradeoff_e_not_k}
      \alpha^*=\begin{cases}
              \frac{\mathcal{M}}{k}, 
         			 & \gamma \in [f_r(\eta-1), +\infty ), \\
              \frac{  \mathcal{M}- \gamma g_r(i)}  {r+ie}  , 
      &        \gamma \in [  f_r(i-1) ,   f_r(i)]
               , i=\eta-1,\ldots 1,\\
               \frac{  \mathcal{M}- \gamma g_r(0)}  {r}   ,  
        &   \gamma \in [ \frac{d \mathcal{M}}{ (\eta+1) d -  e {\eta+1 \choose 2} } , f_r(0) ],
    \end{cases}
\end{align}
where
\begin{align}
\label{f_function_tradeoff}
f_r(i)&=\frac{   2 e d \mathcal{M} }{-k^2-r^2+e (k-r) +2 k d-e^2 (i^2+i)-2i e r }, \\
g_r(i)&=\frac{(\eta-i)(-2 r + e + 2 d - \eta e - e i) }{2 d} .
\end{align}
\end{theorem}
% \begin{IEEEproof}
%See \appref{tradeoff_1}.
% \end{IEEEproof}

\begin{center}
\begin{figure}
\center
\includegraphics[width=.7\linewidth]{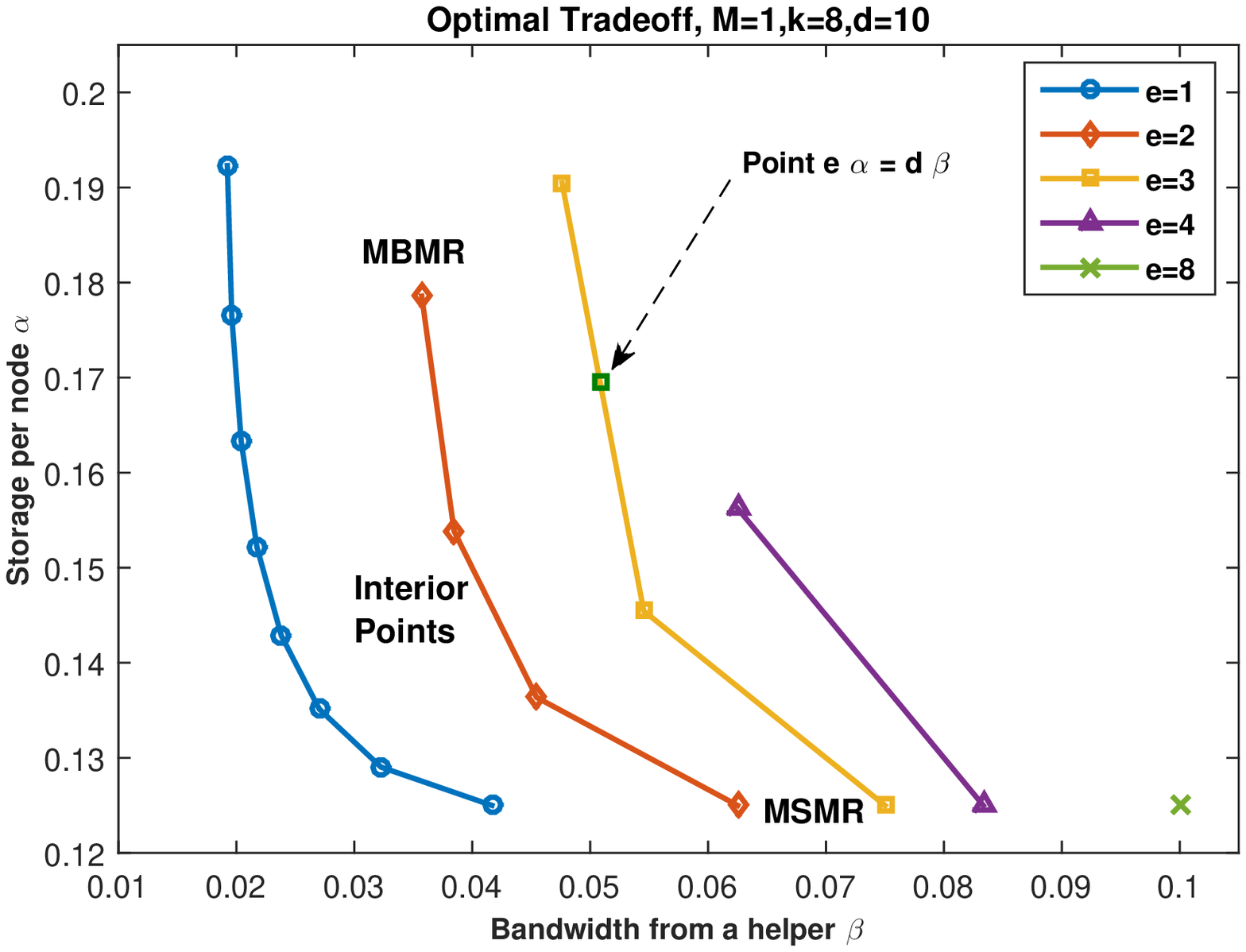}
\caption{Multi-node repair tradeoff: $k=8,d=10, \mathcal{M}=1, e \in \{ 1,2,3,4,8\}$. When $e \nmid k$, the point with $e \alpha= d \beta$ is not the MBMR point.
%\bl{1. Mark extreme points for a single curve?? Currently They are marked for different curves, but not all. 2. On one of the curves with $e \nmid k$, mark the point with $e \alpha= \gamma$, show that it is not the MBMR.}
}
\label{multi_node_example}
 \end{figure}
 \end{center}

The functional repair tradeoff is illustrated in \fref{multi_node_example} for multiple values of $e \in \{1,2,3,4,8 \}$ and $k=8,d=10,$ $\mathcal{M}=1$.
\begin{remark}
In the case of $e|k, e|d$, the following equality holds for all points on the tradeoff
\begin{align*}
\mathcal{M}  &= \sum\limits_{i=0}^{\eta-1} \min (e \alpha, ( d -ie) \beta) 
 \iff \frac{\mathcal{M}}{e}= \sum\limits_{i=0}^{\eta-1} \min (\alpha, (\frac{d}{e}-i) \beta).
\end{align*}
Therefore, the tradeoff between $\alpha$ and $\beta$ is the same as the single erasure tradeoff of a system with reduced parameters given by $\frac{\mathcal{M}}{e}$, $\frac{k}{e}=\eta$ and $\frac{d}{e}$. The expression of the tradeoff in this case can be recovered from \cite{dimakis2010network} with the appropriate parameters.
\end{remark}

We now have the expressions of the two extreme points on the optimal tradeoff. We focus on the case $e< k$, as otherwise the optimal tradeoff reduces to a single point.

\vspace{0.3em}
\noindent {\bf MSMR.} The MSMR point is the same irrespective of the relation between $k$ and $e$, and it is given by 
\begin{align}
\alpha_{\text{MSMR}}=\frac{\mathcal{M}}{k}, \gamma_{\text{MSMR}} = 
\frac{\mathcal{M}}{k}   \frac{e d}{d-k+e} .
\label{MSMR_code_parameters}
\end{align}
\vspace{0.3em}

\noindent {\bf MBMR.}
Interestingly, the MBMR point depends on whether $e$ divides $k$ or not. 
 
$\bullet$ If $k=\eta e$, we obtain
\begin{align}
\label{MBMR_gamma_e_k}
\gamma_{\text{MBMR}}&= \frac{2 e d \mathcal{M}}{-k^2+ek+2 k d}=\frac{d \mathcal{M}}{d \eta- e \binom{\eta}{2}},\\
\label{MSMR_gamma_e_k}
\alpha_{\text{MBMR}}&= \frac{\gamma_{\text{MBMR}}}{e}.
\end{align} 
The amount of information downloaded for repair is equal to the amount of information stored at the $e$ replacement nodes. This property of the MBMR point is similar to the minimum bandwidth point in the single erasure case \cite{dimakis2010network} and also the minimum bandwidth cooperative repair point \cite{closed_form_cooperative_regene}. 

\begin{center}
\begin{figure}
\center
\includegraphics[width=.6\linewidth]{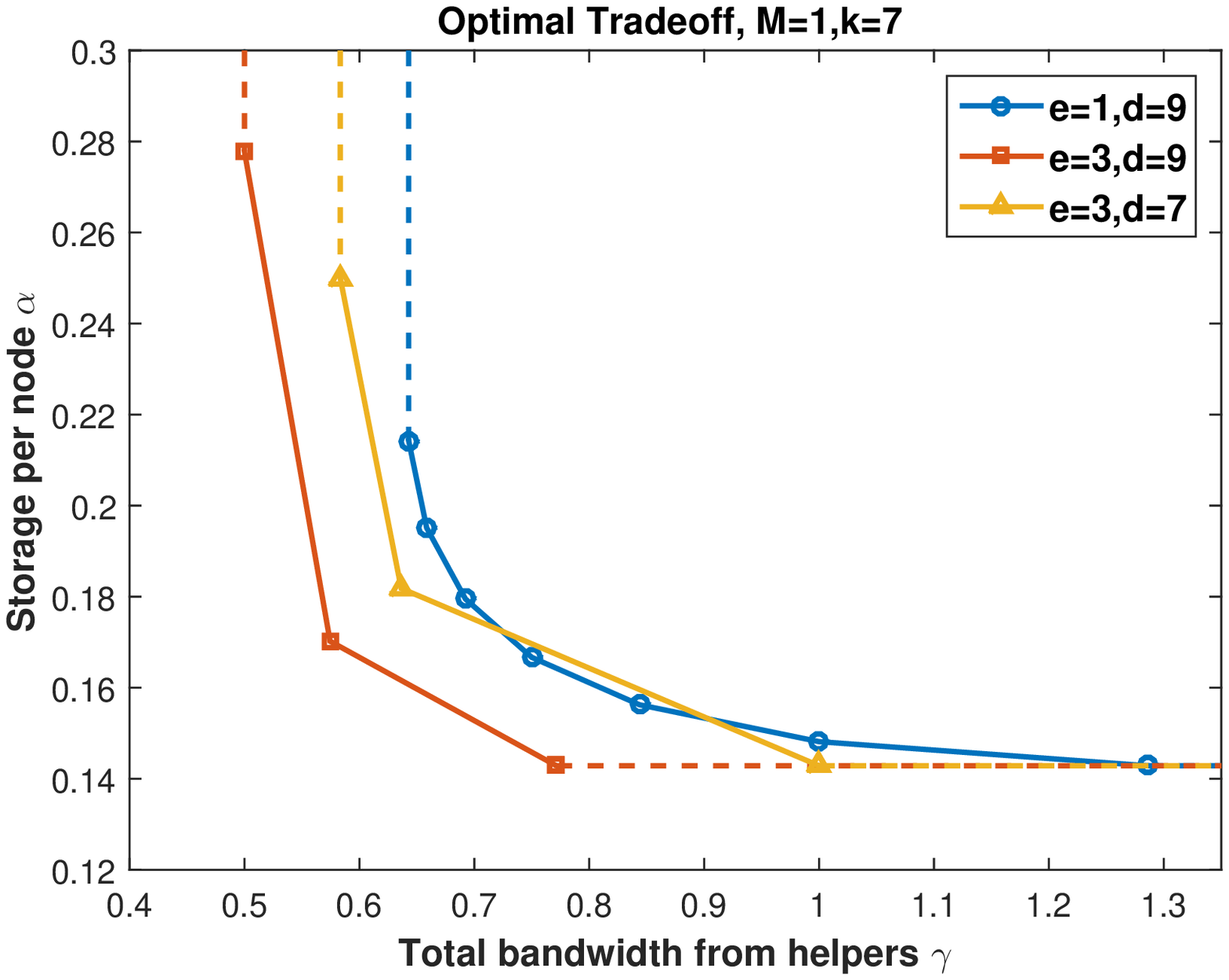}
\caption{Centralized multi-node repair vs separate repair strategy: $k=7, \mathcal{M}=1$. Separate repair strategy uses $d=9$ to repair 3 nodes successively while multi-node repair is plotted for $d=7$ and $d=9$.}
\label{comparison}
 \end{figure}
 \end{center}

$\bullet$ If $k=\eta e + r$, we obtain
\begin{align}
\label{MBMR_gamma_e_not_k}
\gamma_{\text{MBMR}}&=\frac{2 e d\mathcal{M} }{(k-r+e)(2d-k+r)}= \frac{d \mathcal{M}}{d (\eta+1)- e \binom{\eta+1}{2}},\\
\alpha_{\text{MBMR}}&=\gamma_{\text{MBMR}} 
\frac{d+\eta r-e \eta}{r d}.
\label{MSMR_gamma_e_not_k}
\end{align}
This situation is novel for multiple erasures as the $e$ nodes need to store more than the overall downloaded information. This is an extra cost in order to achieve the low value of the repair bandwidth. \fref{multi_node_example} illustrates this situation with $e=3, k=8$.  
%\bl{Point to Figure \ref{multi_node_example}.} 
However, later we will see that for both $e|k$ and $e \nmid k$, the total bandwidth at MBMR is equal to the entropy of the failed nodes (see \lref{entropy_lemma} and \lref{propo_multipl}):
\begin{align}
H(W_{E}) = d \beta = \gamma,
\end{align}
where $E \subset [n]$ is any subset of nodes of size $e$ and $W_{E}$ is the information stored across the nodes in $E$.
\begin{remark}
From the statement of \thref{optimal_cut_result}, we note that if we only consider points between the MSMR and the MBMR points, then the scenario $\mathbf{u}=[r,e,\dots,e]$ always generates the lowest cut. In fact, the scenario $\mathbf{u}=[e,\dots,e,r]$ corresponds to points beyond  the MBMR point, namely, points with $\alpha \geq \alpha_{\text{MBMR}}, \beta = \beta_{\text{MBMR}}$.
 \end{remark}
\begin{remark}
\label{rmk:comparison}
We compare the centralized repair scheme repairing $e$ nodes to a separate strategy repairing each of the $e$ nodes separately using single erasure regenerating codes. 
We fix $k,\alpha$ and $\mathcal{M}$. \\
{\bf Case I:} both strategies use $d$ helpers.
The separate strategy requires a total bandwidth given by $e d \beta_1$, while the centralized repair requires $d \beta_e$, where the subscript indicates the number of erasures repaired at a time. For simplicity, we assume that $e \mid k$. The case $e \nmid k$ can be treated in a similar way. For points on the multi-node repair tradeoff, we have
\begin{align*}
\mathcal{M}=\sum\limits_{j=0}^{\eta-1} \min( e \alpha, (d-je) \beta_e).
\end{align*}
Consider a point with the same $\alpha$ and $d$ on the single erasure tradeoff, we write

\begin{align*}
\mathcal{M} =\sum\limits_{j=0}^{\eta-1} \min( e \alpha, (d-je) \beta_e) =\sum\limits_{j=0}^{k-1} \min( \alpha, (d-j) \beta_1)  =  \sum\limits_{j=0}^{\eta-1} \sum\limits_{i=0}^{e-1} \min( \alpha, (d-i-je) \beta_1)  &\le \sum\limits_{j=0}^{\eta-1} e \min( \alpha, (d-je) \beta_1) \\
&= \sum\limits_{j=0}^{\eta-1}  \min( e \alpha, (d-je) e \beta_1).
\end{align*}
It follows that $\beta_e \le e \beta_1$ with equality if and only if $e=1$.
%Moreover, we note that $\beta_{e,\text{min}}=\frac{2 e \mathcal{M}}{-k^2+ e k + 2 kd } \le e \beta_{1,\text{min}}=\frac{2 e \mathcal{M}}{-k^2+   k + 2 kd }  $. \bl{??what is $\beta_{\min}$?? Why need it?} 
Therefore, for any storage capacity $\alpha$, multi-node repair requires strictly less bandwidth than a separate strategy for the same number of helpers $d$. \\
{\bf Case II: } multi-node repair uses $d-e+1$ helpers, and separate repair uses $d$ helpers. In this case, the original number of available nodes that can serve as helpers is assumed to be $d$, and $e \ge 1$ erasures occur within the available nodes. 
Then a separate strategy may require a smaller bandwidth for some values of $\alpha$, as illustrated by \fref{comparison}. However, as $d$ is sufficiently large, we observe numerically that multi-node repair with $d-e+1$ helpers performs better than a separate strategy for all values of $\alpha$.
Moreover, for the MSMR point, the separate repair bandwidth is $ed\beta_{1, \text{MSR}}= e \frac{\mathcal{M}}{k} \frac{d}{d-k+1}$, and centralized repair bandwidth is
$(d-e+1)\beta_{e,\text{MSMR}}=\frac{\mathcal{M}}{k}   \frac{e (d-e+1)}{d-k+1}$. It follows that a centralized repair is always better that a lazy repair strategy, specifically, for $e>2$,
\begin{align}
\frac{ (d-e+1)\beta_{e,\text{MSMR}} }{ed\beta_{1, \text{MSR}}}= \frac{d-(e-1)}{d}<1.
\end{align} 
 \end{remark}
 
\section{Exact-repair MSMR codes constructions}\label{S3}
In the remainder of the paper, we study exact repair. In this section, we first analyze the case $e \geq k$ and then construct MSMR codes when $e<k$. In later sections, we study the feasibility of MBMR codes and the interior points under exact repair for $e<k$.
\subsection{Construction when $k \le e$}\label{construction}
In the case of $k \le e$, the optimal tradeoff reduces to a single point, so our MSMR construction in this section is also an MBMR code. The optimal parameters satisfy $\alpha=\frac{\mathcal{M}}{k}, \beta=\frac{\mathcal{M}}{d}$ and $\gamma=\mathcal{M}$. We note that the overall repair bandwidth $d \beta$ and the reconstruction bandwidth $k\alpha$ are the same. Therefore, one can achieve $\alpha$ and $\gamma$ by dividing the data into $k$ symbols and encoding them using an $(n,k)$ MDS code (for example, a Reed-Solomon code). The repair can be done by downloading the full content of any $k$ out of $d$ helpers while not using $d-k$ helpers. Such repair is asymmetric in nature. 
We describe one alternative approach for achieving the repair with equal contribution from $d$ helpers.
\begin{enumerate}
\item	Divide the original file into $k d$ symbols (that is $\mathcal{M}=k d$) and encode them using an $(nd,kd)$ MDS code.

\item	Store the encoded symbols at $n$ nodes, such that each node stores $\alpha=d$ encoded symbols.

\item	For reconstruction, from any $k$ nodes, we obtain $k d$ different symbols. By virtue of the MDS property, we can reconstruct the data.

\item	For repair, each helper node transmits any $\beta=\frac{\mathcal{M}}{d}=k$ symbols. The replacement nodes receive $d k$ different coded symbols, which are sufficient to reconstruct the whole data and thus regenerate the missing symbols.
\end{enumerate}
\begin{remark}
The above procedure works for a specific predetermined $d$. However, it can be generalized to support any value of $d$ satisfying $k \le d \le n-e$. For instance, let $\delta=  \text{lcm}(k,k+1,k+2,\ldots,n-e)$ (lcm denotes the least common multiple). Assume $\mathcal{M}=k \delta$. The file of size $\mathcal{M}$ is then encoded using an $(n \delta, k \delta)$ MDS code. Each node stores $\alpha= \frac{\mathcal{M}}{k}= \frac{ k \delta}{k}= \delta$ coded symbols. For repair with $d$ helpers, for any $k \le d \le n-e$, each node transmits any $\beta= \frac{\mathcal{M}}{d}= \frac{k \delta}{d}$ coded symbols for his node. Similarly, it can be seen that reconstruction is always feasible. 
%Note that the constraint of the field size arises from the need for an $(n \delta, k \delta)$ MDS code. The field size needs to be no less than $n \delta$, e.g. Reed Solomon codes.
%, where $r=n-k$. Note that $r \geq e$ as we should have $k \le n-e$ to be able to repair.
\end{remark}
%%%%%%%%%%%%%%%%%%%%%%%%%%%%%%%%%%%%%%%%%%%%%%%%%%%%%%%%%%%%%%%%%%%%%%%%%%%%%%%%%%%%%%%%%%%%%%%%%%%%%%%%%%%%%%%%%%%%%%%%%%%%%%%%%%%%%%%%%%%%%%%%%%%%%%%%%%%%%%%%%%%%%%%%%%%%%%%%%%%%%%%%%%%%%%%%%%%%%%%%%
\subsection{Minimum storage codes framework}
\label{approach_s1ection}
In the following subsections, we discuss an explicit MSMR code construction method using existing MSR codes designed for single failures for $k>e$. We first describe the general framework, and then present two specific codes.

The framework described in this section has been developed in \cite{li2015enabling} for numerical simulations. We present it here in a formal and analytical way. Consider an instance of an exact linear $(n,k,d,\alpha,\beta )$ MSR code, where $\beta=\frac{\alpha}{d-k+1}$. Consider $e$ nodes, indexed by $f_1,\ldots,f_e$, and other distinct $d-e+1$ nodes, indexed by $h_{1},\ldots,h_{{d-e+1} }$, such that $d-e+1 \geq k$. Let $\mathcal{H} =\{f_1,\ldots,f_e,h_{1},\ldots,h_{{d-e+1} }\}$ and define $\mathcal{H} _{f_j}= \mathcal{H} \backslash \{ f_j\}$. 
Consider the single-node repair algorithm corresponding to failed node $f_j$ and helper nodes $\mathcal{H}_{f_j}$. We denote by $\mathbf{s}_{h,f_j}^{\mathcal{H}  }$ the information sent by node $h$ to repair node $f_j$, for helpers $h \in \mathcal{H}_{f_j}$. We drop the superscript $\mathcal{H} $ when it is clear from the context. The size of $\mathbf{s}_{h,f_j}$ is $\beta$ symbols.

Now we construct an $(n,k,d-e+1,e,\alpha, e\beta)$ MSMR code.
Upon failure of the $e$ nodes $f_1,\ldots,f_e$, the centralized node carrying the repair connects to the set of $d-e+1$ helpers $h_{1},\ldots,h_{{d-e+1} }$. Each helper node $h_{i}$ transmits $e \beta$ symbols given by
$
\cup_{j =1  }^e  \{ \mathbf{s}_{h_i,f_j}  \}.
$
One can check that the parameters of an MSMR code in \eref{MSMR_code_parameters} are satisfied with equality. 
%Under this framework, two equivalent approaches can be carried out to to decode the failed nodes:
%\subsubsection{Direct approach}
%Let $\mathbf{w}_{f_i}$ denotes the storage content of node $i$. Then, one can combine $
%\cup_{i=1}^{d-e+1} \{ \cup_{j =1  }^e  \{ s_{h_i,f_j}  \} \}$ to form a system of linear equations given by
%
%\begin{align}
%A \begin{bmatrix}
%x_{f_1}\\
%\vdots \\
%x_{f_e}
%\end{bmatrix} =b,
%\label{direct_approach}
%\end{align}
%where $A$ is a known matrix and $b$ is a known vector. Thus, one can recover the lost nodes by solving \eref{direct_approach} when $A$ is non-singular.
%\subsubsection{Indirect approach}

The approach consists in using the underlying MSR repair procedure for each of the $e$ failed nodes. Note that $\mathbf{s}_{h_i,f_j} $ can be obtained from the $d-e+1$ helpers, for $i \in [d-e+1]$. To this end, %for every set of $e$ failed nodes $\{f_1,\dots,f_e\}$, and every set of $d-e+1$ helper nodes $\{h_1,\dots,h_{d-e+1}\}$, 
the MSR repair procedure requires $\mathbf{s}_{f_i,f_j }^{\mathcal{H}}$ for all $\{ (i,j):  i,j\in [e], i \neq j \}$, which we treat as unknowns. Let ${E}_{i,j}(\cdot)$ denote the encoding function used to encode the information sent from node $h_i$ to node $f_j$. Also, let ${D}_i(\cdot)$ denote the decoding function used by the MSR code to repair node $f_i$ given information from $d$ helpers. Then, we write
\begin{align}
\mathbf{s}_{f_i, f_j}^{\mathcal{H}}&= {E}_{i,j}(\mathbf{w}_{f_i}) \nonumber\\
&={E}_{i,j}( {D}_i( \mathbf{s}_{h,f_i}^{\mathcal{H}}, h \in \mathcal{H}_{f_i}) ),
\label{indirect_approach}
\end{align}
\noindent where $\mathbf{w}_j$ denotes the content of node $j$, $i,j\in [e], i \neq j$. 
%\bl{??What if $\beta>1$?? The number of equations/unknowns should be multiplied by $\beta$, where we treat $s_{i,j}$ as a column vector of length $\beta$, and $E_{i,j}$ returns a vector as well??} 
Equation \eref{indirect_approach} generates $e(e-1)\beta$ linear equations in $e(e-1)\beta$ unknowns. Let $\mathbf{s}$ be a vector containing the unknowns $\mathbf{s}_{f_i,f_j }$. Then, we seek to form a system of linear equations as
\begin{equation}
A \mathbf{s} =\mathbf{b},
\label{indirect_approach_2}
\end{equation}

\noindent where $A$ is a known $(e(e- 1)\beta \times e(e- 1)\beta)$ matrix and $b$ is a known $(e(e- 1)\beta\times 1)$ vector.
If $A$ is non-singular, one can thus recover $\mathbf{s}$. Then, the centralized node can recover the failed node $\mathbf{w}_{f_i}$ as $\mathbf{w}_{f_i}= {D}_i( \mathbf{s}_{h,f_i}, h \in \mathcal{H}_{f_i}) $. We adopt the above framework throughout the section. 
%%%%%%%%%%%%%%%%%%%%%%%%%%%%%%%%%%%%%%%%%%%%%%%%%%%%%%%%%%%%%%%%%%%%%%%%%%%%%%%%%%%%%%%%%%%%%%%%%%%%%%%%%%%%%%%%%%%%%%%%%%%%%%%%%%%%%%%%%%%%%%%%%%%%%%%%%%%%%%%%%%%%%%%%%%%%%%%%%%%%%%%%%%%%%%%%%%%%%%%%%%%%%%%%%%%%%%%%%%%%%%%%%%%%%%%%%%%%%%%%%%%%%%%%%%%%%%%%%%%%%%%%%%%%%%%%%%%%%%%%%%%%%%%%
%\begin{remark}
%
%If the failures are not simultaneous, one may use a lazy repair strategy in which every single failure is repaired using the maximum number of helpers . In the this scenario, repairing $e$ consecutive failures generates a bandwidth given by $\gamma_{lazy}= e \frac{\mathcal{M}}{k} \frac{n-1}{n-k}$. Repairing the $e$ failures simultaneously by contacting $d_e$ helpers results in 
%$\gamma_{\text{MSMR}}=\frac{\mathcal{M}}{k}   \frac{e d_e}{d_e-k+e}$. It follows that a centralized multi-repair is better that a lazy repair if $d_e$ is sufficiently high. For instance, we write
%\begin{align}
% \gamma_{\text{MSMR}} \le \gamma_{lazy}& \iff \frac{d_e}{d_e-k+e} \le \frac{n-1}{n-k} \nonumber\\&\iff d_e \geq \frac{(n-1)(k-e)}{k-1}.
% \label{lazy_repair_strategy}
%\end{align}
%Using the approach described above for repairing $e$ erasures, the number of helpers contacted for repairing $e$ helpers is $d_e=d-e+1$. One can check that \eref{lazy_repair_strategy} is satisfied. This implies that a simultaneous repair of multiples failures is beneficial. Moreover, the gain in bandwidth is given by 
%\begin{align}
%\frac{ \gamma_{\text{MSMR}}}{\gamma_{lazy}}= \frac{n-e}{n-1}.
%\end{align} 
%\end{remark}
\begin{remark}
While the described framework applies to codes with arbitrary rates, we focus in the sequel on low-rate codes. High-rate MSMR constructions have been presented in \cite{ye2017explicit}. However, in the low-rate regime, our constructions perform better. 
%\bl{Low-rate MSR constructions generally have a smaller storage size $\alpha$ compared to high-rate constructions. ?? Is \cite{ye2017explicit} the best sub-packetization? The order of $\alpha$ should be $r^{n/r} = k^2$?? } 
For instance, for a target MSMR code with rate $\frac{1}{2}$, the construction in \cite{ye2017explicit} yields a storage size $\alpha=k^{2k-1}$, while applying the above approach to IA codes \cite{suh2011exact} or to PM codes \cite{Rashmi_Product_Matrix} results in a smaller storage size $\alpha=k $ and $\alpha=k-1$, respectively.
%The above approach can be applied to any single erasure MSR code. High-rate MSMR with large subpacketization have been constructed in the literature \cite{ye2017explicit,wang2016optimal}. Our focus is on the low-rate MSR codes, as these codes have low subpacketization size and lower encoding and decoding complexities.
\end{remark}
\subsection{Product-matrix codes}\label{PM_codes}
In this subsection, we construct MSMR codes for any $e$ erasures based on product-matrix (PM)  codes \cite{Rashmi_Product_Matrix}. The PM framework allows the design of MBR codes for any value of $d$ and the design of MSR codes for $d \geq 2 k-2$. Moreover,
the PM construction offers simple encoding and decoding and ensures optimal repair of all nodes. Product-matrix MSR codes are a family of scalar MSR codes, i.e., $\beta=1$. We first focus on the case $d=2k-2$.
Under this setup, $\alpha=d-k+1=k-1$. The codeword is represented by an $(n\times \alpha)$ code matrix $C$ such that its $i^{\text{th}}$ row corresponds to the $\alpha$ symbols stored by the  $i^{\text{th}}$ node. The code matrix is given by
\begin{equation}
C = \Psi M,
\Psi = \begin{bmatrix}
\Phi & \Lambda \Phi
\end{bmatrix}, 
M =\begin{bmatrix}
S_1\\S_2  
\end{bmatrix},
\end{equation}
\noindent where $\Psi$ is an $(n \times d)$ encoding matrix and $M$ is a $(d \times \alpha)$ message matrix. $S_1$ and $S_2$ are $(\alpha \times \alpha)$ symmetric matrices constructed such that the $ {\alpha+1 \choose 2}  $ entries in the upper-triangular part of each of the two matrices are filled up by ${\alpha+1 \choose 2} $ distinct message symbols. $\Phi$ is an $(n\times \alpha)$ matrix and $\Lambda$ is an $(n \times n)$ diagonal matrix. The elements of $\Psi$ should satisfy:
\begin{enumerate}
\item	any $d$ rows of $\Psi$ are linearly independent;
\item	any $\alpha$ rows of $\Phi$ are linearly independent;
\item	the $n$ diagonal elements of $\Lambda$ are distinct.
\end{enumerate}
The above conditions may be met by choosing $\Psi$ to be a Vandermonde matrix, in which case its $i^{\text{th}  }$ row is given by 
$\mathbf{\psi}_i^t= \begin{bmatrix}
1 & \lambda_i & \cdots & \lambda_i^{d-1}
\end{bmatrix}$. It follows that $\Lambda= \text{diag} \{ \lambda_1^\alpha,\ldots,\lambda_n^\alpha \}$. In the following, we assume that $\Psi$ is a Vandermonde matrix. 
%In the following, we first recall the repair procedure of a single erasure from \cite{Rashmi_Product_Matrix}, then we extend it to multiple erasures.

\vspace{0.3em}
\noindent {\bf Repair of a single erasure in PM codes.}
The single erasure repair algorithm \cite{Rashmi_Product_Matrix} is reviewed below. Let $\mathbf{w}_i^t$ denote the content stored at a failed node. Let $\mathbf{\phi}_i^t$ be the $i^\text{th}$ row of $\Phi$. Then, $\mathbf{w}_i^t=\mathbf{\psi}_i^t M = \begin{bmatrix}
\mathbf{\phi}_i^t & \lambda_i^\alpha \mathbf{\phi}_i^t
\end{bmatrix} M=\mathbf{\phi}_i^t S_1 +  \lambda_i^{\alpha} \mathbf{\phi}_i^t S_2 $. Let $\mathcal{H}_i=\{h_1,\ldots,h_{d} \}$ denote the set of $d$ helpers. Each helper $h$ transmits $s_{h,i} = \mathbf{w}_h^t \mathbf{\phi}_i =   \mathbf{\psi}^t_{h} M \mathbf{\phi}_i$ to the replacement node, who obtains $\Psi_{\mathcal{H}_i} M \mathbf{\phi}_i$, where $\Psi_{\mathcal{H}_i}^t= \begin{bmatrix}
\mathbf{\psi}_{h_1}& 
\cdots   &
\mathbf{\psi}_{h_d}
\end{bmatrix}.$
%\begin{align*}
%\Psi_{\mathcal{H}_f}= \begin{bmatrix}
%\mathbf{\psi}^t_{h_1}\\ 
%\vdots   \\
%\mathbf{\psi}^t_{h_d}
%\end{bmatrix}.
%\end{align*}
Note that $\Psi_{\mathcal{H}_i}$ is invertible by construction. Thus, using the symmetry of $S_1$ and $S_2$, we obtain 
$
(M \mathbf{\phi}_i)^t= \begin{bmatrix}
\mathbf{\phi}_i^t S_1  & \mathbf{\phi}_i^t S_2  
\end{bmatrix}$. We can then reconstruct $\mathbf{w}_i^t=\mathbf{\phi}_i^t S_1 + \lambda_i^\alpha \mathbf{\phi}_i^t S_2 $.

\vspace{0.3em}
\noindent {\bf Repair of multiple erasures in PM codes.}
%In decoding multiple erasures in PM codes, we use the indirect approach described in \sref{approach_section}. 
Given the symmetry of PM codes, we can assume w.l.o.g that nodes in $\mathcal{E}= \{1,\ldots,e \}$ have failed. Define $\mathcal{E}_i=\mathcal{E} \backslash \{i \}$.  Let $\mathcal{H}=\{1,\ldots,d+1\}$. The centralized node connects to helper node $h \in \{ e+1,\ldots,d+1 \}$, and obtains $\{ \mathbf{\psi}_h^t M \mathbf{\phi}_j, j \in \mathcal{E} \}$.

Let $\mathbf{s}= [s_{1,2}, s_{2,1},\ldots, s_{1,e},s_{e,1},\ldots, s_{e-1,e},s_{e,e-1}]^t$.
Our goal is to express explicitly $A$ and $\mathbf{b}$ as in \eref{indirect_approach_2}.

Consider the repair of node $i \in \mathcal{E}$ by the set of helpers in $\mathcal{H}_i=\mathcal{H} \backslash \{i \}$. From the previous subsection, we write 

\begin{align}
\mathbf{w}_i &= \begin{bmatrix}I_\alpha & \lambda^{\alpha}_i  I_\alpha\end{bmatrix} \Psi_{\mathcal{H}_i}^{-1} \mathbf{s}_{\mathcal{H}_i}, \text{ such that }\\
\Psi_{\mathcal{H}_i}^t &= \begin{bmatrix}
\mathbf{\psi}_1  & 
\cdots  &
\mathbf{\psi}_{i-1}   & 
\mathbf{\psi}_{i+1}   & 
\cdots   &
\mathbf{\psi}_{d+1}  
\end{bmatrix},\\
\mathbf{s}_{\mathcal{H}_i}^t&=\begin{bmatrix}
s_{1,i} &
\cdots&
s_{i-1,i} &
s_{i+1,i} &
\cdots&
s_{d+1,i}
\end{bmatrix}.
\end{align}
\noindent It follows that

\begin{align}
s_{i,j} &= \mathbf{\phi}_j^t \mathbf{w}_i \\
&= \begin{bmatrix}\mathbf{\phi}_j^t & \lambda^{\alpha}_i  \mathbf{\phi}_j^t\end{bmatrix} \Psi_{\mathcal{H}_i}^{-1}( \sum_{l \in \mathcal{H}_i } s_{l,i} \mathbf{e}_{l,i} )\\
&=    \sum_{l \in \mathcal{E}_i }  ( \begin{bmatrix}\mathbf{\phi}_j^t & \lambda^{\alpha}_i  \mathbf{\phi}_j^t \end{bmatrix} \Psi_{\mathcal{H}_i}^{-1} \mathbf{e}_{l,i} ) s_{l,i}  +  \sum_{l=e+1 }^{d+1}  ( \begin{bmatrix}\mathbf{\phi}_j^t & \lambda^{\alpha}_i  \mathbf{\phi}_j^t \end{bmatrix} \Psi_{\mathcal{H}_i}^{-1} \mathbf{e}_{l,i} ) s_{l,i},
\label{unknowns_equation}
\end{align}

\noindent Here, for $l \in [d+1] \backslash \{i \}$, we use the column standard basis $\mathbf{e}_l$ and define 
\begin{equation}
\mathbf{e}_{l,i} \triangleq 
  \begin{cases} 
      \mathbf{e}_l, & l<i , \\
      \mathbf{e}_{l-1}, & l > i.
   \end{cases}
\end{equation}
\noindent Note that the second term in \eref{unknowns_equation} is known from the helpers. Moreover, to compute \eref{unknowns_equation}, one may use the inverse of Vandermonde's matrix formula \cite{Knuth:1997:ACP:260999}. Let $h \in \{ 1,\ldots, d\} $, we have

\begin{align}
(\Psi_{\mathcal{H}_i}^{-1} \mathbf{e}_{l,i} )_h  =
\frac{\gamma_h(l,i)}{ \prod_{m \in \mathcal{H}_i \backslash \{ l\} } (\lambda_l-\lambda_m) } 
= \frac{\gamma_h(l,i)}{ \sum_{j=1}^{d} \gamma_j(l,i) \lambda^{j-1} },
 \label{vandermonde_inverse}
\end{align}

%  \sum\limits_{
%\substack{
%1 \le m_1 < \ldots < m_{d-h}\le d+1 ,\\ m_1,\ldots,m_{d-h} \in \mathcal{H}_i \backslash \{ l\}
%}
%} \lambda_{m_1}\ldots \lambda_{m_{d-h}
%}.
\noindent where the subscript $h$ in $(\cdot)_h$ means the $h$-th entry, and

\begin{equation}\label{eq_gamma}
\gamma_h(l,i)=  (-1)^{d-h}
\sum\limits_{ m_1 < \ldots < m_{d-h} \in \mathcal{H}_i \backslash \{ l\} } \lambda_{m_1}\ldots \lambda_{m_{d-h}}.
\end{equation}
\noindent As $\prod\limits_{m \in \mathcal{H}_i \backslash \{ l\} } (\lambda-\lambda_m)= \sum_{h=1}^d \gamma_h(l,i) \lambda^{i-1}$, we obtain

\begin{align} \label{eq_A_entry}
\begin{bmatrix}\mathbf{\phi}_j^t & \lambda^{\alpha}_i  \mathbf{\phi}_j^t \end{bmatrix} \Psi_{\mathcal{H}_i}^{-1} \mathbf{e}_{l,i}
= \frac{ \sum_{h=1}^\alpha ( \gamma_h(l,i) +    \lambda^\alpha_i \gamma_{h+\alpha} (l,i) )\lambda_j^{h-1} }{ \sum_{h=1}^{d} \gamma_h(l,i) \lambda_l^{h-1} }.
\end{align}

Therefore, one can construct $A$ and $\mathbf{b}$ in \eref{indirect_approach_2} as follows:
\begin{itemize}
\item	The entries of $\mathbf{b}$ are indexed with $(i,j)$, corresponding to $s_{i,j}$. The entry of $\mathbf{b}$ at index $(i,j)$ is given by \\$ \sum_{l=e+1 }^{d+1}  ( \begin{bmatrix}\mathbf{\phi}_j^t & \lambda^{\alpha}_i  \mathbf{\phi}_j^t \end{bmatrix} \Psi_{\mathcal{H}_i}^{-1} \mathbf{e}_{l,i} ) s_{l,i}$.
\item	Index the $e(e-1$) rows (and columns respectively) of A with $(i,j)$. $A$ has zero in all entries except: For every row in $A$ indexed by $(i,j)$:
\begin{itemize}
\item	 the entry at column indexed by $(i,j)$ is -1.

\item	for $l \in \mathcal{E}_i$, the entry at column indexed by $(l,i)$ is given by $\begin{bmatrix}\mathbf{\phi}_j^t & \lambda^{\alpha}_i  \mathbf{\phi}_j^t \end{bmatrix} \Psi_{\mathcal{H}_i}^{-1} \mathbf{e}_{l,i}$ as in \eqref{eq_A_entry}.
\end{itemize}
\end{itemize} 
%Once $A$ is constructed, if $A$ is non-singular, then $s$ is obtained as $\mathbf{s}= A^{-1} b$.
%\subsubsection{2 Erasures case}
%we state the result in the following theorem.

For clear presentation, we first prove the existence of product-matrix MSMR codes for 2 erasures, and then prove the result for general $e$.
\begin{theorem} \label{thm_PM_2_erasures}
There exists $(n,k,2k-3,2,k-1,2)$  product-matrix MSMR codes, defined over a large enough finite field, such that any two erasures can be optimally repaired.  
\end{theorem}
\begin{IEEEproof}
In this case, the matrix $A$ is given by 

\begin{equation}
A=\begin{bmatrix}
-1 &
\begin{bmatrix}
 \mathbf{\phi}_2^t & \lambda^{\alpha}_1  \mathbf{\phi}_2^t 
\end{bmatrix}  \Psi_{\mathcal{H}_1}^{-1} \mathbf{e}_{2,1} \\
\begin{bmatrix}
 \mathbf{\phi}_1^t & \lambda^{\alpha}_2  \mathbf{\phi}_1^t   
\end{bmatrix}\Psi_{\mathcal{H}_2}^{-1} \mathbf{e}_{1,2}  & -1
\end{bmatrix}.
\end{equation}

\noindent From \eref{vandermonde_inverse},  noting that $\mathcal{H}_1 \backslash \{ 2\}= \mathcal{H}_2 \backslash \{ 1\}$, we obtain 

\begin{align}
|A|  
 &= 1-  \begin{bmatrix}
 \mathbf{\phi}_2^t & \lambda^{\alpha}_1  \mathbf{\phi}_2^t   
\end{bmatrix} \Psi_{\mathcal{H}_1}^{-1} \mathbf{e}_{2,1}
 \begin{bmatrix}
 \mathbf{\phi}_1^t & \lambda^{\alpha}_2  \mathbf{\phi}_1^t  
\end{bmatrix}\Psi_{\mathcal{H}_2}^{-1} \mathbf{e}_{1,2} \\
&=
%1- \frac{1}{\prod\limits_{m \in \mathcal{H}_1 \backslash \{ 2\} } (\lambda_2-\lambda_m)   } 
%\frac{1}{  \prod\limits_{m \in \mathcal{H}_2 \backslash \{ 1\} } (\lambda_1-\lambda_m) } \nonumber\\
1 - \frac{ ( \sum\limits_{h=1}^\alpha \lambda_2^{h-1} ( \gamma_h(1,2)+    \lambda^\alpha_1 \gamma_{h+\alpha}(1,2) ) ) }{   \sum\limits_{h=1}^d \lambda_2^{h-1} \gamma_h(1,2)  \sum\limits_{h=1}^d \lambda_1^{h-1} \gamma_h(1,2)  }    
 (
  \sum\limits_{h=1}^\alpha \lambda_1^{h-1} (   \gamma_h(1,2)+   \lambda^\alpha_2 \gamma_{h+\alpha}(1,2) ))\\
&\triangleq 1-   \frac{N(\lambda_1,\ldots,\lambda_{d+1})}{D(\lambda_1,\ldots,\lambda_{d+1})}  .
\end{align}
\noindent$|A|$ can be viewed as a rational function of $(\lambda_1,\ldots,\lambda_{d+1})$, as $N $ and $D$  
 are polynomials in $(\lambda_1,\ldots,\lambda_{d+1})$. We want to show that the following polynomial is not zero:

%\begin{footnotesize}
%\begin{align}
%&P(\lambda_1, \ldots,\lambda_{d+1})
%\nonumber\\&\triangleq  ( \prod\limits_{m \in \mathcal{H}_1 \backslash \{ 2\} } (\lambda_2-\lambda_m)  )
% ( \prod\limits_{m \in \mathcal{H}_2 \backslash \{ 1\} } (\lambda_1-\lambda_m)  )  |A|
%\label{polynomial_expression}
%\end{align}
%\end{footnotesize}

\begin{align}
P(\lambda_1, \ldots,\lambda_{d+1}) &\triangleq  D(\lambda_1, \ldots,\lambda_{d+1}) |A|\\
&= D(\lambda_1, \ldots,\lambda_{d+1})- N(\lambda_1, \ldots,\lambda_{d+1}).
\label{polynomial_expression}
\end{align}
%&=   ( \prod\limits_{m \in \mathcal{H}_1 \backslash \{ 2\} } (\lambda_2-\lambda_m)  )
%  ( \prod\limits_{m \in \mathcal{H}_2 \backslash \{ 1\} } (\lambda_1-\lambda_m)  ) \nonumber\\
%  &-  \sum\limits_{h=1}^\alpha \lambda_2^{h-1} ( (-1)^{d-h} \gamma_h+  (-1)^{d-h-\alpha} \lambda^\alpha_1 \gamma_{h+\alpha} ) )
% (
%\sum\limits_{h=1}^\alpha \lambda_1^{h-1} ( (-1)^{d-h} \gamma_h+ (-1)^{d-h-\alpha} \lambda^\alpha_2 \gamma_{h+\alpha} ). 

%As $d=2 \alpha$, we note that $\gamma_{h+\alpha}=\gamma_{\alpha-h}$. 
%Expanding $N$ and $D$, one can observe that all the terms in each polynomial have the same sum degree of $2(d-1)$. 
\noindent Let $y_\alpha= (-1)^{\alpha} \lambda_3 \cdots \lambda_{\alpha+2}$, $y_{\alpha-1}= (-1)^{\alpha-1} \lambda_3 \cdots \lambda_{\alpha+1}$. 
%Then, it can be seen that $D$ contains the term $y_\alpha y_{\alpha-1} ( \lambda_1^{\alpha} \lambda_2^{\alpha-1} +  \lambda_2^{\alpha} \lambda_1^{\alpha -1})$.
%, which is the unique term in $N$ in $(\lambda_1^2,\ldots,\lambda_{\alpha+1}^2,\lambda_{\alpha+2})$. 
%Similarly, $N$ contains $y_\alpha y_{\alpha-1} ( \lambda_1^{d-1}+ \lambda_2^{d-1}  )$. 
Then, it can be seen that $P$ contains the term
%It follows the unique term in $(\lambda_1,\ldots,\lambda_{\alpha+2})$ in $P$ is given by 

\begin{align*}
y_\alpha   y_{\alpha-1}  (\lambda_1^{d-1} +\lambda_2^{d-1} - \lambda_1^{\alpha} \lambda_2^{\alpha-1} -  \lambda_2^{\alpha} \lambda_1^{\alpha -1} ),
\end{align*}
\noindent which is not zero. 
%It can be seen that term $\lambda_1^{d-1} \lambda_2^{d-1}$ in $P(\lambda_1, \ldots,\lambda_{d+1})$ appears only once and thus cannot be canceled. For instance, the maximum degree of $\lambda_1$ in the second term of \eref{polynomial_expression} is $\alpha$.
Hence, $P(\lambda_1, \ldots,\lambda_{d+1})$ is a non-zero polynomial.
The PM construction, when based on a Vandermonde matrix, requires $\lambda_1^\alpha \neq \lambda_2^\alpha$ \cite{Rashmi_Product_Matrix}, or equivalently, $g(\lambda_1,\lambda_2) \triangleq \lambda_1^\alpha- \lambda_2^\alpha \neq 0$. 
Let $Q(\lambda_1,\dots,\lambda_n)$ denote the polynomial obtained by varying the set of helpers and failure patterns, taking the product of all corresponding polynomials $P$, and also multiplied by all $g$ for all pairs of two nodes. Then, $Q$ is not identically zero. By Combinatorial Nullstellensatz \cite{alon1999combinatorial}, we can find assignments of
the variables $\{\lambda_1,\ldots, \lambda_n \}$ over a large enough finite field, such that the 
polynomial is not zero. Equivalently, we can guarantee the successful optimal repair of any two erasures among the $n$ storage nodes.
\end{IEEEproof}
\begin{theorem} \label{thm_PM_general}
There exists $(n,k,2k-e-1,e,k-1,e)$ product-matrix MSMR codes, defined over a large enough finite field, such that any $e$ erasures can be optimally repaired.  
\end{theorem}
\begin{IEEEproof}
Entries in each column indexed by $s_{i,j}$ in $A$ is either $-1$ or some other $(e-1)$ non-zero entries whose denominator is the same and given by $\prod\limits_{m \in \mathcal{H} \backslash \{i \}} (\lambda_i -\lambda_m) $. We multiply this common denominator to all entries in the column $s_{i,j}$, for all pairs $i \neq j$. When $\lambda_i$'s are chosen to be distinct, this does not change the singularity of $A$. Denote this transformed matrix by $B$. Using \eqref{eq_A_entry}, the entry of $B$ in row $(i,j)$ and column $(l,m)$ is a polynomial in $\lambda_1,\dots,\lambda_{d+1}$:

\begin{align}
 B_{(i,j),(l,m)}  
= &\begin{cases}
 -\sum_{h=1}^d \gamma_h(i,j) \lambda_i^{h-1}, & l=i,m=j, \\
 \sum_{h=1}^\alpha \big( \gamma_h(l,i) \lambda_j^{h-1} + \gamma_{h+\alpha}(l,i) \lambda_i^\alpha \lambda_j^{h-1} \big), & m=i, \\
 0, & \textrm{otherwise.}
\end{cases} \nonumber
\end{align} 
\noindent Notice that $e+\alpha-1 \le k-1+\alpha-1= d-1$.
Let $y= (-1)^{\alpha-1} \lambda_{e+1} \cdots \lambda_{e+\alpha-1}$, which is a term in $\gamma_{\alpha+1}(i,j)$ for all $(i,j)$ by \eqref{eq_gamma}. We observe that there is a single term $\pm y\lambda_i^\alpha$ in the polynomial $B_{(i,j),(l,m)}$ for the non-zero entries of $B$.

Recall that the Leibniz formula for determinant of a $(m \times m)$ matrix $B$ is given by

\begin{equation}
|B|= \sum\limits_{\sigma} \text{sgn}  (\sigma)  \prod\limits_{i} b_{\sigma(i),i},
\label{determinant_expression}
\end{equation} 
\noindent where $\sigma$ is a permutation from the permutation group $S_m$,  sgn is the sign function of permutations, and $b_{i,j}$ is the entry $(i,j)$ of $B$.

{\bf Claim 1.} The term $T=\prod_{i=1}^{e} (y\lambda_i^\alpha)^{e-1}$ in $|B|$ has a non-zero coefficient.
 
Claim 1 implies that $|B|$ is not a zero polynomial.  Then, proceeding as in the proof in Theorem \ref{thm_PM_2_erasures}, by Combinatorial Nullstellensatz \cite{alon1999combinatorial}, we can find assignments of
the variables $\{\lambda_1,\ldots, \lambda_n \}$ over a large enough finite field, such that the 
code guarantees optimal repair of any set of $e$ erasures.

Next, we prove Claim 1. Note that the term $T$ can be created if and only if we take the single term $\pm y\lambda_i^\alpha$ in the non-zero entries of $B$ (depending on the permutation $\sigma$). Therefore, it is easy to see that the coefficient of term $T$ in $|B|$ is the determinant of the following $(e(e-1) \times e(e-1))$ matrix $C$
%$ \in \mathbb{R}^{e(e-1) \times e(e-1)}$:
\begin{align}
C_{(i,j),(l,m)}  
= \begin{cases}
 -1, & l=i,m=j, \\
 1, &  m=i, \\
 0, & \textrm{otherwise.}
\end{cases}
\end{align} 
One can verify that $C$ is diagonalizable, and the eigenvalues satisfy:
\begin{itemize}
\item Eigenvalue $e-2$ has multiplicity 1, with the corresponding (right) eigenvector $(1,1,\dots,1)^t$.
\item Eigenvalue $-2$ has multiplicity $e-1$, with the corresponding eigenspace $\{(x_{1,2},\dots,x_{e,e-1})^t: x_{1,j}=x_{1,2}, \forall j\in[e] \backslash\{1\},
x_{i,j}=x_{i,1},\forall i\in[e]\backslash\{1\}, \forall j\in[e] \backslash\{i\},  $ $ x_{1,2}+ \sum_{i=2}^e x_{i,1}=0  \}$ of dimension $e-1$.
\item Eigenvalue $-1$ has multiplicity $e(e-2)$, with the corresponding eigenspace $\{(x_{1,2},\dots,x_{e,e-1})^t: \sum_{1 \le i \le e, i \neq j} x_{i,j}=0, \forall j \in [e]\}$ of dimension $e(e-2)$.
\end{itemize}
To ensure that $|C| \neq 0$, a sufficient condition is to require the finite field to have a characteristic greater than 2 such that the elements $\{e-2,-2,-1,0 \}$ are pairwise distinct. In this case, the eigenvalues of $C$ are non-zero, and $|C| \neq 0$. Therefore, Claim 1 is proved and the theorem statement follows. 
We note that the sufficient condition on the finite field applies only to our proof 
%of \thref{thm_PM_general} 
and is not necessary for the existence of PM codes. Indeed, as it will be shown in Example \ref{example: PM}, we can construct PM codes with optimal multi-node repair property over finite fields of characteristic 2.
\end{IEEEproof}
\begin{remark}
\label{all_e_PM}
There exists product-matrix MSMR codes, defined over a large enough finite field, that simultaneously repair any $e \in [n-k]$ erasures with optimal bandwidth. 
Indeed, let $\widetilde{Q}=\prod_{e=2}^{n-k} Q_e$, where $Q_e$ is the polynomial corresponding to the code constraints for $e$ erasures. Recall that the reconstruction process for PM codes requires that $ \alpha_i^\alpha- \alpha_j^\alpha \neq 0 $ for $\alpha_i \neq \alpha_j$. Let $g(\lambda_1,
\ldots,\lambda_n)= \prod_{  1 \le i< j \le n }(\lambda_j^\alpha -\lambda_i^\alpha)$. 
Let $Q(\lambda_1,
\ldots,\lambda_n)= g(\lambda_1,
\ldots,\lambda_n) \widetilde{Q}(\lambda_1,
\ldots,\lambda_n) $. 
By \thref{thm_PM_2_erasures} and \thref{thm_PM_general}, $Q$ is not zero and the result follows by Combinatorial Nullstellensatz.
\end{remark}
\begin{example} \label{example: PM}
Consider the product-matrix code with $n=11,k=6,d=10$, $\alpha=5$. The code is defined over $\mathbb{F}_{2^6}$ with $\Lambda= \text{diag}\{\lambda_1^\alpha,\ldots,\lambda_{11}^\alpha\}$ and  $\lambda_i=g^{i-1}$ with $g$ being the generator of the multiplicative group of $\mathbb{F}_{2^6}$. Recall that with the above choice of $\lambda_i$, any field of size at least $n \alpha=55$ is sufficient to meet the PM code requirements \cite{Rashmi_Product_Matrix}.  
We first consider repair of $e=2$ erasures. One can check that out of the ${11 \choose 2}=55$ possible 2 failure patterns, 2 patterns are not recoverable according to \eqref{indirect_approach_2}: $\mathcal{E} \in \{ \{1,2\},\{10,11\} \} $. Considering the same code structure, for $e=3$ erasures, one observes that out of the ${11 \choose 3}=165$ possible 3 failure patterns, 5 patterns are not recoverable: $\mathcal{E} \in \{ \{ 1 ,    2 ,   11\},\{2,3,7\},\{2,4,8\},\{3,4,7\},\{5,9,10\} \} $. It is worth noting that a lazy repair strategy can be beneficial in the following way: if nodes 10 and 11 failed, i.e., $\mathcal{E}=\{10,11\} $, then, one can optimally repair any 3 erasures  $\mathcal{E}
\in \{  \{i, 10,11\}, i\neq 10, i\neq 11\}$.
Finally, as suggested by \thref{thm_PM_2_erasures} and \thref{thm_PM_general}, we find that increasing the underlying field size to $\mathbb{F}_{2^8}$ suffices to ensure optimal repair of all two and three erasure patterns in this scenario.
\end{example}

\begin{remark}
Following the code shortening procedure described in \cite{Rashmi_Product_Matrix}, we construct an $(n,k,d-e+1,e,k-1,e)$ product-matrix MSMR code $\mathcal{C}$ with optimal repair for any $e \in [n-k]$ erasures such that $2 k -2 \le d \le n-1$. First, as described in Remark \ref{all_e_PM}, we consider an $(n+ (d-2k+2),k+(d-2k+2),d+(d-2k+2)-e+1, e, d-k+1,1)$ product-matrix MSMR code $\mathcal{C}^{'}$ in systematic form with varying $e \in [n-k]$. Note that the code $\mathcal{C}'$ exists because the parameters satisfy Theorem \ref{thm_PM_general}. The first $(d-2k+2)$ systematic nodes of $\mathcal{C}^{'}$ are set to zeros. Then, the target code $\mathcal{C}$ is formed by deleting the first $ (d-2k-2)$ rows in each code matrix of $\mathcal{C}^{'}$. It can be seen that the repair procedure for $e$ erasures in $\mathcal{C}$ can be done by invoking that of the original code $\mathcal{C}^{'}$, which leads to the result.
%
%Recall that in \cite{Rashmi_Product_Matrix}, an $(n,k,d,d-k+1,1)$ MSR code with parameters $ 2 k -2 \le d \le n-1$, is constructed from a systematic MSR code with parameters $(n+ (d-2k-2),k+(d-2k-2),d+(d-2k-2),d-k+1,1)$ and then applying shortening the code (setting the first $\alpha (d-2k-2)$ of the message symbols to zero). It follows that using the above procedure, our construction guarantees the existence of $(n,k,d-e+1,e,k-1,e)$ product-matrix MSMR codes, defined over a large enough finite field, guaranteeing optimal repair of any $e \in [n-k]$ erasures such that $2 k -2 \le d \le n-1$.
\end{remark}
\subsection{Interference alignment codes}\label{IA_codes}
In this subsection, we give explicit code coefficient conditions for optimal MSMR codes from IA codes \cite{suh2011exact} for $e=2, 3, 4$ erasures, and for any $e \le k$ erasures from only the systematic (or only the parity) nodes. Moreover, we show the existence of MSMR codes for any $e \le k$ erasures.

The scalar MSR IA code construction is based on interference alignment techniques. The code is systematic and defined over a finite field $\mathbb{F}_q$ with optimal repair bandwidth for the case $\frac{k}{n}\le \frac{1}{2}$ and $d \geq 2 k-1$. We focus on the case $n=2k, d=2k-1, \beta=1$.
%\subsection{case $n=2 k, d=2 k -1$}
In this scenario, the storage size is $\alpha= d-k+1=k$. 

{\bf Notation.} For an invertible matrix $B$, we define its inverse transpose to be $B'\triangleq (B^{-1})^{t}$. The columns of $B^{'}$ constitute the dual basis of the column vectors of $B$. Recall that $B_{i,j}$ denotes the $(i,j)$-th element of matrix $B$. We use the following symbols to denote the transmission of information during repair operations. 
\begin{itemize}
\item	$s_{i,j}$: from systematic node $i$ to parity node $j$.
\item	$r_{i,j}$: from systematic node $i$ to systematic node $j$.
\item	$\bar{s}_{i,j} $: from parity node $i$ to systematic node $j$.
\item	$\bar{r}_{i,j}$: from parity node $i$ to parity node $j$.
\end{itemize}

The IA code is constructed as below. Consider $k$ linearly independent vectors $\{ \mathbf{v}_1,\ldots, \mathbf{v}_k \}$, $\mathbf{v}_i \in \mathbb{F}_q^k, i \in [k]$. Let 
\begin{align}
V&=\begin{bmatrix}
\mathbf{v}_1,  \ldots,   \mathbf{v}_k
\end{bmatrix} ,U= \kappa^{-1} V^{'} P,
\label{code_parameters}
\end{align}
\noindent where every submatrix of the $(k \times k)$ matrix $P$ is invertible and $\kappa$ is an arbitrary non-zero constant in $\mathbb{F}_q$ satisfying $\kappa^2 -1\neq 0$. Let $\mathbf{w}_l, l\in [k]$ denote the content of systematic node $l$ and $\bar{\mathbf{w}}_i$ the content of parity node $i$ $,i\in [k]$. Let $\mathbf{u}_i,\mathbf{v}_i,\mathbf{u}_i^{'}, \mathbf{v}_i^{'}$ be the $i$-th column of $U,V,U',V'$, respectively. Then, by the construction in \cite{suh2011exact},
\begin{equation}
\bar{\mathbf{w}}_i^{t} = \sum\limits_{j=1}^k \mathbf{w}_j^t G_j^{(i)}, 
G_j^{(i)}  =   \mathbf{u}_i  \mathbf{v}_j^{t} +   P_{j,i} I ,
\end{equation}
\noindent such that the matrix $G_j^{(i)}$ indicates the encoding submatrix for parity node $i$, associated with information unit $j$,
and $I$ is the identity matrix of size $(k \times k)$ . 
%\bl{??The notation $B'$ for a matrix is the inverse transpose, but the notation $w'$ is for parity nodes. Is this confusing??}

\vspace{0.3em}
\noindent{\bf Repair of a systematic node.}
Assume that systematic node $l$ fails. The general repair procedure is described in \cite{suh2011exact}. In this section, we explicitly develop the exact expression of $\mathbf{w}_l$ as it is needed later in repairing multiple erasures. Each systematic node $j \in [k]\backslash \{l \}$ transmits $r_{j,l}=\mathbf{w}_j^t \mathbf{v}_l^{'}$. Each parity node $i \in [k]$ transmits $\bar{s}_{i,l} = {\bar{\mathbf{w}}_i }^{ t} \mathbf{v}_l^{'}$. Noting that $G_j^{(i)} \mathbf{v}_l^{'}=\mathbbm{1}_{\{ j=l\}} \mathbf{u}_i + P_{j,i}  \mathbf{v}_l^{'}$,
it follows that
\begin{equation}
\bar{s}_{i,l} 
= \mathbf{w}_l^t  (\mathbf{u}_i+ P_{l,i}\mathbf{v}_l^{'} ) + \sum\limits_{j \in [k] \backslash \{ l\}} P_{j,i} r_{j,l} .
\end{equation}
\noindent Canceling the interference from systematic nodes, and arranging the contributions of parity nodes in matrix form, we write 
\begin{align}
\begin{bmatrix}
\bar{s}_{1,l} -\sum\limits_{j \in [k] \backslash \{ l\}}  P_{j,1} r_{j,l}\\ 
\vdots\\ 
\bar{s}_{k,l} -\sum\limits_{j \in [k] \backslash \{ l\}}  P_{j,k} r_{j,l}\\ 
\end{bmatrix}
&=
\begin{bmatrix}
\mathbf{u}_1^t+ P_{l,1} \mathbf{v}_l^{'t}  & \\ 
\vdots\\ 
\mathbf{u}_k^t+ P_{l,k} \mathbf{v}_l^{'t} 
\end{bmatrix} \mathbf{w}_l,
%&=(U^t + \begin{bmatrix}
%  P_{l,1}\\
%\vdots\\
%P_{l,k}
%\end{bmatrix}\mathbf{v}_l^{'t})  \mathbf{w}_l
%\\
%&=( U^t+ P^t \mathbf{e}_l \mathbf{e}_l^t V^{'t}  ) \mathbf{w}_l=  \frac{1}{\kappa}  P^t (I +\kappa \mathbf{e}_l \mathbf{e}_l^t  ) V^{-1}  \mathbf{w}_l,
\end{align}
\noindent where the last equality is obtained by substituting $U$ by its expression in \eref{code_parameters}. Using the Sherman-Morrison formula, 
for an invertible square matrix $A$ of size $(k \times k)$ and vectors $\mathbf{u},\mathbf{v}$ of length $k$,
\begin{align}
(A+\mathbf{u}\mathbf{v}^t)^{-1} = A^{-1} - \frac{A^{-1}\mathbf{u}\mathbf{v}^t A^{-1}}{1+\mathbf{v}^t A^{-1} \mathbf{u}},
\end{align}
we obtain that $(\frac{1}{\kappa}  P^t (I +\kappa \mathbf{e}_l \mathbf{e}_l^t  ) V^{-1})^{-1}= U^{'} -  \frac{\kappa^2}{1+\kappa } V \mathbf{e}_l \mathbf{e}_l^t P^{'}$, it follows that

\begin{equation}
\mathbf{w}_l
=  ( U^{'} -  \frac{\kappa^2}{1+\kappa } V \mathbf{e}_l \mathbf{e}_l^t P^{'}  ) \begin{bmatrix}
\bar{s}_{1,l} -\sum\limits_{j \in [k] \backslash \{ l\}}  P_{j,1} r_{j,l}\\ 
\vdots\\ 
\bar{s}_{k,l} -\sum\limits_{j \in [k] \backslash \{ l\}}  P_{j,k} r_{j,l}\\ 
\end{bmatrix}.
\label{repair_systematic_expression}
\end{equation}
\vspace{0.3em}
\noindent{\bf Repair of a parity node.}
The repair of a parity node is optimally achieved through the duality property of IA codes resulting in a structure that is also conducive to interference alignment. Indeed, inverting the roles of parity and systematic nodes, it follows from \cite{suh2011exact} that 
\begin{equation}
\mathbf{w}_i = \sum\limits_{j=1}^k \bar{\mathbf{w}}_j^{t} G_j^{'(i)}; G_j^{'(i)} =  \frac{1}{1-\kappa^2} (\mathbf{v}_i^{'} \mathbf{u}_j^{'t} - \kappa^2 P_{i,j}^{' } I).
\end{equation}
\noindent Assume parity node $l$ fails, then systematic node $i$ transmits $s_{i,l}=\mathbf{w}_i^t \mathbf{u}_l$ and parity node $j$ sends $\bar{r}_{j,l}=\bar{\mathbf{w}}_j^{t} \mathbf{u}_l$. Note that $G_j^{'(i)} \mathbf{u}_l=\frac{1}{1-\kappa^2}(-\kappa^2 u_j + \mathbbm{1}_{ \{ j=l \}} \mathbf{v}_i^{'})$. It follows that
\begin{equation}
s_{i,l}
%&=  \sum\limits_{j=1}^k \bar{\mathbf{w}}_i^{t}\frac{1}{1-\kappa^2} (\mathbf{v}_i^{'} \mathbf{u}_j^{'t} - \kappa^2 P_{i,j}^{'} I) \mathbf{u}_l
% \\
%  &= \bar{\mathbf{w}}_l^{t} \frac{1}{1-\kappa^2} (\mathbf{v}_i^{'}- \kappa^2 p_i^{'(l)} \mathbf{u}_l) + \sum\limits_{j \in [k] \backslash \{ l\}} %%\frac{-\kappa^2}{1-\kappa^2}  P_{i,j}^{'} (\bar{\mathbf{w}}_j^{t}\mathbf{u}_l) 
%  \\
   = \frac{1}{1-\kappa^2} \bar{\mathbf{w}}_l^{t}  (\mathbf{v}_i^{'}- \kappa^2 P_{i,l}^{'} \mathbf{u}_l) + \sum\limits_{j \in [k] \backslash \{ l\}} \frac{-\kappa^2}{1-\kappa^2}  P_{i,j}^{'} \bar{r}_{j,l}.
\end{equation}
\noindent Combining information from different helpers, we obtain after simplification
\begin{align}
\begin{bmatrix}
s_{1,l} +\frac{ \kappa^2}{1-\kappa^2} \sum\limits_{j \in [k] \backslash \{ l\}}   P_{1,j}^{'}  \bar{r}_{j,l}\\
\vdots\\
s_{k,l} +\frac{ \kappa^2}{1-\kappa^2} \sum\limits_{j \in [k] \backslash \{ l\}}    P_{k,j}^{'} \bar{r}_{j,l}
\end{bmatrix} &=
\frac{1}{1-\kappa^2} 
\begin{bmatrix}
\mathbf{v}_1^{'t} - \kappa^2 P_{1,l}^{'} \mathbf{u}_l^{t} \\
\vdots\\
\mathbf{v}_k^{'t} - \kappa^2 P_{k,l}^{'} \mathbf{u}_l^{t}
\end{bmatrix} \bar{\mathbf{w}}_{l}= \frac{\kappa}{1-\kappa^2}   P^{'} (I- \kappa \mathbf{e}_l \mathbf{e}_l^{t})U^t \bar{\mathbf{w}}_l,
\end{align}
\noindent where the last equality is obtained by replacing $V^{-1}= \kappa P^{'} U^t$. Inverting the system of equations and using the Sherman-Morrison formula, we obtain

\begin{equation}
\bar{\mathbf{w}}_l= ( (1-\kappa^2)V+ (1+\kappa) U^{'} \mathbf{e}_l \mathbf{e}_l^{^t} P^t )\begin{bmatrix}
s_{1,l} +\frac{ \kappa^2}{1-\kappa^2} \sum\limits_{j \in [k] \backslash \{ l\}}   P_{1,j}^{'}  \bar{r}_{j,l}\\
\vdots\\
s_{k,l} +\frac{ \kappa^2}{1-\kappa^2} \sum\limits_{j \in [k] \backslash \{ l\}}    P_{k,j}^{'} \bar{r}_{j,l}
\end{bmatrix}.
\label{repair_parity_expression}
\end{equation}
\vspace{0.3em}

\noindent\textbf{Repair of multiple erasures.} The goal is to construct the system of linear equations as in \eref{indirect_approach_2}. We need to derive the equations relating the information transferred across the failed systematic and parity nodes according to \eqref{indirect_approach}. Consider a systematic node $l \in [k]$ and a parity node $m \in [k]$, from \eref{repair_systematic_expression}, we write
\begin{align}
s_{l,m}&= \mathbf{u}_m^t \mathbf{w}_l\\
&= ( \mathbf{u}_m^t U^{'} - \frac{\kappa^2}{1+\kappa } \mathbf{u}_m^t    V \mathbf{e}_l \mathbf{e}_l^t P^{'}  ) \begin{bmatrix}
\bar{s}_{1,l} -\sum\limits_{j \in [k] \backslash \{ l\}}  P_{j,1} r_{j,l}\\ 
\vdots\\ 
\bar{s}_{k,l} -\sum\limits_{j \in [k] \backslash \{ l\}}  P_{j,k} r_{j,l}\\ 
\end{bmatrix}
\\
&= ( e_m^t -  \frac{\kappa }{1+\kappa }   P_{l,m}\mathbf{e}_l^t P^{'} ) \begin{bmatrix}
\bar{s}_{1,l} -\sum\limits_{j \in [k] \backslash \{ l\}}  P_{j,1} r_{j,l}\\ 
\vdots\\ 
\bar{s}_{k,l} -\sum\limits_{j \in [k] \backslash \{ l\}}  P_{j,k} r_{j,l}
\end{bmatrix}  \label{second_equality} \\
&= ( e_m^t -  \frac{\kappa }{1+\kappa }   P_{l,m} \mathbf{e}_l^t P^{'} )
(\sum\limits_{j \in [k]  }  \bar{s}_{j,l} \mathbf{e}_{j}   -\sum\limits_{j \in [k] \backslash \{ l\}}
 r_{j,l} P^t \mathbf{e}_{j}  )\label{third_equality} \\
 &= (1 - \frac{\kappa}{1+\kappa}  P_{l, m } P_{l, m }^{'})
 \bar{s}_{m,l} -   \sum\limits_{j \in [k] \backslash \{ m\} } 
(\frac{\kappa}{1+\kappa}  P_{l, m } P_{l, j }^{'} )\bar{s}_{j,l}    
- \sum\limits_{j \in [k] \backslash \{ l\}} 
P_{j,m}   r_{j,l}.
\label{coupling_expression_systematic_to_parity}  
\end{align}
\noindent Here \eref{second_equality} is obtained by noting that $U^t V= \frac{1}{\kappa} P^t$, and
\eref{coupling_expression_systematic_to_parity} follows using $P^{'} P^{t}=I$.
\noindent Similarly, consider two systematic nodes $l_1,l_2 \in [k], l_1 \neq l_2$, starting from 
 \eref{repair_systematic_expression} and noting that $ V^{'t} U^{'}= \kappa  P^{'}$, we obtain after simplification
%\begin{align}
%r_{l_1,l_2}&= \mathbf{v}_{l_2}^{'t} \mathbf{w}_{l_1} \\
%&= \mathbf{v}_{l_2}^{'t}  ( U^{'} -  \frac{\kappa^2}{1+\kappa } V \mathbf{e}_{l_1} \mathbf{e}_{l_1}^t P^{'}  ) \begin{bmatrix}
%\bar{s}_{1,{l_1}} -\sum\limits_{j \neq l_1}  p_{j,1}  r_{j,l_1}\\ 
%\vdots\\ 
%\bar{s}_{k,{l_1}} -\sum\limits_{j \neq l_1}  p_{j,k} r_{j,l_1}\\ 
%\end{bmatrix}\\
%&=  \kappa \mathbf{e}_{l_2}^t P^{'} \begin{bmatrix}
%\bar{s}_{1,{l_1}} -\sum\limits_{j \neq l_1}  p_{j,1} r_{j,l_1}\\ 
%\vdots\\ 
%\bar{s}_{k,{l_1}} -\sum\limits_{j \neq l_1}  p_{j,k} r_{j,l_1}\\ 
%\end{bmatrix}\\
% &= \sum\limits_{j \in [k] } ( \kappa   P^{'}_{l_2,j} )  
% \bar{s}_{j,l_1} - \kappa r_{l_2,l_1}.
%\label{coupling_expression_systematic_to_systematic}
%\end{align}
\begin{align}
r_{l_1,l_2}&= \mathbf{v}_{l_2}^{'t} \mathbf{w}_{l_1}= \sum\limits_{j \in [k] } ( \kappa   P^{'}_{l_2,j} )  
 \bar{s}_{j,l_1} - \kappa r_{l_2,l_1}.
\label{coupling_expression_systematic_to_systematic}
\end{align}
\noindent Proceeding in a similar way, for a systematic node $l \in [k]$ and a parity node $m \in [k]$, starting from \eref{repair_parity_expression}, we obtain
%\begin{align}
%&\bar{s}_{m,l} = \mathbf{v}_l^{'t} \bar{\mathbf{w}}_m\\
%&=\mathbf{v}_l^{'t}   ( (1-\kappa^2)V+ (1+\kappa) U^{'} \mathbf{e}_m \mathbf{e}_m^{^t} P^t ) 
%\begin{bmatrix}
%s_{1,m} +\frac{ \kappa^2}{1-\kappa^2} \sum\limits_{j\neq m}   P_{1,j}^{'}  \bar{r}_{j,m}\\
%\vdots\\
%s_{k,m} +\frac{ \kappa^2}{1-\kappa^2} \sum\limits_{j\neq m}    P_{k,j}^{'} \bar{r}_{j,m} 
%\end{bmatrix}
%\\
%&= ( (1-\kappa^2)\mathbf{e}_l^t+ (1+\kappa) \kappa P^{'}_{l,m} \mathbf{e}_m^{^t} P^t ) 
%\begin{bmatrix}
%s_{1,m} +\frac{ \kappa^2}{1-\kappa^2} \sum\limits_{j\neq m}   P_{1,j}^{'}  \bar{r}_{j,m}\\
%\vdots\\
%s_{k,m} +\frac{ \kappa^2}{1-\kappa^2} \sum\limits_{j\neq m}    P_{k,j}^{'} \bar{r}_{j,m}
%\end{bmatrix}\\
%&=(1-\kappa^2 + \kappa(1+\kappa) P_{l,m}^{'} P_{l,m}) s_{l,m}  
%+ \sum\limits_{j \in [k] \backslash \{l \}}( \kappa(1+\kappa) P_{l,m}^{'}  P_{j,m} ) s_{j,m} 
%+ \sum\limits_{j \in [k]} ( \kappa^2 P^{'}_{l,j}) \bar{r}_{j,m}.
%\label{coupling_expression_parity_to_systematic} 
%\end{align}
\begin{align}
\bar{s}_{m,l} = \mathbf{v}_l^{'t} \bar{\mathbf{w}}_m=(1-\kappa^2 + \kappa(1+\kappa) P_{l,m}^{'} P_{l,m}) s_{l,m}  
+ \sum\limits_{j \in [k] \backslash \{l \}}( \kappa(1+\kappa) P_{l,m}^{'}  P_{j,m} ) s_{j,m} 
+ \sum\limits_{j \in [k]} ( \kappa^2 P^{'}_{l,j}) \bar{r}_{j,m}.
\label{coupling_expression_parity_to_systematic} 
\end{align}
\noindent Finally, consider two parity nodes $m_1,m_2 \in [k], m_1 \neq m_2$, starting from \eref{repair_parity_expression}, we obtain
%\begin{align}
%\bar{r}_{m_1,m_2}&= \mathbf{u}_{m_2}^{t} \bar{\mathbf{w}}_{m_1} \\
%&=\mathbf{u}_{m_2}^{t}  ( (1-\kappa^2)V+ (1+\kappa) U^{'} \mathbf{e}_{m_1} \mathbf{e}_{m_1}^{^t} P^t ) 
%   \begin{bmatrix}
%s_{1,m_1} +\frac{ \kappa^2}{1-\kappa^2} \sum\limits_{j\neq m_1}   P_{1,j}^{'}  \bar{r}_{j,m_1}\\
%\vdots\\
%s_{k,m_1} +\frac{ \kappa^2}{1-\kappa^2} \sum\limits_{j\neq m_1}    P_{k,j}^{'} \bar{r}_{j,m_1}
%\end{bmatrix}
%\\
%&=  \frac{1-\kappa^2}{\kappa} \mathbf{e}_{m_2}^t P^t  \begin{bmatrix}
%s_{1,m_1} +\frac{ \kappa^2}{1-\kappa^2} \sum\limits_{j\neq m_1}   P_{1,j}^{'}  \bar{r}_{j,m_1}\\
%\vdots\\
%s_{k,m_1} +\frac{ \kappa^2}{1-\kappa^2} \sum\limits_{j\neq m_1}    P_{k,j}^{'} \bar{r}_{j,m_1}
%\end{bmatrix}\\
%&= \sum\limits_{ j\in [k]}   
%( \frac{1-\kappa^2}{\kappa} P_{j,m_2}) s_{j,m_1}
%+  \kappa \bar{r}_{m_2,m_1}.
% \label{coupling_expression_parity_to_parity}
%\end{align}
\begin{align}
\bar{r}_{m_1,m_2}&= \mathbf{u}_{m_2}^{t} \bar{\mathbf{w}}_{m_1}= \sum\limits_{ j\in [k]}   
( \frac{1-\kappa^2}{\kappa} P_{j,m_2}) s_{j,m_1}
+  \kappa \bar{r}_{m_2,m_1}.
 \label{coupling_expression_parity_to_parity}
\end{align}
\noindent The details of deriving \eref{coupling_expression_systematic_to_systematic},
\eref{coupling_expression_parity_to_systematic} and \eref{coupling_expression_parity_to_parity} can be found in 
\appref{Derivations_IA}. Equations \eref{coupling_expression_systematic_to_parity},  
\eref{coupling_expression_systematic_to_systematic},
\eref{coupling_expression_parity_to_systematic} and \eref{coupling_expression_parity_to_parity} can thus be used to derive $A$ and $b$ as defined in \eref{indirect_approach_2}.
 
In the following theorem, we show that the IA code already provides optimal repair for systematic (respectively parity) failures, without the need to modify the coding matrices.
\begin{theorem} \label{thm_IA_1_group} 
In the interference alignment MSR code \cite{suh2011exact}, it is possible to optimally repair any set of $e \le k$ systematic (respectively parity) failures.
\end{theorem}
\begin{IEEEproof}
Assume w.l.o.g that nodes $\{ 1,\ldots,e \}$ have failed. Let $\mathbf{s}=\begin{bmatrix}
r_{1,2}, r_{2,1},\ldots,    r_{e-1,e},r_{e,e-1}
\end{bmatrix}^t$. Then, from \eref{coupling_expression_systematic_to_systematic}, it follows that $A$ is a block-diagonal matrix given by 

\begin{equation}
A= \begin{bmatrix}
1 &  \kappa   \\ 
 \kappa &1   \\ 
&  &   \ddots \\
&  &    &   1 &  \kappa  \\ 
&  &   &    \kappa & 1
\end{bmatrix}.
\end{equation}
\noindent It follows that $|A|= (1-\kappa^2)^{\frac{e(e-1)}{2}} \neq 0$ as $\kappa^2 \neq 1$ by design. The same procedure applies to any set of $e$ failures among parity nodes using equation \eqref{coupling_expression_parity_to_parity}.
\end{IEEEproof}
%\subsection{1 systematic failure and one parity failure}
\begin{theorem} \label{thm_IA_2_erasures}  
The interference alignment MSR code achieves optimal simultaneous repair of one systematic node $l$ and one parity node $m$ if $P_{l,m} (P^{-1})_{m,l}  \neq 1$.
\end{theorem}
\begin{IEEEproof}
Assume that systematic node $l$ and parity node $m$ failed. Let $\mathbf{s}=[
s_{ l,m}, \bar{s}_{m,l} ]^t$.  
From \eref{coupling_expression_systematic_to_parity}, we obtain

\begin{equation}
s_{l,m}= ( 1 -  \frac{\kappa }{1+\kappa }   P_{l,m} P_{l,m}^{'} ) \bar{s}_{m,l}
+ c_{1},
\label{first_equation_two_erasure}
\end{equation}
where $c_1$ is a known quantity independent of $\mathbf{s}$.
\noindent Similarly, from \eref{coupling_expression_parity_to_systematic}, we obtain 
\begin{equation}
\bar{s}_{m,l} = (1-\kappa^2 + \kappa(1+\kappa) P_{l,m} P_{l,m}^{'} ) s_{m,l} + c_{2},
\label{second_equation_two_erasure}
\end{equation} 
where $c_2$ is a known quantity independent of $\mathbf{s}$.
It follows that $A$, as defined in \eref{indirect_approach_2}, is given by

\begin{equation}
A=\begin{bmatrix}
-1 & 1 -  \frac{\kappa }{1+\kappa }   P_{l,m} P_{l,m}^{'}   \\ 
 1-\kappa^2 + \kappa(1+\kappa) P_{l,m} P_{l,m}^{'} & -1
\end{bmatrix}.
\end{equation}
\noindent After simplification, we have $|A| \neq 0 \iff \kappa^2 (P_{l,m} P_{l,m}^{'} -1 )^2 \neq 0 \iff P_{l,m} (P^{-1})_{m,l}  \neq 1$, as $\kappa \neq 0$.
\end{IEEEproof}
Combining Theorems \ref{thm_IA_1_group} and \ref{thm_IA_2_erasures} we know that $(2 k,k,2 k-2,2,k,2)$ MSMR codes for 2 erasures can be constructed through IA codes. We point out that Theorems \ref{thm_IA_1_group} and \ref{thm_IA_2_erasures} have been derived in \cite{IA_code} for cooperative repair, using a different technique. Recall that MSCR codes are in particular MSMR codes \cite{rawat2016centralized}. However, their technique cannot be extended to more than two node failures including systematic and parity nodes \cite{IA_code}.
\begin{theorem} \label{thm_IA_3_erasures}
The interference alignment MSR code achieves optimal simultaneous repair of:

\begin{itemize}
\item	two systematic failures $l_1,l_2$ and one parity failure $m$  if 
$1-P_{l_1,m} (P^{-1})_{m,l_1}-P_{l_2,m} (P^{-1})_{m,l_2} \neq 0$,

\item	one systematic failure $l$ and two parity failures $m_1,m_2$  if 
$1-P_{l ,m_1} (P^{-1})_{m_1,l}-P_{l ,m_2} (P^{-1})_{m_2,l } \neq 0$,

\item	three systematic failures $l_1,l_2,l_3$ and one parity failure $m$ if 
\begin{align}
1-P_{l_1,m} (P^{-1})_{m,l}-P_{l_2,m} (P^{-1})_{m,l_2}-P_{l_3,m} (P^{-1})_{m,l_3} \neq 0,
\end{align}

\item	one systematic failure $l$ and three parity failures $m_1,m_2,m_3$   if 
\begin{align}
1-P_{l ,m_1} (P^{-1})_{m_1,l}-P_{l ,m_2} (P^{-1})_{m_2,l }-P_{l ,m_3} (P^{-1})_{m_3,l} \neq 0,
\end{align}

\item	two  systematic failures $l_1,l_2$ and two parity failures $m_1,m_2$ if
\begin{align}
&1- P_{l_1 ,m_1} (P^{-1})_{m_1,l_1}-P_{l_1 ,m_2} (P^{-1})_{m_2,l_1 } - P_{l_2 ,m_1} (P^{-1})_{m_1,l_2}-P_{l_2 ,m_2} (P^{-1})_{m_2,l_2 } \nonumber\\
&+ P_{l_1 ,m_1} (P^{-1})_{m_1,l_1} P_{l_2 ,m_2} (P^{-1})_{m_2,l_2 }   + P_{l_1 ,m_2} (P^{-1})_{m_2,l_1 } P_{l_2 ,m_1} (P^{-1})_{m_1,l_2} \nonumber\\
& -P_{l_1 ,m_1} (P^{-1})_{m_1,l_2 } P_{l_2 ,m_2} (P^{-1})_{m_2,l_1} 
 -P_{l_1 ,m_2} (P^{-1})_{m_2,l_2 } P_{l_2 ,m_1} (P^{-1})_{m_1,l_1} \neq 0.
\end{align}
\end{itemize}

\end{theorem}
\begin{IEEEproof}
The proof follows along similar lines as \thref{thm_IA_2_erasures} by constructing $A$  using \eref{coupling_expression_systematic_to_parity},  
\eref{coupling_expression_systematic_to_systematic},
\eref{coupling_expression_parity_to_systematic} and \eref{coupling_expression_parity_to_parity}. The explicit expression of $|A|$ can then be obtained for example by using the Symbolic Math Toolbox of MATLAB, from which the above conditions can be readily obtained (the MATLAB source code can be found in \cite{online_code}). 
\end{IEEEproof}

Combining Theorems \ref{thm_IA_1_group} and \ref{thm_IA_3_erasures} we know that $(2 k,k,2 k-e,e,k,e)$ MSMR codes for $e=3,4$ erasures can be constructed through IA codes.
\begin{remark}
Deriving an exact condition under which the recovery of multiple failures for large $e$ is not straightforward. However, we suspect that the general formula is given by the following expression

\begin{align}
&|A| = k^{2 s p} (1 - k^2) ^{\binom{s}{2}+\binom{p}{2}}  \left(1-
\sum\limits_{\substack{  L \subset S,  J \subset P,\\ |L|=|J| \le \min(s,p) } } 
\sum\limits_{ \sigma \in \Pi_{L,J}}
\sum\limits_{ \sigma^{'} \in \Pi_{L,J}}
(\text{sgn} (\sigma)    \prod_{i \in L} P_{i,\sigma(i)} ) 
(\text{sgn} (\sigma^{'})    \prod_{j \in J} P^{'}_{j,\sigma^{'}(j)} ) \right)^e,
\label{claim}
\end{align}

\noindent where $\Pi_{L,J} $ is the group of permutations between the two sets $L$ and $J$ ($L$ and $J$ are ordered in increasing order),
%, and element of order $h$  in $L$ is mapped to element of order $h$ in $J$)
 and sgn($\sigma$) refers to the sign of a permutation $\sigma$, counting the number of inversions in $\sigma$, and given by
\begin{align}
 \textrm{sgn}(\sigma)= (-1)^{\sum\limits_{i_1< i_2 \in L} \mathbbm{1}_{\sigma(i_1) > \sigma(i_2)}}.
\end{align} 
For example, if $L= \{1,2,3 \}$, $J= \{2,3,4 \}$ and $\sigma(1)=3,\sigma(2)=4,\sigma(3)=2 $. Then, sgn($\sigma$)=1. 
%\bl{??notation $I$ used for identity matrix. Should we change to a different letter??}
 
One can check that the formulas in Theorems \ref{thm_IA_1_group}, \ref{thm_IA_2_erasures} and \ref{thm_IA_3_erasures} satisfy \eref{claim}. A general proof of \eref{claim} is still open.
\end{remark}
\begin{example}
Consider the IA code with $n=8,k=4,d=7$, $\alpha=4, \beta=1$. 
The code is defined over the finite field $\mathbb{F}_{2^5}$ with $g$ being the generator of its multiplicative group. Let $P$ being a Vandermonde matrix given by
\begin{align}
P =   \begin{bmatrix}
1        &   1      &     1     &      1\\
           1         &  g     &      g^2     &      g^3\\
            1         &  g^2     &      g^4      &    g^6\\
            1       &    g^3      &    g^6     &     g^9
\end{bmatrix}.
\end{align}
Using Theorems \ref{thm_IA_1_group}, \ref{thm_IA_2_erasures} and \ref{thm_IA_3_erasures}, one can check that any two, three and four erasures can be repaired optimally using our repair framework. 
%\bl{?? need to explain what the numbers mean in the matrix. In Example \ref{example: PM} you used powers of primitive element instead of numbers.??}
\end{example}
In the following theorem, we provide an existence proof of IA MSMR codes for multiple erasures.
\begin{theorem} \label{thm_IA_general}
There exists $(2k,k,2k-e,e,k,e)$ interference alignment MSMR codes, defined over a large enough finite field, such that any $e \le k$ erasures can be optimally repaired.  
\end{theorem}
 
\begin{IEEEproof}
From \thref{thm_IA_1_group}, we know that if the errors are all either systematic or parity nodes, then efficient repair is possible. Thus, we only need to analyze the case of a mixture of systematic and parity failures. 

Consider $e \le k$ failures consisting of $q$ systematic nodes and $p$ parties nodes, indexed by the sets $\mathcal{Q}$ and $\mathcal{P}$. W.lo.g, assume that $\mathcal{Q}=[q]$ and $\mathcal{P}=[p]$. Let 
$\mathbf{s}$ denote the vector of unknowns such that pairs $( r_{i,j},r_{j,i}), ( \bar{r}_{i,j},\bar{r}_{j,i})$ and $( s_{i,j},\bar{s}_{j,i})$ are grouped together. Using
\eref{coupling_expression_systematic_to_parity},  
\eref{coupling_expression_systematic_to_systematic},
\eref{coupling_expression_parity_to_systematic} and \eref{coupling_expression_parity_to_parity}, we construct $A$ as in \eref{indirect_approach_2}. Denote the determinant of $A$ as $F(\kappa,P_{i,j}, P_{i,j}^{'}, i \in \mathcal{Q},j \in \mathcal{P} ) \triangleq |A| $. The rows and columns of $A$ are indexed by $\{ r_{i,j},s_{i,j},\bar{r}_{i,j},\bar{s}_{i,j}  \}$.
Let $M_{i,j}$ denote the minor in $A$ corresponding to $A_{i,j}$. Similarity, $N_{i,j}$ denotes the minor in $P$ corresponding to $P_{i,j}$. As $P'=(P^{-1})^t$, $P_{i,j}^{'}=\frac{(-1)^{i+j}N_{j,i}}{|P|}$, one observes that $F$ is a rational function in $(\kappa,P_{i,j},(i,j)  \in \mathcal{Q} \times \mathcal{P})$.

\textbf{Claim 2.} $F$ is  not identically zero for any $q,p \ge 0, q+p=e \le k$. 

If Claim 2 holds, then the theorem is proved due to the following argument. By symmetry, any $e$-erasure pattern corresponds to a non-zero rational function $F$. Recall from \cite{suh2011exact} that the reconstruction process requires that every submatrix of $P$ is invertible. This can be translated into a polynomial constraint given by $g(P_{i,j} , i \in \mathcal{Q},j \in \mathcal{P} ) \neq 0$. Let $T\triangleq g\prod_{e \textrm{ erasures}}F$. Here the product is over all possible $e$ erasures, and the rational function $F$ depend on the erasure pattern. Then, it follows that $T$ is a non-zero rational polynomial in $(\kappa, P_{i,j}, (i,j) \in [k] \times [n-k])$. By Combinatorial Nullstellensatz \cite{alon1999combinatorial}, we can find assignments of
the variables $(\kappa, \{P_{i,j}\})$ over a large enough finite field, such that the 
code guarantees optimal recovery of any set of $e$ erasures.

Next, we prove Claim 2. We assume first that $q \le \frac{k}{2}$. Let 
\begin{align}
P_{i,j}= 0\  \forall  \ (i,j)  \in \mathcal{Q} \times \mathcal{P}.
\label{condition_P}
\end{align}
Note that one can always construct a (normalized) invertible matrix $P$ satisfying \eref{condition_P}, so we can assume $|P|=1$. Thus $F$ is a polynomial. We will show
\begin{align}
  F(\kappa,P_{i,j}=0,P'_{i,j}, (i,j)\in\mathcal{Q}\times \mathcal{P}) 
=  F(\kappa,P_{i,j}=0,P'_{i,j}=0, (i,j)\in\mathcal{Q}\times \mathcal{P}) 
 \neq & 0, 
\end{align}
which implies
\begin{align}
 F(\kappa,P_{i,j},P'_{i,j}, (i,j)\in\mathcal{Q}\times \mathcal{P}) \neq 0.
\end{align}

To this end, we first prove that $F(\kappa,P_{i,j}=0,P'_{i,j}, (i,j)\in\mathcal{Q}\times \mathcal{P})$, viewed as a polynomial of $(\kappa,\{P'_{i,j}\})$, does not depend on $\{P_{i,j}^{'}\}$.
From \eref{coupling_expression_systematic_to_systematic} and
\eref{coupling_expression_parity_to_systematic}, one can check that $P_{i,j}^{'}$ appears in $A$ at entries given by
\begin{itemize}
\item	$A_{r_{l,i},\bar{s}_{j,l}}$ for $l \in \mathcal{Q} \backslash \{ i\}$,
\item	$A_{\bar{s}_{m,i},\bar{r }_{j,m}}$ for $m \in \mathcal{P} \backslash \{j\}$.
\end{itemize}
For any $l \in \mathcal{Q} \backslash \{ i\}$, consider the two columns in $A$ indexed by $ r_{l,i} $ and $ r_{i,l} $. Both columns have non-zero entries only at rows indexed with $ r_{l,i} $ and $ r_{i,l}$. Then, after removing entries at row $r_{l,i}$, it follows that both columns become linearly dependent, as both columns are scalar multiples of the same standard basis vector. Thus, $M_{r_{l,i},\bar{s }_{j,l}}=0$. 

The example in \eref{example_erasure} illustrates the case of two systematic failures, given by systematic nodes 1 and 2, and one parity failure, given parity node 1. In this case $s=\begin{bmatrix}
s_{11}, \bar{s}_{11} , r_{12},r_{21},s_{21}, \bar{s}_{12}
\end{bmatrix}$. Setting $P_{i,j}=0$ for all $i=1,2,j=1$ and looking at the submatrix of $A$ by removing row $r_{1,2}$ and column $\bar{s}_{2,1}$ in \eref{example_erasure}, it can be seen that columns $r_{1,2}, r_{2,1}$ are dependent, hence
 its corresponding minor $M_{r_{1,2},\bar{s}_{2,1}}=0$.
 
\begin{figure*}[!]
\begin{equation}
 A=
  \begin{blockarray}{*{6}{c} l}
    \begin{block}{*{6}{>{$\footnotesize}c<{$}} l}
      $s_{1,1}$ & $\bar{s}_{1,1} $ & $r_{1,2}$ & $r_{2,1}$ & $s_{2,1} $ & $\bar{s}_{2,1}$ & \\
    \end{block}
    \begin{block}{[*{6}{c}]>{$\footnotesize}l<{$}}
     -1&1-\frac{\kappa P_{11} P_{11}^{'}}{\kappa+1}  &0  &- P_{21}  &0  &0& $s_{1,1}$ \\
     1-\kappa^2+ \kappa (\kappa+1)P_{11} P_{11}^{'} &-1  &0  & 0 &\kappa (\kappa+1)P_{21} P_{11}^{'}  &0 & $\bar{s}_{1,1}$ \\
     0&  \kappa P_{21}^{'}& -1 & -\kappa & 0 & 0& $r_{1,2}$ \\
      0& 0 &  -\kappa & -1& 0 &  \kappa P_{11}^{'} & $r_{2,1}$ \\
     0&  0& - P_{11} &  0& -1 & 1-\frac{\kappa P_{21} P_{21}^{'}}{\kappa+1} & $s_{2,1}$ \\
    \kappa (\kappa+1)P_{11} P_{21}^{'}& 0 &0  &0&  1-\kappa^2+ \kappa (\kappa+1)P_{21} P_{21}^{'} & -1& $\bar{s}_{2,1}$ \\
    \end{block}
  \end{blockarray}
  \label{example_erasure}
\end{equation}
\end{figure*}
Similarly, for any $m \in \mathcal{P} \backslash \{ j\}$, consider the two rows in $A$ indexed by $\bar{r}_{j,m}$ and $\bar{r}_{m,j}$. Both rows have non-zero entries only at columns indexed with $\bar{r}_{j,m}$ and $\bar{r}_{m,j}$. Then, after removing entries at column $\bar{r}_{j,m}$, it follows that both rows become linearly dependent. Thus, $M_{\bar{s}_{m,i},\bar{r}_{j,m}}=0$. 

The minors in $A$ of all terms corresponding to $P_{i,j}^{'}$ are thus equal to zero. Therefore, w.l.o.g, one can assume that $P_{i,j}^{'}=0, \forall (i,j)  \in \mathcal{Q} \times \mathcal{P}$. It follows that $A$ is block-diagonal matrix such that
\begin{itemize} 
\item	Row/column pairs $( r_{i,j},r_{j,i})$ correspond to $\begin{bmatrix}
-1 & -\kappa  \\
-\kappa & -1
\end{bmatrix}$,
\item Row/column pairs $( \bar{r}_{i,j},\bar{r}_{j,i})$ correspond to $\begin{bmatrix}
-1 & \kappa  \\
 \kappa & -1
\end{bmatrix}$,

\item Row/column pairs $( s_{i,j},\bar{s}_{j,i})$ correspond to $\begin{bmatrix}
-1 & 1 \\
1-\kappa^2  &-1
\end{bmatrix}$.
\item Other entries are 0.
\end{itemize}
Therefore, $|A|=\kappa^{2 q p} (1-\kappa^2)^{  {q \choose 2}+{p \choose 2}} \neq 0$, as $\kappa \neq 0$ and $\kappa^2 \neq 1$. 

Assume now that $q > \frac{k}{2}$. Then, $p \le \frac{k}{2}$. Proceeding similarly, one can show that if $P_{ij}^{'}=0,  \forall (i,j)  \in \mathcal{Q} \times \mathcal{P}$, then, all terms $P_{ij}$ have no impact on $|A|$ and one obtains similarly $|A|=\kappa^{2 r p} (1-\kappa^2)^{  {r \choose 2}+{p \choose 2}}$.
\end{IEEEproof}
%%%%%%%%%%%%%%%%%%%%%%%%%%%%%%%%%%%%%%%%%%%%%%%%%%%%%%%%%%%%%%%%%%%%%%%%%%%%%%%%%%%%%%%%%%%%%%%%%%%%%%%%%%%%%%%%%%%%%%%%%%%%%%%%%%%%%%%%%%%%%%%%%%%%%%%%%%%%%%%%%%%%%%%%%%%%%%%%%%%%%%%
\section{Non-existence of exact MBMR regenerating codes}\label{MBMR_codes}
%Exact regenerating codes are of interest in practice. Exact regenerating codes achieving the MSMR point have been constructed \cite{cadembe_bound,wang2016optimal,ye2017explicit,rawat2016centralized}. 
Recall that the MBMR point is defined as the minimum bandwidth point on the functional tradeoff. In this section, we explore the existence of linear exact MBMR regenerating codes for $1 <e<k$. Unlike the single erasure repair problem \cite{Rashmi_Product_Matrix} and the cooperative repair problem \cite{wang2013exact}, we prove that linear exact regenerating codes do not exist. Following \cite{wang2013exact,Rashmi_Product_Matrix}, we proceed by investigating subspace properties that
linear exact MBMR codes should satisfy. Then, we prove that the derived properties over-constrain the system. 
\subsection{Subspace viewpoint}
Linear exact regenerating codes can be analyzed from a viewpoint based on subspaces. A linear storage code is a code in which every stored symbol is a linear combination of the $\mathcal{M}$ symbols of the file. Let $\mathbf{f}$ denote an $\mathcal{M}$-dimensional vector containing the source symbols. Then, any symbol $x$ can be represented by a vector $\mathbf{h}$ satisfying $x=  \mathbf{f}^{t} \mathbf{h} $ such that $\mathbf{h}  \in \mathbb{F}^\mathcal{M}$, $\mathbb{F}$ being the underlying finite field. 
%This observation/characterization is also true in the classical coding setup in which each node simply stores one vector $\mathbf{h}$.
The vectors $\mathbf{h}$ define the code. A node storing $\alpha$ symbols can be considered as storing $\alpha$ vectors. Node $i$ stores $\mathbf{h}_1^{(i)} \ldots \mathbf{h}_\alpha^{(i)}$. It is easy to see that linear operations performed on the stored symbols are equivalent to the same operations performed on the these vectors: $  \sum\limits \gamma_i \mathbf{f}^{t} \mathbf{h}_i = \mathbf{f}^{t}( \sum\limits \gamma_i  \mathbf{h}_i )$, $\gamma_i \in \mathbb{F}$. Thus, each node is said to store a subspace of dimension at most $\alpha$. We write $W_A$ to denote the subspace stored by all nodes in the set $A$, $A \subseteq [n]$.
For repair, each helper node passes $\beta$ symbols. Equivalently, each node passes a subspace of dimension at most $\beta$. We denote the subspace passed by node $j$ to repair a set $R$ of $e$ nodes by $S_j^R$. The subspace passed by a set of nodes $A$ to repair a set $R$ of $e$ nodes is denoted by $S_A^R = \sum_{j \in A} S_j^R$, where the sum denotes the sum of subspaces. 

{\bf Notation.} The notation $\bigoplus_{j}X_j$ denotes the direct sum of subspaces $\{X_j\}$. For a general exact regenerating code, which can be nonlinear, we use by abuse of notation $W_A$, $S_A^R$ to represent the random variables of the stored information in nodes $A$, and of the transmitted information from helpers $A$ to failed nodes $R$.
Properties that hold using entropic quantities for a general code do hold when considering linear codes. 
For instance, consider two sets $A$ and $B$. Then, we note the following
\begin{align}
H(W_A) &\to \dim(W_A),\\
H(W_A| W_B) &\to \dim(W_A)- \dim(W_A \cap W_B), \\
I(W_A, W_B) &\to \dim(W_A \cap W_B),
\end{align}
where the symbol $\to$ means \textit{translates to}. When results hold for general codes, we only prove for the entropy properties, and the proof for the subspace properties of linear codes is omitted. All results on entropic quantities are for general codes, and all results on subspaces are for linear codes. Moreover, all results in this section refer to properties of \emph{optimal exact} multi-node repair codes with $k>e$ (constructions for $k \le e$ are presented in Section \ref{construction}), some of which are specific to MBMR codes and will be noted.
%By abuse of notation, $W_i$ denotes the subspace stored bu node $i$ and also the random variable corresponding to the information stored by node $i$ 

In this section, we assume that the codes are \emph{symmetric}. Namely, the entropy (or subspace) properties do not depend on the indices of the nodes. Note that one can always construct a symmetric code from a non-symmetric code \cite{elyasi2015linear}, hence our assumption does not lose generality. We now start by proving some properties that exact regenerating codes, satisfying the optimal functional tradeoff, should satisfy. 
%The following property is valid for all optimal exact regenerating codes, not necessarily MBMR codes. 
We note that the following property is also presented in \cite[Lemma 4]{wang2016optimal}.
\begin{lemma}
\label{inequality_term}
Let $B \subseteq [n]$ be a subset of nodes of size $e$, then for an arbitrary set of nodes $A$, such that $0 \le |A| \le d, B \cap A= \emptyset$,
\begin{align}
H(W_B|W_A) \le  H(W_B|S_A^B) \le \min( e \alpha, (d- |A|) \beta). 
\end{align}
\end{lemma}

\begin{IEEEproof}
%If nodes $B$ and some other $e-|B|$ nodes are erased, 
If nodes $B$ are erased, consider the case of having nodes $A$ and nodes $C$ as helper nodes, $|C|=d-|A|$. Then, the exact repair condition requires
\begin{align}
0&=H(W_B|S_A^B,S_C^B)\nonumber\\
&=H(W_B|S_A^B)-I(W_B,S_C^B | S_A^B)\nonumber\\
&\geq H(W_B|S_A^B)- H(S_C^B) \nonumber\\
& \geq H(W_B | S_A^B)- (d-|A|) \beta.
%& \geq H(W_B | W_A)- (d-|A|) \beta.
\end{align}
Moreover, we have $H(W_B|S_A^B) \le H(W_B) \le |B| \alpha$,  $ H(W_B|W_A) \le  H(W_B|S_A^B)$, and the results follows. %The proof was also given in \cite{wang2016optimal}.
\end{IEEEproof}

In the next two subsections, we focus on the cases where $e \mid k$ and $e \nmid k$, respectively.
\subsection{Case $e \mid k$}
Note that in this case since $e<k$, we have $k \ge 2e$.
Recall from Theorem \ref{optimal_cut_result} that points on the optimal tradeoff satisfy
\begin{align} \label{eq_tradeoff_ek}
\mathcal{M}=  \sum\limits_{j=0}^{\eta-1} \min (e \alpha, (d-je) \beta).
\end{align}
Points between and including MSMR and MBMR satisfy
\begin{align}
\frac{d-k+e}{e}\beta \le \alpha \le \frac{d}{e} \beta.
\end{align}
%prove another property that is satisfied by all exact regenerating codes.
\begin{lemma}{(Entropy of data stored): }
\label{entropy_lemma}
Consider points on the optimal tradeoff. For an arbitrary set $L$ of storage nodes of size $e$, and a disjoint set $A$ such that $|A|=e m < k$ for some integer $m$,
\begin{align}
H(W_{L})&=e \alpha,\\
H ( W_{L} | W_A  )&= \min (e \alpha, (d- e m) \beta).
\end{align}
For linear codes,
\begin{align}
\dim(W_{L})&=e \alpha,\\
\dim ( W_{L} )-\dim ( W_{L} \cap W_A  )  & = \min (e \alpha, (d- e m) \beta).
\end{align}
Hence,  the contents of any group of $e$ nodes are independent. In particular, for a set $A$ of nodes, $1 \le |A| \le e$, $H(W_A) = |A| \alpha$.
\end{lemma}
\begin{IEEEproof}
By reconstruction requirement, we write

\begin{align}
\mathcal{M}&= H(W_{[k]})\nonumber \\
&=H(W_{[e]}) +\sum\limits_{j=1}^{\eta-1} H(W_{\left\{ ej+1,\ldots, e (j+1) \right\} } | W_{[j e]}) \nonumber\\
\label{use_lemma}
& \le \min(e \alpha, d \beta) + \sum\limits_{j=1}^{\eta-1} \min(e \alpha, (d-e j) \beta) \nonumber\\
%\label{eq_tradeoff_ek2}
&=\mathcal{M}, 
\end{align}
\noindent where the inequality follows from \lref{inequality_term}. Thus, all inequalities must be satisfied with equality. 
\end{IEEEproof}
%\begin{remark}
%\lref{entropy_lemma} states that the contents of any group of $e$ nodes are independent. In particular, for each node $i$, we have $H(W_i)=\alpha$. For a set $A$ of nodes, $|A| \le e$, $H(W_A) = |A| \alpha$. 
%\end{remark}

\begin{corollary}
\label{intersection_dimension}
At the MBMR point, for any set $L$ of size $e$ and disjoint set $A$ of size $|A|=e m <k$, we have 
$$\dim ( W_{L} \cap W_A  )= e m  \beta.$$   
\end{corollary}
\begin{IEEEproof}
By Lemma \ref{entropy_lemma} and $e\alpha = d\beta$,
\begin{align}
\dim ( W_{L} )-\dim ( W_{L} \cap W_A  )  & = \min (e \alpha, (d- e m) \beta)=(d- e m) \beta.
\end{align}
Using the fact that $\dim ( W_{L} )= e \alpha= d \beta$, we obtain the result.
\end{IEEEproof}
\begin{lemma}
\label{direct_sum_property}
For any set $E$ of size $e$, and a disjoint set $A$ of size $d$, the MBMR point satisfies
\begin{align}
W_E=  \bigoplus_{j \in A} S_j^E, \dim( S_j^E)= \beta.
\end{align}
Hence, the subspaces $ S_j^E$ and $ S_{j'}^E$ are linearly independent. For every set $Q \subseteq A$,  $\dim(S_Q^E)= |Q| \beta.$ Moreover, each subspace $S_j^E$ has to be in $W_E$, namely, $ S_j^E \subseteq W_E$.
\end{lemma}
\begin{IEEEproof}
For exact repair, we need $W_E \subseteq \sum\limits_j S_j^E$. Thus, 
\begin{align}
d \beta = e \alpha = \dim(W_E) \le \dim( \sum\limits_j S_j^E) \le d \beta .
\end{align}
Thus, every inequality has to be satisfied with equality. 
%This implies that the subspaces $S_j^E$ are mutually linearly independent. Moreover, $\dim(S_j^E)= \beta.$
\end{IEEEproof}  
\begin{lemma}
\label{general_intersection_dimension}
At the MBMR point, for any set $E$ of $e$ nodes and any other disjoint set $Q$ of size $|Q| \le k-e$, we have
\begin{align}
\label{subspace_intersection}
S_Q^E= W_E \cap W_Q, \dim(W_E \cap W_Q) = \dim(S_Q^E)= |Q| \beta.
\end{align}
\end{lemma}
\begin{IEEEproof}
Consider $Q$ nodes such that $|Q|\le k-e$ helping in the repair of a set $E$ of $e$ nodes. Let $J$ contains $Q$ such that $|J|=k-e$. Denote $Q^c = J \backslash Q$. 
From \cororef{intersection_dimension}, we have $\dim(W_E \cap W_J)= (k-e) \beta$.  On the other hand, from \lref{direct_sum_property}, we have $\dim (S_J^E) = (k-e)\beta$ and $S_J^E \subseteq W_E$. Moreover, by definition, $S_J^E \subseteq W_J$. Thus, $S_J^E \subseteq  W_E \cap W_J$. As the dimensions match, it follows that $S_J^E =  W_E \cap W_J$. Note that $S_A^E \subseteq  W_E \cap W_A$ holds for any subset $A$ of size $|A| \le d$. Now, we write
\begin{align}
S_J^E=W_E \cap W_J&= W_E \cap (W_Q+ W_{Q^c}) \nonumber\\
  &\supseteq	W_E \cap  W_Q   + W_E \cap    W_{Q^c}  \nonumber\\&\supseteq S_Q^E+   S_{Q^c}^E= S_J^E.
\end{align}
This implies that all    inclusion inequalities have to be satisfied with equality and the result follows.
\end{IEEEproof}
%\newpage
%\subsubsection{Impossibility result}
The next lemma plays an important role in establishing the non-existence of exact MBMR codes. It only holds true when $e \geq 2$, which conforms with the existence of single erasure MBMR codes.
\begin{lemma}
\label{intersection_property}
Consider the MBMR point. When $e\geq 2$, for any set of $e+2 \le k$ nodes, labeled 1 through $e+2$, it holds that
\begin{align}
\dim(  W_{e+2} \cap W_{[ e+1]}) 
 =\dim(  W_{e+2} \cap  W_{[ e]})=\beta.
\end{align}
\end{lemma}
\begin{IEEEproof}
We have
\begin{align}
 \dim( W_{[ e+2]}) 
&=\dim( W_{[ e]})+ \dim(W_{e+1}+W_{e+2}) \nonumber\\
&\qquad - \dim(   W_{[ e]}  \cap (W_{e+1}+W_{e+2}))\nonumber\\&
= e \alpha + 2 \alpha - 2 \beta,
\end{align} 
where the second equality follows from \lref{entropy_lemma}, \lref{general_intersection_dimension}. % and the fact that any set of $e$ nodes are linearly independent. 
On the other hand, we write
\begin{align}
\dim( W_{[e+2]})
&=\dim( W_{[e]}) \nonumber\\& +
  \dim(W_{e+1}) - \dim(  W_{e+1} \cap  W_{[e]}) 
 \nonumber\\& +
  \dim(W_{e+2}) - \dim(  W_{e+2} \cap  W_{[e+1]} )\nonumber\\
 &= e \alpha + 2 \alpha -\beta- \dim(  W_{e+2} \cap   W_{[e+1]}).
\end{align}
The lemma follows from equating both equations.
\end{IEEEproof}
%\lref{intersection_property} does not hold in the case of $e=1$, for which exact-regenerating MBR codes can be constructed \cite{rashmi2009explicit}.
%
\begin{theorem}
\label{non_existence_MBMR_1}
Exact linear regenerating MBMR codes do not exist when $2 \le e<k$ and $e \mid k$.
\end{theorem}
\begin{IEEEproof}
Assuming that there exists an exact-repair regenerating code, we consider the first $e$ nodes. Then, these nodes store linearly independent vectors. We write, for $i=1,\ldots,e$,
$W_i=\begin{bmatrix}
 V_{i1} & V_{i2}   
 \end{bmatrix}$ where $V_{i,1}$ contains $\beta$ linearly independent columns and $ V_{i,2}$ contains the remaining $(\alpha-\beta)$ basis vectors for node $i$. Now, consider node $e+1$. We have $\dim(  W_{e+1} \cap  W_{[e]})=\beta$ by \lref{intersection_property}. That means that node $e+1$ contains $\beta$ columns, linearly dependent on the columns from the first $e$ nodes. %Since any set of $e$ nodes among the first $e+1$ nodes should be linearly independent, 
 Since the first $e$ nodes should be linearly independent, w.l.o.g, we can assume that the $\beta$ dependent vectors of node $e+1$, denoted by $V_{e+1,1}$, is of the form
 \begin{align}
 \label{repair_form}
 V_{e+1,1} = \sum_{i=1}^e   V_{i,1} \mathbf{x}_i,
 \end{align}
such that $ \mathbf{x}_i   \neq \mathbf{0}_{\beta \times 1} \ \forall i=1,\ldots,e$.
Now, consider node $e+2$. From \lref{intersection_property}, node $e+2$ contains $(\alpha-\beta)$ vectors linearly independent from vectors in nodes $1$ through $e+1$. The remaining basis vectors of node $e+2$ (which are linearly independent of the $(\alpha-\beta)$ vectors) are denoted by $V_{e+2,1}$. Now, to repair any set of $e$ nodes from the set of first $e+1$ nodes, node $e+2$ can only pass $V_{e+2,1}$. Otherwise, \lref{general_intersection_dimension} will be violated. Then, this implies that $V_{e+2,1} \subseteq W_{J}$, for all ${J} \subseteq \left\{1,\ldots,e+1 \right\}$ such that $|{J}|=e$. Then, it can be seen that $V_{e+2,1}$ can only be of the same form in \eref{repair_form}
\begin{align}
 V_{e+2,1} = \sum_{i=1}^e   V_{i,1} \mathbf{y}_i, \text{ such that } \mathbf{y}_i \neq \mathbf{0}_{\beta \times 1} \ \forall i=1,\ldots,e.
 \end{align}
Similar reasoning applies to node $i$ for $i=e+3,\ldots,k+1$ to conclude that $V_{i,1}$ can be written as in \eref{repair_form}.

Now, assume the first $e$ nodes fail. Then, node $i$ can only pass $V_{i,1}$ for $i=e+1,\ldots k+1$. We recall from \lref{general_intersection_dimension} that $S_i^{[e ]}= W_i \cap W_{ [e ]}$. The total number of vectors passed by these nodes is $(k-e+1) \beta \geq (e+1) \beta$. On the other hand, from \eref{repair_form}, all $V_{i,1}$ are generated by $e \beta$ vectors. Thus, the set $\left\{V_{i,1}, i=e+1,\ldots,k+1 \right\}$ must be linearly dependent, which contradicts the linear independence property of the passed subspaces passed for repair, as stated by \lref{direct_sum_property}.
\end{IEEEproof}
%%%%%%%%%%%%%%%%%%%%%
\subsection{Case $e \nmid k$}

Recall that from the analysis of \thref{optimal_cut_result}, for $e = \eta e + r$, $1 \le r \le e-1$, 
at the MBMR point, two scenarios generate the same minimum cut:  
$$\mathbf{u}_1=[r, e,\ldots,e    ]  \textrm{ and } \mathbf{u}_2=[ e,\ldots,e , r ]. $$ 
Equivalently, we have
\begin{align}
&\mathcal{M}=f(\mathbf{u}_1)=
f(\mathbf{u}_2),
\label{second_eq_MBMR}
\end{align}
where $f()$ is defined as in \eref{f_definition}.

Moreover, all points between and including MSMR and MBMR on the tradeoff satisfy
\begin{align}\label{eq119}
&\mathcal{M}=\min(r \alpha,d \beta)+ \sum\limits_{i=0}^{\eta-1} \min (e \alpha, (d-r-ie) \beta) = f(\mathbf{u}_1),
\end{align}
%Points between and including MSMR and MBMR satisfy
\begin{align}
&\frac{d-k+e}{e} \beta \le \alpha \le \frac{d + \eta r - \eta e}{r} \beta.
\end{align}
Properties satisfied by exact regenerating codes developed in the previous section extend to the case $e \nmid k$ with slight modifications. We state the properties without detailed proofs as the techniques are the same.
\begin{lemma}
\label{dimension_of_stored_data_2nd_case}
Consider points on the optimal tradeoff.
For an arbitrary set $R$ of storage nodes of size $r$, and a set $A$ such that $|A|=j e  +r < k$ for some integer $j \le \eta-1$, for all exact-regenerating codes operating on the functional tradeoff, it holds that

\begin{align}
H(W_{R})&=r \alpha,\\
H( W_{E} | W_A  )  & = \min (e \alpha, (d-r- j e   ) \beta).
\end{align}
For linear codes, 
\begin{align}
\dim(W_{R})&=r \alpha,\\
\dim ( W_{E})- \dim( W_{E} \cap W_A  )  & = \min (e \alpha, (d-r- j e   ) \beta).
\label{genral_property_exact_codes_case_2}
\end{align}
\end{lemma}

\begin{IEEEproof}
The result can be derived by proceeding as in \lref{entropy_lemma} and using the fact that $\mathcal{M}=f(\mathbf{u}_1)$ from \eref{eq119}.
\end{IEEEproof}

\begin{remark}
in the case of $e \nmid k$, a set of $e$ nodes are no longer linearly independent. This is expected as $e \alpha > d \beta$. Instead, it can be seen from \lref{dimension_of_stored_data_2nd_case} that any set of $r$ nodes are linearly independent.
\end{remark}

\begin{lemma}
\label{propo_multipl}
For exact-regenerating codes operating at the MBMR point, given sets $E,A,R$ and $B$ such that $|E|=e$, $E$ and $A$ are disjoint, $R$ and $B$ are disjoint, $|A|=je $ with $j \le \eta-1$, $|R|=r$ and $|B|=\eta e$, it holds that 
\begin{align}
&H(W_E )= d  \beta,\\
&H(W_E | W_A)=  (d-je) \beta,\\
&H(W_R | W_B)= (d-\eta e) \beta.
\end{align}
For linear codes, 
\begin{align}
&\dim(W_E )= d  \beta,\\
\label{second_equ}
&\dim(W_E)-\dim(W_E\cap W_A)=  (d-je) \beta,\\
\label{third_equ}
&\dim(W_R)-\dim(W_R \cap W_B)= (d-\eta e) \beta.
\end{align}
\end{lemma}
\begin{IEEEproof}
The result can be derived by proceeding as in \lref{entropy_lemma} and using the fact that $\mathcal{M}=f(\mathbf{u}_2)$ from \eref{second_eq_MBMR} and $e \alpha \geq d \beta, r\alpha \ge (d-ae)\beta$.
\end{IEEEproof}

It is easy to see that \lref{direct_sum_property} and \lref{general_intersection_dimension} hold true for the case $e \nmid k$, and for conciseness we do not repeat these lemmas.
The following lemma is used to derive the contradiction in our non-achievability result.
\begin{lemma}
\label{intersection_property_2}
Let  $k= \eta e +r$, then exact linear MBMR point is not achievable when $d>k$. When $d=k$,
%then at the MBMR point, 
for any set of $r+1$ nodes, it holds that
\begin{align}
\dim(W_{r+1} \cap W_{[ r]})=\beta.
\end{align}
\end{lemma}
\begin{IEEEproof}
%We note first that if $r+1=e$, the result follows from \lref{general_intersection_dimension}. 
We have
\begin{align}
d \beta&= \dim(W_{[ e]})  \\&=  \sum\limits_{i= 1}^e \dim(W_i)-\dim(W_i\cap W_{[ i-1]} ) \\
&= e \alpha - \sum\limits_{i= r+1}^e  \dim(W_i\cap W_{[ i-1]} ),
\end{align}
where the last equality follows from the fact that the first $r$ nodes are linearly independent. Thus, it follows that
\begin{align}
\sum\limits_{i= r+1}^e  \dim(W_i\cap W_{[ i-1]})= e \alpha - d \beta  =(e-r) (\alpha - \eta \beta).
\label{sum_intersect}
\end{align}
Now we write 
\begin{align}
(e-r) (\alpha - \eta \beta)&=\sum\limits_{i= r+1}^e  \dim(W_i\cap W_{[ i-1]}) \\&\geq \sum\limits_{i= r+1}^e  \dim(W_i\cap W_{[ r]})\\&=(e-r)  \dim(W_{r+1}\cap W_{[ r]}),
\end{align}
where the last equality follows using symmetry. Then, it follows that
\begin{align}
 \dim(W_{r+1}\cap W_{[ r]}) \le  \alpha - \eta \beta. 
 \label{inequality_r_1}
\end{align}
Combining \eref{sum_intersect} and \eref{inequality_r_1}, we obtain
 \begin{align}
\sum\limits_{i=r+2}^e   \dim(W_i \cap W_{[ i-1]})  \geq (e-r-1) (\alpha- \eta \beta).
\label{lower_bound}
\end{align}
On the other hand, we have
\begin{align}
\sum\limits_{i=r+2}^e   \dim(W_i \cap W_{[ i-1]})  &\le \sum\limits_{i=r+2}^e   \dim(W_i \cap W_{\mathcal{E}_i} ) \\
&= (e-r-1)  \beta,
\label{upper_bound}
\end{align}
where $\mathcal{E}_i $ is a set of $e$ nodes containing the first $i-1$ nodes and arbitrary $e-i+1$ nodes, excluding node $i$, and the equality follows from \lref{general_intersection_dimension}.
Combining \eref{lower_bound} and \eref{upper_bound}, it follows
\begin{align}
\label{semi_step}
(e-r-1) (\alpha- \eta \beta) \le \sum\limits_{i=r+2}^e   &\dim(W_i \cap W_{[ i-1]})   \le (e-r-1)  \beta.
\end{align}
It follows that $ \alpha- \eta \beta=  \frac{d -ae}{r}\beta \le \beta.$
The last inequality holds only when $d=k$ and $ \alpha- \eta \beta=\beta$. Indeed, when $d>k$, we have $\alpha-\eta\beta > \beta$. Therefore, we only consider the case $d=k$. %for which $\alpha=(a+1) \beta$ and $\alpha- a \beta= \beta$. 
Hence, it follows from \eref{semi_step} that
\begin{align}
\sum\limits_{i=r+2}^e   \dim(W_i \cap W_{[ i-1]}) 
= (e-r-1)  \beta.
\end{align}
Using \eref{sum_intersect}, we obtain $ \dim(W_{r+1} \cap W_{[ r]}) =    \beta$.
\end{IEEEproof}

%From the proof of the above lemma, we see that exact linear MBMR codes may only be feasible when $d=k$. The next theorem states that in fact such codes do not exist. 
\begin{theorem}
\label{non_existence_MBMR_2} 
Exact linear regenerating MBMR codes do not exist when $e<k$ and $e\nmid k$.
\end{theorem}
\begin{IEEEproof}
%\vspace{-.1cm}
From Lemma \ref{intersection_property_2}, exact linear MBMR codes may only be feasible when $d=k$. Next we show that in fact such codes do not exist.
Consider repair of the set of nodes $E$ containing nodes $1$ through $e$. Consider helper node $i$. As $\dim(W_i \cap W_{[ r]})=\dim(W_i \cap W_{[ e]})=\beta$, it follows that $W_i \cap W_{[ r]}= W_i \cap W_{[ e]}=S_i^{E}$. Then, each helper node sends vectors in the span of $W_{[ r]}$. Thus, the span of all sub-spaces $S_i^{[ e]}$ is included in the span of $W_{[ r]}$: $\sum\limits_i  S_i^{E} \subseteq W_{[ r]}$. This implies that  $\dim(\sum\limits_i S_i^{E})\le \dim(W_{[ r]}) $. Namely, we should have $d \beta \le r \alpha$: this is a contradiction as $d \beta > r \alpha$.
\end{IEEEproof}
%%%%%%%%%%%%%%%%%%%%%%%%%%%%%%%%%%%%%%%%%%%%%%%%%%%%%%%%%%%%%%%%%%%%%%%%%%%%%%%%%%%%%%%%%%%%%%%%%%%%%
\subsection{ Minimum bandwidth cooperative regenerating codes as centralized multi-node repair codes }\label{MBCR_codes}
In \cite{rawat2016centralized}, the authors argued that MBCR codes can be used as centralized multi-node repair regenerating codes. We recall that MBCR codes are characterized with
\begin{align}
(\alpha_{\text{MBCR}},\gamma_{\text{MBCR}}) = (\frac{\mathcal{M} (2 d +e-1)}{k(2d-k+e)}, \frac{\mathcal{M}  2 d e}{k(2d-k+e)}).
\end{align}
In the case of $e \mid k $, it is shown that MBCR codes achieve the MBMR bandwidth, i.e, $\gamma_{\text{MBCR}}= \gamma_{\text{MBMR}}$. In the case of $e \nmid k $, by imposing a certain entropy accumulation property on the entropy of any group of $r$ nodes, \cite{rawat2016centralized} showed that the bandwidth achieved by MBCR codes is optimal. It is important to note here that, from \eref{MBMR_gamma_e_not_k}, it can be checked that the entropy constraint condition results in $\gamma_{\text{MBCR}} >\gamma_{\text{MBMR}}$ for $e \nmid k$. Moreover, in both cases, it is not clear whether MBCR lies on the exact tradeoff of  centralized repair.
The next theorem determines the cases in which MBCR codes meet the centralized functional repair tradeoff (but does not correspond to the minimum bandwidth point on the functional curve). As a consequence, for such cases, MBCR codes meet the exact repair tradeoff as well.

\begin{theorem}
\label{MBCR_achievability}
Assume $1 < e \le k $, then, minimum bandwidth cooperative regenerating codes meet the centralized functional repair tradeoff if and only if $k \equiv 1 \mod e$.
\end{theorem}
\begin{IEEEproof}
When $e \mid k$, from \eref{MBMR_gamma_e_k} and \eref{MSMR_gamma_e_k}, it follows that $\gamma_{\text{MBMR}} = \gamma_{\text{MBCR}}$ and $\alpha_{\text{MBCR}} = \alpha_{\text{MBMR}}+ \frac{\mathcal{M}(e-1)}{k(2d-k+e)}$. Thus, 
  $\alpha_{\text{MBMR}} < \alpha_{\text{MBCR}}$. When  $e \nmid k$, from \eref{f_function_tradeoff} and \eref{MSMR_gamma_e_not_k},
%from \eref{f_function_tradeoff} and \eref{MSMR_gamma_e_not_k}, it one can heck that
%$
%f_r(0) =\frac{   2 e d \mathcal{M} }{k (2 d - k +e)   -e r - r^2}, 
%$. 
one can check that
$
\gamma_{\text{MBMR}} <  \gamma_{\text{MBCR}} < f_r(0) .
$
Using \eref{tradeoff_e_not_k}, it follows that the optimal storage size corresponding to $\gamma_{\text{MBCR}}$ and achieving the centralized functional repair tradeoff is given by 
\begin{align}
\alpha^*(\gamma_{\text{MBCR}})&=    \frac{  \mathcal{M}- \gamma_{\text{MBCR}} g_r(0)}  {r} 
\\&=   
 \frac{\mathcal{M} (2 d + e-r)}{k(2d-k+e)}\\
 &= \alpha_{\text{MBCR}}- \frac{\mathcal{M} (r-1)}{k(2d-k+e)}.
\end{align}
Therefore, $\alpha^*(\gamma_{\text{MBCR}}) \le \alpha_{\text{MBCR}}$ with equality if and only if $r=1$.
\end{IEEEproof}

\begin{center}
\begin{figure}
\center
\includegraphics[width=.6\linewidth]{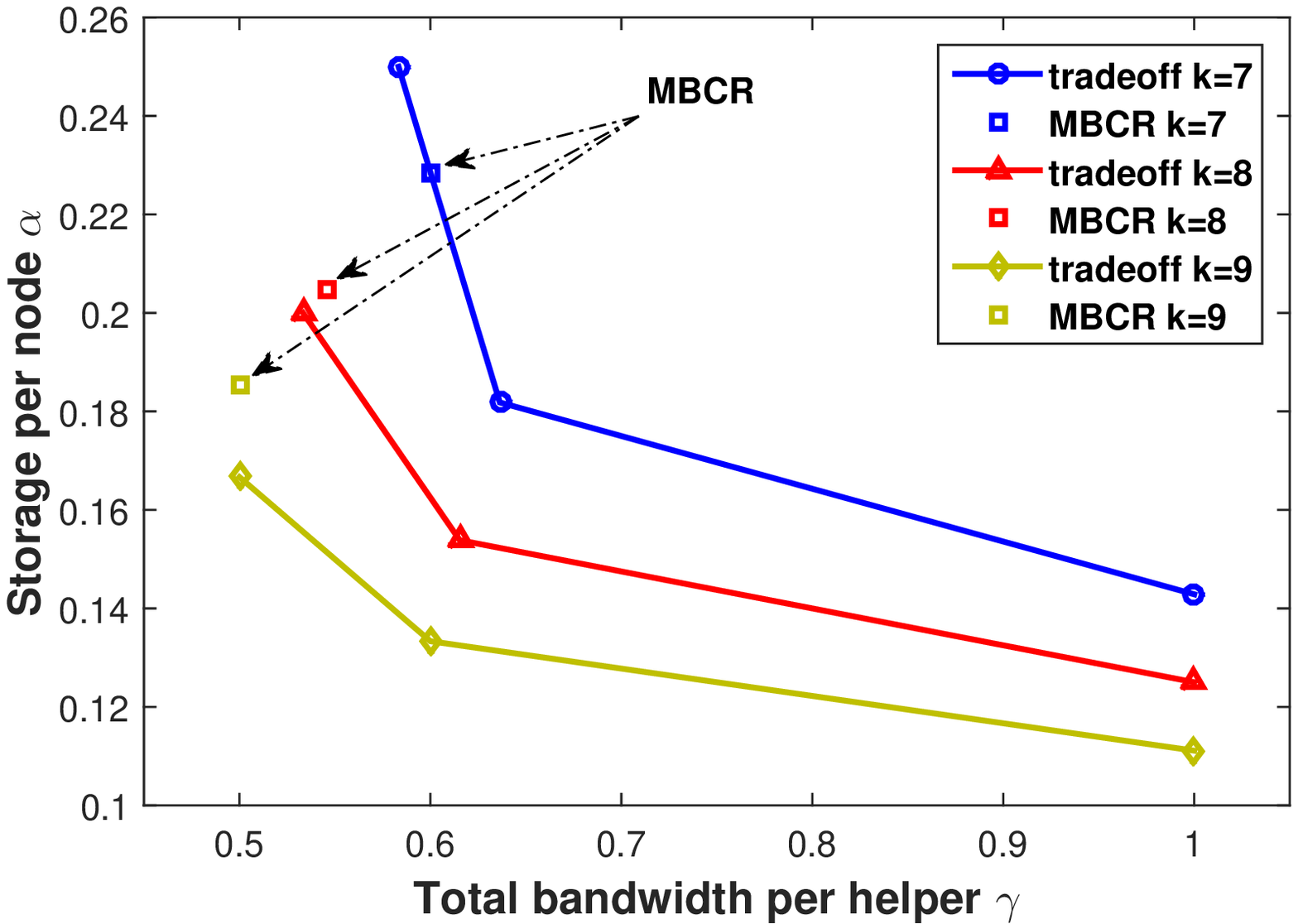}
\caption{Optimal repair tradeoffs for fixed $e=3$ and different $k \in \{7,8,9 \}$. The MBCR point lies on the tradeoff only in the case of $k=7$. 
%\bl{??can you use a different marker for the MBCR points? Or can you use arrows to point to MBCR points? It is  difficult to differentiate them from tradeoff?? }
}
\label{mbcr}
 \end{figure}
 \end{center}
 
\fref{mbcr} illustrates the functional tradeoff for fixed $e=3, \mathcal{M}=1$ and multiple values of $k \in \{7,8,9 \}$ such that $d=k$. As proved in Theorem \ref{MBCR_achievability}, MBCR codes are optimal centralized repair codes only when $r=1$, which corresponds to $k=7$ in \fref{mbcr}. 
When $e \mid k$, MBCR codes achieve the same bandwidth as MBMR codes, but have a higher storage cost. 
%In \sref{MBMR_codes}, we investigate the achievability of MBMR codes under exact repair.
\begin{remark}
\label{achievability_Interior}
Theorem \ref{MBCR_achievability} proves that, when $e \nmid k, r=1$, MBCR codes achieve an interior point on the functional tradeoff that lies near the MBMR point. We note that the existence of this exact-repair interior point does not contradict the infeasibility result in \sref{interior_points}, where we assume $e|k$.
\end{remark}
%%%%%%%%%%%%%%%%%%%%%%%%%%%%%%%%%%%%%%%%%%%%%%%%%%%%%%%%%%%%%%%%%%%%%%%%%%%%%%%%%%%%%%%%%%%%%%%%%%%%%%%%%%%%%%%%%%%%%%%%%%%%%%%%%%%%%%%%%%%%%%%%%%
\section{Infeasibility of the exact- repair interior points}\label{interior_points}
In this section, we study the infeasibility of the interior points on the optimal \emph{functional-repair} tradeoff for $e \mid k, e \mid d, 2 e<k$, similarly to \cite{non_achievability}. We note that all interior points satisfy $(d-k+e)  \beta \le  e \alpha \le d \beta$. This can be written as $(d'-\eta+1)\beta \le \alpha \le d' \beta$, where $d^{'}=\frac{d}{e}$ and $\eta=\frac{k}{e} $. This is similar to the single erasure case with reduced parameters. The proof techniques in this section follow along similar lines as \cite{non_achievability} and some of the proofs are relegated to the appendix.

\vspace{0.3em}
{\bf Parameterization of the interior points.}
Let $\alpha=(d'-p) \beta -\theta$, namely $e \alpha=(d -e p) \beta -e \theta$ with $p \in \{0,1,\ldots,\eta-1 \}$, $\theta \in [0, \beta)$ such that $\theta=0$ if $p=\eta-1$.
Points on the functional tradeoff satisfy 
\begin{align*}
\mathcal{M} = e \sum\limits_{i=0}^{\eta-1} \min (\alpha, (d'-i) \beta).
\end{align*}
\subsection{Properties of exact-repair codes}
We present a set of properties that exact-repair codes, satisfying the optimal functional tradeoff, must satisfy. 
\begin{lemma} 
\label{prop_1}For a set $A$ of arbitrary nodes of size $e j$, a set $L$ of nodes of size $e$ such that $L \cap A = \emptyset $, we have 
\begin{align}
 I(W_L, W_A)= 
 \begin{cases}
0, &\qquad j \le p ,\\
  e((j-p) \beta - \theta ), &\qquad p < j <\eta,	\\
 e \alpha,     &\qquad    j \geq \eta.
\end{cases}
\end{align}
\end{lemma}
\begin{IEEEproof}
See \appref{prop_1_proof}.
\end{IEEEproof}
%\begin{IEEEproof}
%First, we note that when $j \geq \eta$, $I(W_L,W_A)=H(W_L)-H(W_L|A)=H(W_L)= e \alpha$. In the following, we assume $j <\eta$. We write
%\begin{align}
%I(W_L, W_A)&=H(W_L)-H(W_L|A) \nonumber\\
%& =e \alpha - \min (e \alpha, (d-j e) \beta) \label{eq124} \\
%&=e (\alpha-\min(\alpha,(d'-j)\beta))\nonumber\\
%&= e (\alpha- (d'-j)\beta)^+ \nonumber\\
%&= e ((j-p)\beta-\theta)^+,\nonumber
%\end{align} 
%\noindent where we use the notation $(x)^+ \triangleq \max(x,0)$. 
%Here \eqref{eq124} follows from Lemma \ref{entropy_lemma}.
%\end{IEEEproof}
%\begin{remark}
%Note here that only when $e\mid k$ does \eqref{eq124} hold, when $e \nmid k$, $H(W_L)<e\alpha$. In fact, as pointed out in Remark \ref{achievability_Interior}, the infeasibility result does not hold for $e \nmid k$.
%\end{remark}
\begin{corollary}\label{cor:H_W_given_S} 
For an arbitrary set $L$ of size $e$, and a disjoint set $A$ such that $|A|=e m < k$ for some integer $m$, we have
\begin{align}
H(W_L|S_A^L)=H(W_L|W_A)=\min (e \alpha, (d-e m)\beta).
%= e \min (  \alpha, (d'-m)\beta).
\end{align}
\end{corollary}
\begin{IEEEproof}
From \lref{inequality_term}, we have $H(W_L|S_A^L) \le \min (e \alpha, (d-e m)\beta)$. On the other hand, from \lref{entropy_lemma},
\begin{align}
H(W_L|S_A^L) \geq H(W_L|W_A)=  \min (e \alpha, (d-e m)\beta) .
\end{align}
Thus, $H(W_L|S_A^L)=H(W_L|W_A)=\min (e \alpha, (d-e m)\beta).$
\end{IEEEproof}
\begin{lemma}\label{lem:contribution_of_nodes}
In the situation where node $m$ is an arbitrary helper node assisting in the repair of a second set of arbitrary nodes $L$ of size $e$, we have
\begin{align}
H(S_m^L)=\beta,
\end{align}
irrespective of the identity of the other $d-1$ helper nodes. Moreover, for set $B$ of size $|B|\le d-k+e$ with $B \cap L= \emptyset$, we have 
\begin{align}
H(S_B^L)=|B| \beta.
\label{contribution_of_nodes}
\end{align}
\end{lemma}
\begin{IEEEproof}
See \appref{lem:contribution_of_nodes_proof}.
\end{IEEEproof}
%\label{lem:contribution_of_nodes}
%\begin{IEEEproof}
%%Similar to the proof of \lref{general_intersection_dimension}.
%%
%Partition the set of $d$ helpers into $A$ and $B$ such that $|A|=k-e$ and $|B|=d-k+e$, such that $m \in B$. We have $H(W_L|S_A^L)=\min(e \alpha, (d-k+e) \beta)=(d-k+e) \beta$, as $e \alpha \geq (d-k+e) \beta$	for all points on the tradeoff. Moreover, exact repair requires $H(W_L|S_A^L, S_B^L)=0$. Thus, $H(S_B^L) \geq (d-k+e) \beta$. This implies $H(S_B^L) = (d-k+e) \beta$. Moreover, it must hold that $H(S_m^L) = \beta$ in addition to $S_{m}^L$ and $S_{m'}^L$ being independent if $m \neq m'$. Moreover, by choosing $M \subseteq B$, one obtains $H(S_M^L)=e \beta$.
%\end{IEEEproof}

\vspace{0.3em}{\bf Helper node pooling.}
Consider a set $F$ consisting of a collection of $f \le d+e$ nodes ($f$ is a multiple of $e$), and a subset $R$ of the set $F$ consisting of $e r'$ nodes, $r' \ge 1$. A helper node pooling scenario is a scenario where upon failure of any $e$ nodes $L \subseteq R$, the $d$ helper nodes include all the $f-e$ remaining nodes in $F$. The remaining $d-f+e$ helper nodes are fixed given $L$. %denoted by $\mathcal{V}(L)$. 
Consider a subset of nodes $M \subseteq F \backslash R$. Partition the nodes in $R$ into arbitrary but fixed sets $
R_1,R_2,\ldots, R_{r'}, 
$
each of size $e$. Denote by $S_M^R=(S^{R_1}_M, \dots S^{R_{r'}}_M)$ the collective transmitted information from helper nodes $M$ to repair $R_1, \dots, R_{r'}$, respectively.  %Let $|R|= r' e$.
\begin{lemma}
\label{proof_helper_node_pooling}
In the helper node pooling scenario where
$\min(\eta,\frac{f}{e}) > p+2 \geq r'$, for any set of $e$ arbitrary nodes $M \subseteq F\backslash R$, we have
\begin{align}
H(S_M^R) \le e ( 2 \beta - \theta ).
\end{align}
\end{lemma}
\begin{IEEEproof}
See \appref{proof_helper_node_pooling_proof}.
\end{IEEEproof}
%\begin{IEEEproof} 
%If the statement holds true for some $f,r'$, then it also holds true for all $f' \geq f$ and $r'' \le r'$. Thus, for the proof, we only need to consider $F= R\cup   M , |F|=f=e(p+3), |R|=r'e = (p+2)e, |M| = e$.
% 
%Consider repair of an arbitrary set of $e$ nodes $L \subseteq R$, where the set of helpers include $M$ and the $e(p+1)$ remaining nodes in $R$. Then, we write
%\begin{align}
%I(S_M^L;W_R)&= I(S_M^L;W_L, W_{R-{L} })\nonumber\\
%&= I(S_M^L; W_{R-{L} })+ I(S_M^L;W_L| W_{R-{L} })\nonumber \\
%&\geq I(S_M^L;W_L| W_{R-{L} }) \nonumber\\
%& = H(W_L| W_{R-{L} })- H(W_L| W_{R-{L} }, S_M^L)\nonumber\\
%& \geq H(W_L| W_{R-{L} })- H(W_L| S_{R-{L} }^L, S_M^L)\nonumber\\
%&= \min(e \alpha, (d-e(p+1))\beta)-  \min(e \alpha,( d-e(p+2))\beta) \label{eq137}\\
%&=   (d-e(p+1))\beta -  ( d-e(p+2))\beta = e \beta.\nonumber
%\end{align}
%Here \eqref{eq137} follows from Lemma \ref{entropy_lemma} and Corollary \ref{cor:H_W_given_S}.
%Then, we obtain
%\begin{align}
%H(S_M^L|W_R)= H(S_M^L)-I(S_M^L;W_R) \le e \beta - e \beta =0.
%\end{align}
%Hence, $H(S_M^L| W_R)=0.$ Since $L$ is arbitrary, it follows that $H(S_M^{R} | W_R)=0$.
%It follows from \lref{prop_1} that 
%\begin{align*}
%H(S_M^R)= I(S_M^R; W_R) \le I(W_M;W_R)=e (2 \beta- \theta).
%\end{align*}
%Hence the proof is completed.
%\end{IEEEproof}
\begin{lemma}
\label{prop_5}
In the helper node scenario where $\min\{ \eta,\frac{f}{e} \} > p+1 \geq r'$, for an arbitrary set of $e$ nodes $M \subseteq F \backslash R$, and an arbitrary pair of set of $e$ nodes $L_1,L_2$, it must be that
\begin{align}
H(S_M^{L_1}|S_M^{L_2}) \le e \theta,
\end{align}
and hence
\begin{align}
H(S_M^R) \le e (\beta + (r'-1)\theta).
\end{align}
\end{lemma}
\begin{IEEEproof}
See \appref{prop_5_proof}.
\end{IEEEproof}
\subsection{Non-existence proof}
For interior points, $1 \le p \le \eta-2$.
First, we consider the interior points for which $e\alpha$ is a multiple of $\beta$. That is: $e \alpha= (d- e p) \beta, \theta=0$. %with $p$ lying in the range $1 \le p \le a-2$.
\begin{theorem}
Exact-repair codes do not exist for the interior points with $\theta=0$.
\label{non_achivability_1}
\end{theorem}
\begin{IEEEproof}
Consider a sub-network $F$ consisting of $d+e$ nodes. 
The parameters satisfy the condition in Lemma \ref{prop_5}.
Note that by the regeneration property for any set of $e$ nodes $L \subseteq F$, $H(W_L|S_{F-L}^L)=0$. Moreover, for distinct $M,L_1,L_2 \subseteq F$, with $\theta=0$, we have $H(S_{M}^{L_1}|S_{M}^{L_2})=0 $.
%Let $d'= \floor*{\frac{d+e}{e}}$. We partition the nodes into $F'=d'+1_{\{e \nmid k\}}$ groups, where all the groups, except the last one, contain e nodes, while the last group contains ($d\mod e$) nodes.
We partition the nodes in $F$ into groups of size $e$, denoted $L_i, i=1,2,\dots,d'+1$. Then, we write
\begin{align}
 \mathcal{M}\le H(W_F) \le H(   S_{F-L_1}^{L_1} ,\dots,  S_{F-L_{d'+1}}^{L_{d'+1}}     )
 &= H(   S^{F-L_1}_{L_1} ,\dots,  S^{F-L_{d'+1}}_{L_{d'+1}}     )\nonumber\\
 & \le \sum\limits_{i=1}^{d'+1}  H(    S_{L_i}^{F-L_i}       ) \nonumber\\
&\le  \sum\limits_{i=1}^{d'+1}   e \beta \label{eq162}\\
&= (d+e) \beta,\nonumber
\end{align}
where the inequality \eqref{eq162} follows from \lref{prop_5}.
On the other hand,
\begin{align}
\mathcal{M}= \sum_{i=0}^{d'-1} \min(e \alpha,(d-ie)\beta) =  \sum_{i=0}^{d'-1} \min((d-e p)\beta,(d-ie)\beta)
&= 2(d-e p) \beta + \sum_{i=2}^{d'-1}  \min((d-e p)\beta,(d-ie)\beta) \nonumber\\
& \geq  2(d-e p) \beta +   (\eta-2) e \beta \nonumber\\
& \geq 2 e \beta + (d-e p) \beta +   (\eta-2) e \beta \nonumber\\
&= (d-2 e)   \beta +    (k -2 e-ep)  \beta\nonumber\\
& \geq  (d-2 e)   \beta,
\end{align}
\noindent where we assume $1 \le p \le \eta-2$ (non-MSMR point). Thus, $e p + 2e \le k    \le d$. Both bounds are contradictory, thus proving the impossibility result in the case of $\theta=0$.
\end{IEEEproof}
\begin{theorem}
\label{non_achivability_2} 
For any given values of $\mathcal{M}$, exact-repair regenerating codes do not exist for the parameters lying in the interior of the storage-bandwidth tradeoff when $\theta \neq 0$, except possibly for the case $p+2 =\eta$ and $\theta \geq  \frac{d-ep -e}{d- ep} \beta.$
\end{theorem}
\begin{IEEEproof}
See \appref{non_achivability_2_app}.
\end{IEEEproof}
%%%%%%%%%%%%%%%%%%%%%%%
\section{Adaptive multi-node repair for MBR codes}
\label{adaptive_MBR}
%From \eqref{MSMR_gamma_e_k} and \eqref{MSMR_gamma_e_not_k} one can see that when the number of helpers $d$ (or the number of failures $e$) changes, the storage size of the MBMR code also changes. Namely, a code cannot be a MBMR code for different $d$ (or different $e$) under functional repair, where $e \ge 1$. 
%Moreover,
In this section, we study multi-node repair for MBR codes, allowing a varying number of helpers and a varying number of failures. 
In \sref{MBMR_codes}, we proved that MBMR codes are not achievable for linear exact repair codes, when $2\le e <k$. When $e=1$, exact MBMR codes are MBR codes and their existence is well established in the literature \cite{Rashmi_Product_Matrix}. 
\emph{Adaptive} regenerating codes possess the extra feature that the number of helpers involved in the repair process can be adaptively selected, which provides the storage system with robustness to the network varying conditions \cite{kermarrec2011repairing,aggarwal2014distributed}. Adaptive MSR codes have been constructed in \cite{ye2017explicit}. On the other hand, adaptive MBR codes have been investigated in \cite{mahdaviani2016bandwidth}, in which case optimal repair means that the total repair bandwidth for each number of helpers $d$ is the lowest possible, and is given by $\gamma = \alpha, \forall d_{\text{min}} \le d \le d_{\text{max}}$ (assuming the storage per node contains no redundancy). Here $d_{\text{min}}, d_{\max}$ are between $k$ and $n-1$. It is shown in \cite{mahdaviani2016bandwidth} that adaptive MBR codes, designed for arbitrary $d$, $d_{\text{min}} \le d \le d_{\text{max}}$, are equivalent to MBR codes that are designed for the worst-case number of helpers $d_{\text{min}}$, and they satisfy optimal repair for arbitrary number of helpers $d_{\text{min}} \le d \le d_{\text{max}}$. 
Namely, adaptive MBR codes satisfy  for any $d_{\text{min}} \le d \le d_{\text{max}}$, 
\begin{align}
d\beta=\alpha,
\end{align}
\begin{align}\label{eq_adaptive_alpha}
\alpha= \frac{2  d_{\min} \mathcal{M}}{-k^2+k+2 k d_{\min} }=\frac{d_{\min}  \mathcal{M}}{d_{\min}  k-  \binom{k}{2}},
\end{align}
where the storage size $\alpha$ corresponds to the MBR code with $d_{\min}$ helpers.

A natural question of interest is whether there exists an MBR code that efficiently recover from varying number of failures simultaneously. In this section, we investigate the problem of repairing multiple failures in MBR codes under exact repair, for \emph{varying number of helpers $d$ and varying number of failures $e$}, such that
$d_{\text{min}} \le d \le d_{\max}$,
$1 \le e \le k$, $e+d \le n$.
First, we derive a lower bound on the multi-node repair bandwidth for MBR codes,
which applies to exact and functional codes. We assume that an MBR code is designed for $d$ helpers, and we want to repair $e$ failures. To emphasize the dependency on $e$, denote the total repair bandwidth by $\gamma_{\text{MBR}}(e)$.
\begin{theorem}
\label{MBR_costruction}  
Consider an $(n,k,d,\alpha,\beta )$ MBR regenerating code, the total repair bandwidth $\gamma_{\text{MBR}}(e)$ needed to repair any set of $1 \le e\le k$ nodes satisfies 
\begin{align}
\gamma_{\text{MBR}}(e) \geq   e \alpha - {e \choose 2}    \frac{\alpha}{d}.
\label{min_bdw_mbr}
\end{align}
\end{theorem}

\begin{IEEEproof}
Assume w.l.o.g that the first $e$ nodes are to be repaired. From \cite{rashmi2009explicit}, at the MBR point, for any set of $A$ nodes of size $m<k$ and for $i \notin A$, we have $H(W_i | W_A)= (d-m) \beta$. Therefore,
\begin{align}
H(W_{[e]})&= \sum\limits_{i=1}^e H(H_i | W_{[i-1]})= \sum\limits_{i=1}^e (d-i+1) \beta =
(e d - {e \choose 2}  ) \beta.
\end{align}
Noting that at the MBR, $\alpha= d \beta$, \eref{min_bdw_mbr} follows.
\end{IEEEproof}
We now briefly describe a construction of adaptive MBR codes that simultaneously and efficiently repair single node failures, presented in \cite{mahdaviani2016bandwidth}. Then, we show how to optimally repair multiple failures in this construction.
\subsection{Adaptive single-failure MBR construction}
The construction is based on product matrix codes \cite{mahdaviani2016bandwidth, Rashmi_Product_Matrix}. %We first review the repair of single failure, and then present our scheme for multiple failures. 
Let $\alpha= \prod\limits_{d=d_{\text{min}}}^{d_{\text{max}}}d$  . Define $z=\frac{\alpha}{d_{\text{min}}}$ and construct the $(\alpha \times \alpha)$ data matrix $M$ as 
\begin{align}
M=\begin{bmatrix}  M_1 & O & \cdots &  O \\ 
O & M_2 & \cdots & O     \\
\vdots && \ddots&  \vdots \\
O & \cdots & O & M_z
\end{bmatrix},
\end{align}
where $O$ is a $(d_{\text{min}} \times d_{\text{min}})$ zero matrix and each of the submatrices $M_i$ is filled with information symbols, and is  symmetric and satisfies the structural properties of a product-matrix MBR code for parameters $k$ and $d_{\text{min}} $. For instance, $M_i$ is given by
\begin{align}
M_i= \begin{bmatrix}
N_i& L_i \\
L_i^t & O^{'}
\end{bmatrix},
\end{align}
 where $N_i$ is a symmetric $(k \times k)$ matrix, $L_i$ is $(k \times (d_{\text{min}}-k ))$  matrix, and $O^{'}$ is $(d_{\text{min}}-k) \times (d_{\text{min}}-k)$ zero matrix. let $\Psi$ be an $(z n \times d_{\text{min}})$ Vandermonde matrix, with rows denoted by $\mathbf{\psi}_j^t$, for $1 \le j \le z n$. Then, storage node $l$ is associated with 
\begin{align*}
\mathbf{w}_l^t= \begin{bmatrix}  \mathbf{\psi}_{(l-1)z+1}^t, \ldots, \mathbf{\psi}_{lz}^t\end{bmatrix} M= 
\begin{bmatrix}  \mathbf{\psi}_{(l-1)z+1}^t M_1, \ldots, \mathbf{\psi}_{lz}^t M_z\end{bmatrix}.
\end{align*}  
\vspace{0.3em}
\noindent{\bf Single node repair.}
Denote the set of helpers by $\mathcal{H}$ such that $|\mathcal{H}|=d $ and $ d_{\text{min}} \le d  \le d_{\text{max}}$. Let $\Omega$ be an $(z \times z)$ matrix such that $\Omega^t$ is a Vandermonde matrix. Assume that node $f$ fails. Let $\Omega_d$ an $(\frac{\alpha}{d} \times z)$ containing the first $\frac{\alpha}{d}$ rows of $\Omega$. Moreover, let $\Phi_i$  be an $(\alpha \times z)$ matrix
\begin{align*}
\Phi_i = \begin{bmatrix} \mathbf{\psi}_{(i-1)z+1 }\\
& \ddots& \\
& & \mathbf{\psi}_{i z} 
\end{bmatrix}.
\end{align*}
Each helper node $i_j \in \mathcal{H}$ transmits $\mathbf{s}_{i_j,f}^t=\mathbf{w}_{i_j}^t \Phi_f \Omega_d^t$. After simplification, the replacement node obtains
\begin{align}
\mathbf{w}_{f}^t \begin{bmatrix}
\Phi_{i_1} \Omega_d^t, \ldots, \Phi_{i_d} \Omega_d^t
\end{bmatrix}= \mathbf{w}_{f}^t \Theta_\mathcal{H}
\end{align}
Noting that $\Theta_\mathcal{H}$ is invertible \cite{mahdaviani2016bandwidth}, the replacement node can thus recover $\mathbf{w}_{f}^t$.
\subsection{Adaptive multi-node repair in MBR codes}
We state our result in the following theorem.
\begin{theorem}
\label{multiple_erasures}
Adaptive single-failure MBR regenerating codes with storage per node $\alpha$ and arbitrary number of helpers $d_{\text{min}} \le d \le d_{\text{max}}$, presented in\cite{mahdaviani2016bandwidth}, can simultaneously and optimally repair $e$ failures with $d$ helpers, for all $d_{\text{min}} \le d \le d_{\max}$, 
$1 \le e \le k$, $e+d \le n$.
\end{theorem}
\begin{IEEEproof}
Assume w.l.o.g that the first $e$ nodes failed and $d$ helpers are used, where  $d_{\text{min}} \le d \le d_{\max}$, 
$1 \le e \le k$, $e+d \le n$. Denote the helpers by the set $\mathcal{H}=\{ i_1,\ldots, i_d\}$. First, the repair of node $1$ is done by contacting all the $d$ helpers and downloading $\frac{\alpha}{d}$ symbols from each one of them, using the procedure described for single node repair. Node $2$ is then repaired using only $d_{\text{min}}$ helpers, comprising repaired node $1$ and any other $d_{\text{min}}-1$ helpers in $\mathcal{H}$, such that each helper provides $\frac{\alpha}{d_{\text{min}}}=z$  symbols. The same procedure is then applied repeatedly until recovering the last node $e$ by contacting any $d_{\text{min}}-e+1$ helpers in $\mathcal{H}$ and using contributions from the $e-1$ already repaired nodes. The overall repair bandwidth is given by
\begin{align}
d \frac{\alpha}{d}+ \sum\limits_{i=1}^{e-1} z (d_{\text{min}} -i)= e \alpha - {e \choose 2}    \frac{\alpha}{d_{\text{min}}},
\end{align}
which matches the bound in \eref{min_bdw_mbr}, establishing the optimality of the repair procedure.
\end{IEEEproof}
\begin{remark}
Repairing $e$ failures in an $(n,k,d_{\text{min}},\alpha,\beta )$ MBR code separately requires a bandwidth of size $e \alpha$. However, simultaneously repairing $e$ failures using $d \geq d_{\text{min}} $ reduces the bandwidth by $ {e \choose 2}    \frac{\alpha}{d_{\text{min}}}$  .
\end{remark}
\begin{remark}
The repair procedure of multiple erasures in \thref{multiple_erasures} is asymmetric. However, one can always duplicate the code a sufficient number of times to achieve a symmetric repair strategy (e.g. \cite{elyasi2015linear}). 
\end{remark}

%%%%%%%%%%
\section{Conclusion}\label{conclusion}
We studied the problem of centralized repair of multiple erasures in distributed storage systems. We explicitly characterized the optimal functional tradeoff between the repair bandwidth and the storage size per node. For instance, we obtained the expressions of the extreme points on the tradeoff, namely the minimum storage multi-node repair (MSMR) and the minimum bandwidth multi-node repair (MBMR) points.
In the case of $e \ge k$, we showed that the tradeoff reduces to a single point, for which we provided a code construction achieving it.
We described a general framework for converting single erasure minimum storage regenerating codes to MSMR codes. Then we applied the framework to product-matrix codes and interference alignment codes. 
Furthermore, we proved that the functional MBMR point is not achievable for linear exact repair codes for $1 <e < k$. We also showed that the functional repair tradeoff is not achievable under exact repair, except for maybe a small portion near the MSMR point for $e < k, e\mid k, e \mid d$.
Finally, we presented an MBR code that can adaptively and optimally repair varying number of failures with varying number of helpers.

Open problems include the generalization of the non-existence proof of linear exact-repair MBMR regenerating codes to non-linear codes. It is interesting to determine the storage and bandwidth values of an exact minimum bandwidth regenerating code. Moreover, characterization of the storage-bandwidth tradeoff for exact repair for the interior points is still not known. 
%Last but not least, reducing the subpacketization size for high rate exact-repair MSMR codes is an important direction to study for its practical implications.
%%%%%%%%%%%%%%%%%%%%%%%%%%%%%%%%%%%%%%%%%%%%%%%%%%%%%%%%%%%%%%%%%%%%%%%%%%%%%%%%%%%%%%%%%%%%%%%%%%%%%%%%%%%%%%%%%%%%%%%%%%%%%%%%%%%%%%%%%%%%%%%%%%%%%%%%%%%%%%%%%%%%%%%%%%%%%%%%%%%%%%%%%%%%%%%%
\appendix\label{Appendices}
%\section{Appendices}
\subsection{proof of \lref{alpha_c_lemma}}
\label{General_case}
%Recall that when $e \nmid k$, the optimal scenario is one of $\eta+1$ different scenarios, which differ by the position of the repair group with $r$ nodes. For that, we need to study the behavior of the $\eta+1$ different cuts. 
We first state the following lemma which will be useful in the proof.
\begin{lemma}
\label{final_value}
For fixed $\beta$, the scenario $\mathbf{u}=[ e,\ldots,e, r ]$ achieves the lowest final value of minimum cut:
\begin{align}
\lim\limits_{\alpha \to +\infty} f(\mathbf{u}) \geq \lim\limits_{\alpha \to +\infty} f([ e,\ldots,e, r ]), \forall \mathbf{u} \in \mathcal{P},
\end{align}
where $f(\mathbf{u})$ and $\mathcal{P}$ are defined in \eref{f_definition} and \eref{def_P}, respectively.
\end{lemma}
\begin{IEEEproof}
for a specific cut $\mathbf{u} $, we have
\begin{align}
\lim\limits_{\alpha \to +\infty} f(\mathbf{u})
&= \sum\limits_{i=1}^{g} (d-\sum\limits_{j=1}^{i-1} u_j) \beta\nonumber\\
&= d \beta g -\beta \sum\limits_{i=1}^{g} \sum\limits_{j=1}^{i-1} u_j= g d \beta - \beta \sum\limits_{i=1}^{g-1} u_i (g-i)\nonumber\\
&= \beta (d g - g \sum\limits_{i=1}^{g-1} u_i +   \sum\limits_{i=1}^{g-1} i u_i)= \beta (  (d-k)g + \sum\limits_{i=1}^{g } i u_i).
\end{align}
To obtain the smallest minimum cut value, we need to solve the following problem 
\begin{equation}
\begin{aligned}
& \underset{\mathbf{u},g}{\text{minimize}}
& & (d-k)g + \sum\limits_{i=1}^{g } i u_i \\
& \text{subject to}
& & 1 \le u_i \le e,\\
& & &\sum\limits_{i=1}^{g} u_i= k.
\end{aligned}
\label{MBMR_optimization}
\end{equation}
It can be seen that the solution to \eref{MBMR_optimization} is given by $\mathbf{u}=[ e,\ldots,e, r ]$.
\end{IEEEproof}
%It is worth mentioning at this point that $C_0(\alpha)$ has the highest final value among all other curves $C_j, j\geq 1$. This can be seen from \eref{MBMR_optimization}, as for $g=\eta+1$, the maximum of $\sum\limits_{i=1}^{\eta+1} i u_i$ is obtained by $\mathbf{u}=[e,\ldots,e,r]$. This observation will used later in the proof.
%Moreover, $C_{\eta}$ achieves the lowest final value.

\begin{center}
\begin{figure}
\center
\includegraphics[width=.5\linewidth]{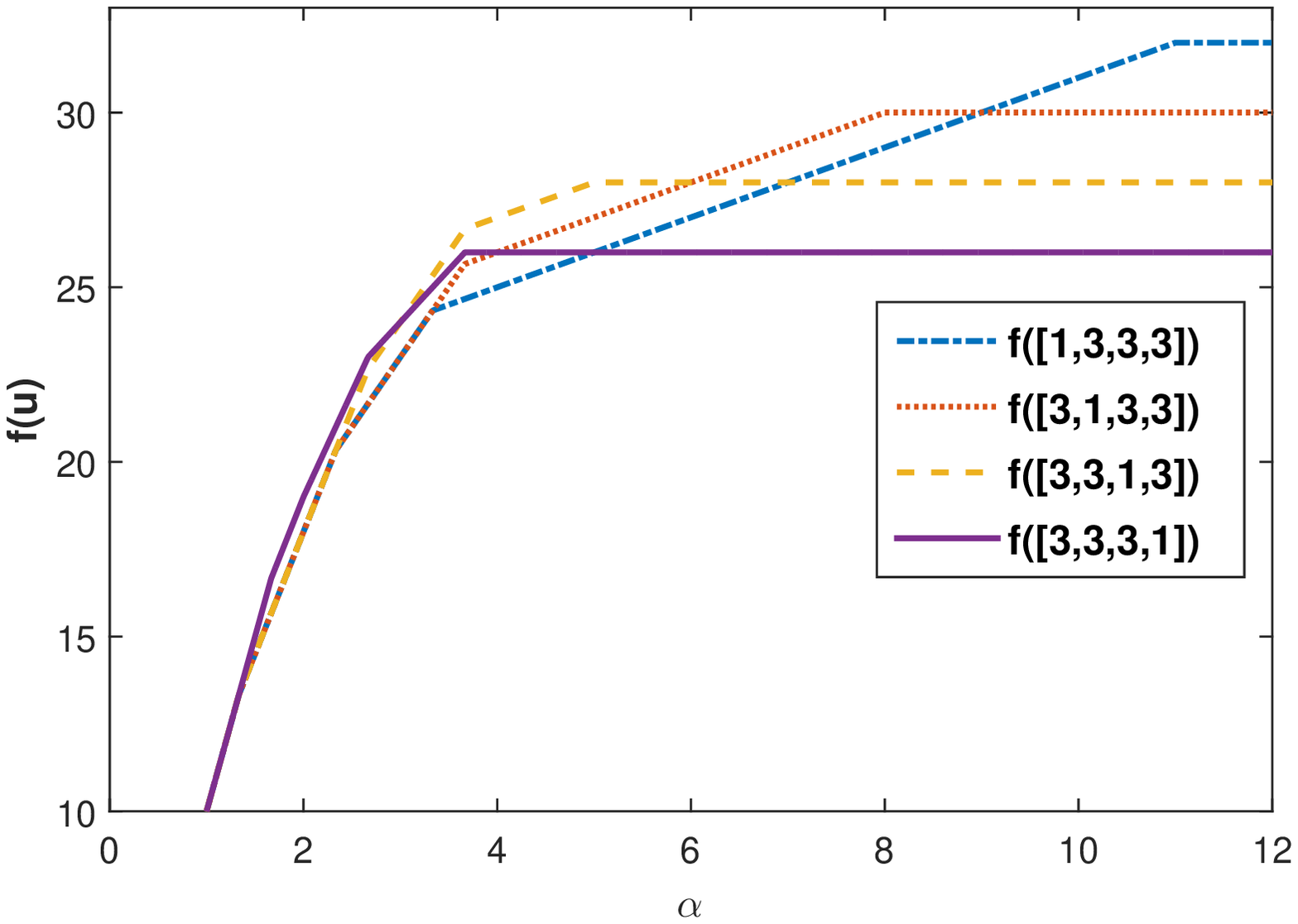}
\caption{Values of the cut function for different vectors $\mathbf{u}$ for $k=9,d=10, \beta=1$}
\label{example_curves}
 \end{figure}
 \end{center}
We now study the different functions $C_j(\alpha)$ for $j=0,\ldots,\eta$. An example of the different functions to be analyzed is given in \fref{example_curves}, with $k=9,d=10, \beta=1$. It is observed that $\mathbf{u}=[1,3,3,3]$ generates the lowest cut before some threshold $\alpha^*=5$, after which the lowest cut is generated by $\mathbf{u}=[ 3,3,3,1]$. In the following, by analyzing $C_j(\alpha)$ for $j=0,\ldots,\eta$, we prove that the above observation holds true in general. 
\paragraph{j=0}
we have 
\begin{align}
C_0(\alpha)&= \min(r \alpha, d \beta)+ \sum\limits_{i=0}^{\eta-1}  \min(e \alpha, (d-r-ie) \beta)
= r \min( \alpha, \frac{d \beta}{r})+ \sum\limits_{i=0}^{\eta-1}  e\min( \alpha, \frac{(d-r-ie) \beta}{e})  .  
\end{align}
%$C_0(\alpha)= \min(r \alpha, d \beta)+ \sum\limits_{i=0}^{\eta-1}  \min(e \alpha, (d-r-ie) \beta)= r \min( \alpha, \frac{d \beta}{r})+ \sum\limits_{i=0}^{\eta-1}  e\min( \alpha, \frac{(d-r-ie) \beta}{e}) $. 
$C_0(\alpha)$ is a piecewise linear function with breakpoints given by $\{ \frac{d-r-(\eta-1)e}{e} \beta, \frac{d-r-(\eta-2)e}{e}\beta , \ldots, \frac{d-r}{e}\beta, \frac{d}{r}\beta \}$. $C_0$ increases from 0 at a slope of $k$. Its slope is then reduced by $e$ by the successive breakpoints and then finally by $r$ until it levels off.
%For $\alpha \le\frac{d-r-(a-1)e}{e} \beta, C_0(\alpha)= k \alpha $. Then, at the second segment, $C_0(\alpha)$ increases by $r \beta$. $C_0(\alpha)$ increases by $e \beta$ in the following segments, which has length of $ \beta$. In last linear increasing segment $[\frac{d-r}{e}\beta,\frac{d}{r}\beta]$, whose width is larger than  $\beta$, $C_0 (\alpha)$ increases at a rate of $e$, before leveling off. Let $\eta_0=\frac{d-r-(\eta-1)e}{e} \beta$ and $b_0=\frac{d}{r}\beta$ be the extreme points of the breakpoints of $C_0$.
\paragraph{$1 \le j \le \eta$}
for each $j$, we have
\begin{align}
C_j(\alpha)&= \sum\limits_{i=0}^{j-1}  \min(e \alpha, (d-ie) \beta)
+\min(r \alpha, (d-j e ) \beta) + 
\sum\limits_{i=j}^{\eta-1}  \min(e \alpha, (d-r-ie) \beta) \nonumber\\
&= \sum\limits_{i=0}^{j-1} e \min( \alpha, \frac{(d-ie) \beta}{e})
 + r \min(  \alpha, \frac{(d-j e ) \beta}{r})+ 
\sum\limits_{i=j}^{\eta-1} e \min(  \alpha, \frac{(d-r-ie) \beta}{e}).
\end{align}

$C_j(\alpha)$ is also piecewise-linear function with non-increasing successive slopes. Its breakpoints are given by  
$$ \{ \frac{d-r-(\eta-1)e}{e} \beta,\ldots,  \frac{d-r-je}{e} \beta,
 \frac{d-(j-1)e}{e} \beta , \ldots, \frac{d}{e} \beta\}\cup\{\frac{d-je}{r} \beta\}.$$ The exact relative position of the breakpoint $\frac{d-je}{r} \beta$ with respect to the other breakpoints of $C_j(\alpha)$ depends on the system's parameters. However, we give a lower bound on $\frac{d- j e}{r} \beta$.
\begin{align}
\frac{d- j e}{r}- \frac{d-r-(j-1)e}{e}&=\frac{ed-rd-re+r^2-j(e^2-r e)}{r e} \nonumber\\
&\geq  \frac{(e-r)d-re+r^2-\eta(e^2-r e)}{r e} \nonumber\\
&\geq   \frac{(e-r)k-re+r^2-\eta(e^2-r e)}{r e}\nonumber\\
&=0,
\end{align}
\noindent where the first inequality follows by noticing that the expression is decreasing in $j$ and letting $j=\eta$, and the second inequality follows as the corresponding expression is increasing $d$.  

\begin{figure*}
\centering
\includegraphics[width=1\textwidth]{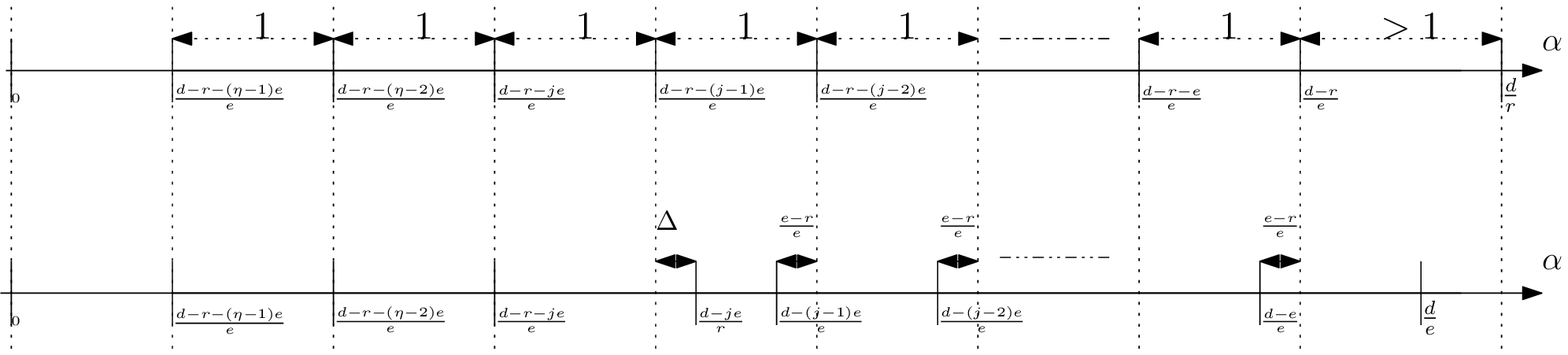}
\caption{relative positions of the breakpoints of $C_0(\alpha)$ and $C_j(\alpha)$ (with $\beta=1$)}.
\label{proof_general_case}
\end{figure*}

\fref{proof_general_case} illustrates the relative positions of all the breakpoints of $C_0(\alpha)$ and $C_j(\alpha), j \geq 1$, where for example $\frac{d- j e}{r} \in [ \frac{d-r-(j-1)e}{e},  \frac{d-r-(j-2)e}{e}]$. We denote by $ C_j(\infty)=\lim\limits_{\alpha \to +\infty}C_j(\alpha)$.
%To our end, we first discuss these scenarios.

\begin{lemma}
\label{alpha_intersect_lemma}
For $1\le j \le \eta$, there exists a point $\alpha_c(j)\in [ \frac{d}{e}, \frac{d}{r}]$ such that 
\begin{equation}
\label{alpha_intersect}
\begin{aligned}
C_0(\alpha_c(j))&=C_j(\alpha_c(j)),\\
C_0(\alpha)  &\le C_j(\alpha) &&\text{if } \alpha \le \alpha_c(j),\\
C_0(\alpha ) & \geq C_j(\alpha) &&\text{if } \alpha \geq \alpha_c(j),\\
C_j(\alpha)&=C_j(\infty) &&\text{if } \alpha \geq \alpha_c(j).
\end{aligned}
\end{equation}

%
%\begin{align*}
%C_0(\alpha_c(j))&=C_j(\alpha_c(j))\\
%C_0(\alpha)  &\le C_j(\alpha) &&\text{if } \alpha \le \alpha_c(j),\\
%C_0(\alpha ) & \geq C_j(\alpha) &&\text{if } \alpha \geq \alpha_c(j),\\
% C_j(\alpha)&=C_j(\infty) &&\text{if } \alpha \geq \alpha_c(j).
%\end{align*}
%Moreover, $C_j(\alpha)$ is constant for all $\alpha \geq \alpha_c(j)$
%$\forall \alpha \le \alpha_c(j)$, we have $C_0(\alpha) \le C_j(\alpha)$ while $\forall \alpha \geq \alpha_c(j)$, $C_0(\alpha) \geq C_j(\alpha)$.
\end{lemma}

\begin{IEEEproof}
W.l.o.g, assume $\beta=1$. First, we note that
\begin{align*}
 C_0(\alpha)=C_j(\alpha)= k \alpha \text{ for } \alpha \le \frac{d-r-(j-1)e}{e}.
\end{align*}
Next, we analyze the behavior of each of the functions $C_0(\alpha)$ and $C_j(\alpha)$ over the successive intervals 
$I_i \triangleq  ( \frac{d-r-ie}{e}, \frac{d-r-(i-1)e}{e} ]$ for $i \in \{j-1,j-2,\ldots,1 \}$. Let $x_i=\frac{d-r-ie}{e}$ and define $s_j(I_i)$ as the slope of $C_j(\alpha)$
 just before  $\alpha=x_{i } $ .
Consider a given interval $I_i=(x_i,x_{i-1}]$, we have
\begin{itemize}
\item	$C_0(\alpha)$ has no breakpoint inside $I_i$. Thus, $C_0(\alpha)$ increases by
\begin{align*}
C_0(x_{i-1})-C_0(x_{i})&=s_0(I_{i })-e.
\end{align*}  

\item	$C_j(\alpha)$ has either one or two breakpoints inside $I_i$.
\begin{enumerate}
\item	in the case of $C_j(\alpha)$ has a single breakpoint inside  $I_i$ (at $\alpha=\frac{d-i e}{e}$), $C_j(\alpha)$ increases by 
\begin{align*}
C_j(x_{i-1})-C_j(x_{i})&=s_j(I_i) \frac{r}{e} + (s_j(I_i)-e) \frac{e-r}{e} =s_j(I_i)-e+r.
\end{align*}

\item	in the case of $C_j(\alpha)$ has two breakpoints inside $I_i$, namely at $\alpha=\frac{d-je}{r}$ and $\alpha=\frac{d-i e}{e}$. Let $\Delta=\frac{d-je}{r}-\frac{d-r-i e}{e}$ (c.f. \fref{proof_general_case}).
Assuming $\frac{d-je}{r} \le\frac{d-i e}{e} $, then, $C_j(\alpha)$ increases by

\begin{align}
C_j(x_{i-1})-C_j(x_{i})
 = (s_j(I_i)-r)(1-\Delta-\frac{e-r}{e}) 
 + \frac{e-r}{e}(s_j(I_i)-r-e)+s_j(I_i) \Delta
 = s_j(I_i)-e + \Delta r. 
\end{align}
\noindent Assuming $\frac{d-je}{r} \geq \frac{d-i e}{e}$, then, $C_j(\alpha)$ increases by 
\begin{align}
 C_j(x_{i-1})-C_j( x_{i}) 
 =\frac{r}{e} s_j(I_i)+ (k-e)(\Delta-\frac{r}{e}) + (s_j(I_i)-r-e) (1 -\Delta)
 = s_j(I_i)-e + \Delta r,
\end{align}
\noindent which shows that the increase does not depend on the relative position of the two breakpoints.
\end{enumerate}
\end{itemize}
Now that we have computed the increase increment of each $C_j$ over $I_i$, we proceed to compare $C_0(\alpha)$ and $C_j(\alpha)$ for $1 \le j \le \eta$.

We discuss two cases:
\paragraph*{Case 1}
Assume $\frac{d-j e}{r} \in I_{j_0}$ for some $j_0 \in [1, j-1 ]$. $j_0$ may not exist, which will be discussed in the second case.
Based on the above discussion, it can be seen that 
\begin{align*}
C_j(\alpha) \geq C_0(\alpha), \text{ for } \alpha \le x_{j_0}.
\end{align*}
This can be seen by noticing that $  \forall i< j_0, s_0(I_i)= s_j(I_i) $ and that 
\begin{align*}
 (C_j(x_{i-1})-C_j(x_{i})) - (C_0(x_{i-1})-C_0(x_{i})) =  r \geq 0.
\end{align*}
%Moreover, it is clear for $\alpha \in I_i$ with $i<j_0$, $C_j$

Over $I_{j_0}$, $C_j$ also dominates $C_0$ at every point as $s_0(I_{j_0}) = s_j(I_{j_0}) $
 and
 \begin{align*}
 (C_j(x_{i-1})-C_j(x_{i})) - (C_0(x_{i-1})-C_0(x_{i})) = \Delta r \geq 0.
\end{align*}
For $i> j_0$, we have $s_0(I_i)-s_j(I_i)=r$. Moreover, over each $I_i, i>j_0$, we have 
 \begin{align*}
 (C_j(x_{i-1})-C_j(x_{i})) - (C_0(x_{i-1})-C_0(x_{i})) = (s_j(I_i) -e+r )-(s_0(I_i)-e)= 0.
\end{align*}
Combining the last equation and the observation that $C_j(x_{j_0-1}) \geq C_j(x_{j_0-1})$, it follows that $C_j$ continues to dominate $C_0$ over the successive intervals $I_i, i> j_0$.
So far, we have shown that
\begin{align*}
C_j(\alpha) \geq C_0(\alpha),\text{ for } \alpha \le \frac{d-r}{e}.
\end{align*}
For $\alpha \geq \frac{d-r}{e}$, we observe that $C_j$ increases with a slope of $e$ and levels off at $\frac{d}{e}$ while $C_0$ increases at smaller slope given by $r$ and levels off at $\frac{d}{r} > \frac{d}{e}$. 
Moreover, we know from \lref{final_value} that $C_0$ levels off at a higher value than that of $C_j$. Thus, there exists $\alpha_c(j) \in [ \frac{d}{e}, \frac{d}{r}]$ that satisfies \eref{alpha_intersect}.
%\begin{equation}
%\label{alpha_intersect}
%\begin{aligned}
%  C_0(\alpha_c(j))&=C_j(\alpha_c(j)),\\
%C_0(\alpha_c(j))&\le C_j(\alpha_c(j)) \forall \alpha \le \alpha_c(j), \\
%C_0(\alpha_c(j))&\geq C_j(\alpha_c(j)) \forall \alpha \geq \alpha_c(j), \\
%C_j(\alpha) &=C_j(\infty) \forall \alpha \geq \alpha_c(j).
%\end{aligned}
%\end{equation}

\paragraph*{Case 2} Assume $\frac{d-r}{e} < \frac{d-j e}{r} \le  \frac{d}{r} $, then, using similar arguments as in the first case, it follows that for $\alpha \le \frac{d-r}{e}$, $C_j(\frac{d-r}{e}) \geq C_0(\frac{d-r}{e})$. At $\alpha=\frac{d-r}{e}$, $C_j(\alpha)$ has a slope of $r+e$, which is higher than that of $C_0$, given by $r$. 
Thus, the slope of $C_j$ remains higher than than of $C_0$ until $C_j$ levels off.
Combining these observations with the fact that $C_0$ levels off at a higher value, it follows that both curves will intersect only once. Moreover, the intersection at a point at which $C_j$ has leveled off i.e., we have $\alpha_c(j) \geq \max(\frac{d}{e}, \frac{d-j e}{r})$. Therefore, \eref{alpha_intersect} holds also in this case.
\end{IEEEproof}

%\begin{IEEEproof}[proof of \lref{alpha_c_lemma}]
Using \lref{alpha_intersect_lemma} and the fact that $C_\eta$ achieves the smallest final value from \lref{final_value}, that is
$
C_\eta(\infty) \le C_j(\infty), j\in [0,\eta-1]
$, it follows that \eref{c_j_compare} holds for any $j\in [0,\eta]$.
%\begin{align*}
%C_0(\alpha)  &\le C_j(\alpha) &&\text{if } \alpha \le \alpha_c(a),\\
%C_a(\alpha ) & \le C_j(\alpha) &&\text{if } \alpha \geq \alpha_c(a).
%\end{align*}
Moreover, as $\alpha_c(\eta) \in [\frac{d }{e},\frac{d }{ r}]$, $\alpha_c(\eta)$ satisfies
\begin{align}
 r \alpha_c(\eta) + \sum\limits_{i=0}^{\eta-1} (d-r-ie) \beta= (\eta+1) \beta d - e \beta \frac{\eta^2+\eta}{2}, 
 \end{align}
 which implies that
 \begin{align}
 r \alpha_c(\eta) + \eta (d-r -\frac{e \eta}{2}+ \frac{e}{2})\beta=(\eta+1) \beta d - e \beta \frac{\eta^2+\eta}{2}.
\end{align} 
Simplifying the last equation yields \eref{alpha_c_a}. 
%\end{IEEEproof}
\subsection{Storage-bandwidth tradeoff expression}
\label{tradeoff_1}
We start with the case $k =\eta e +r$. The optimization trade-off is
\begin{equation}
\begin{aligned}
& \underset{\alpha \geq 0}{\text{minimize}}
& & \alpha \\
& \text{subject to}
& & 
  C(\alpha) \geq  \mathcal{M}.
\end{aligned}
\label{tradeoff_formualtion}
\end{equation} 
The constraint is a piece-wise linear function $C(\alpha) $ is given by

\begin{small}
\begin{align}
     C(\alpha)=\begin{cases}
     (\eta+1)\beta d-e \beta \eta (\eta+1)/2,       & \alpha \geq \alpha_c, \\
      r  \alpha + \sum\limits_{j=0}^{\eta-1} b_j,  & \alpha \in [ \frac{b_0}{e}, \alpha_c], \\         
               (r + i e) \alpha + \sum\limits_{j=i}^{\eta-1} b_j,
              &   \alpha \in [\frac{b_{i} }{e} ,\frac{b_{i-1} }{e} ],  
               \text{ for }  i=1,\ldots ,\eta-1,\\
                k \alpha , &	\alpha \leq   \frac{b_{\eta-1} }{e} ,
            \end{cases}
\end{align} 
\end{small}

\noindent with $\alpha_c=  \frac{d+\eta r - \eta e  }{r} \beta, b_i= (d-r- i e) \beta$ and 
\begin{align}
 \sum\limits_{j=i}^{\eta-1} b_j= \beta (\eta-i) (d-r-\frac{e(\eta-1+i)}{2}) 
= \gamma \frac{(\eta-i)(-2 r + e + 2 d - \eta e - e i) }{2 d} \triangleq \gamma g_r(i),
\end{align}
 such that 
\begin{align*}
g_r(i)=\frac{(\eta-i)(-2 r + e + 2 d - \eta e - e i) }{2 d} .
\end{align*}

The expression $C(\alpha)$  increases from 0 to a maximum value given by $\beta ( (\eta+1)d - \binom{\eta+1}{2})$. To solve \eref{tradeoff_formualtion}, we let $\alpha^*= C^{-1}(\mathcal{M})$ under the condition $\mathcal{M} \leq \beta ( (\eta+1)d - \binom{\eta+1}{2})$.
Therefore, we obtain,
\begin{align}
      \alpha^*=\begin{cases}
              \frac{\mathcal{M}}{k}   ,  
     \qquad\quad \mathcal{M} \in [0, \frac{k b_{\eta-1} }{e }  ] \\
              \frac{  \mathcal{M}- \sum\limits_{j=i}^{\eta-1} b_j}  {r+ie}  ,  
            \mathcal{M} \in [(r + i e) \frac{b_i}{e} + \sum\limits_{j=i}^{\eta-1} b_j, 
              (r + i e) \frac{b_{i-1}}{e} + \sum\limits_{j=i}^{\eta-1} b_j  ], \text{ for } i=\eta-1,\ldots 1,\\
               \frac{  \mathcal{M}- \sum\limits_{j=0}^{\eta-1} b_j}  {r}   ,  
             \mathcal{M} \in [ \frac{b_0 r}{e}+\sum\limits_{j=0}^{\eta-1} b_j, r \alpha_c +\sum\limits_{j=0}^{\eta-1} b_j] ,
    \end{cases}
\end{align}
\noindent with

\begin{align}
 \frac{r b_{i }}{e} + i  b_{i }+ \sum\limits_{j=i }^{\eta-1} b_j 
 &=
\frac{- \eta^2 e^2 + \eta e^2 - 2aer + 2dae - e^2i^2 - e^2 i - 2 e i r - 2 r^2 + 2 d r)}{2 d e} \gamma \nonumber\\
& =\frac{-k^2-r^2+e (k-r) +2 k d-e^2 (i^2+i)-2i e r }{   2 e d } \gamma \nonumber\\
&\triangleq 	\gamma \frac{\mathcal{M}}{f(i)},
\end{align}
\noindent such that 
\begin{align*}
f_r(i)=\frac{   2 e d \mathcal{M} }{-k^2-r^2+e (k-r) +2 k d-e^2 (i^2+i)-2i e r }.
\end{align*}
Therefore, fixing $\mathcal{M}$ and varying $\gamma$, we write
\begin{align}
\label{tradeoff_e_not_divide_k}
      \alpha^*=\begin{cases}
              \frac{\mathcal{M}}{k}   ,  
    & \mathcal{M} \in [0, \frac{k  g_r(\eta-1) \gamma }{e }  ], \\
              \frac{  \mathcal{M}- \gamma g_r(i)}  {r+ie}  ,  
    &       \mathcal{M} \in [  \frac{\gamma \mathcal{M}}{f_r(i)},  \frac{\gamma \mathcal{M}}{f_r(i-1)}],
               \text{ for } i=\eta-1,\ldots 1,\\
               \frac{  \mathcal{M}- \gamma g_r(0)}  {r}   ,  
       &   \mathcal{M} \in [  \frac{\gamma \mathcal{M}}{f_r(0)}, (g_r(0)+\frac{d+ar-ae}{d}) \gamma ].
    \end{cases}
\end{align}

As a function of $\gamma$, after simplifications, we obtain the expression of $\alpha^*$ as in \thref{tradeoff_expression}.
We note that there are $\eta$ piece-wise linear portions on the curve. 
%After simplification, we have 
%$\frac{e \mathcal{M}}{k g_r(\eta-1)}=f_r(\eta-1)=\frac{\mathcal{M}}{k}   \frac{e d}{d-k+e}$.
%After simplification, the minimum bandwidth point $\gamma_{\text{MBMR}}$, given by $\gamma_{\text{MBMR}}=\frac{\mathcal{M}}{g_r(0)+\frac{d+ar-ae}{d}}$ can be simplified to give 
%$
%\gamma_{\text{MBMR}}= \frac{d \mathcal{M}}{d (\eta+1)- e \binom{\eta+1}{2}}.
%$
%which is already expected from \sref{MSMR_MBMR}. 
Moreover, the minimum bandwidth point $\gamma_{\text{MBMR}}$ is given by 
\begin{align}
\gamma_{\text{MBMR}}=\frac{\mathcal{M}}{g_r(0)+\frac{d+ar-ae}{d}}=\frac{d \mathcal{M}}{d (\eta+1)- e \binom{\eta+1}{2}}.
\end{align}
The expression of $\alpha_{\text{MBMR}}$ is given by
\begin{align}
\alpha_{\text{MBMR}}=  \frac{  \mathcal{M}- \gamma_{\text{MBMR}} g(0)}{r} &=\gamma_{\text{MBMR}} 
\frac{d+\eta r-e \eta}{r d}  .
%\\&=
%\gamma_{\text{MBMR}} \frac{d-k+(1+\eta) r}{r d}. 
\end{align} 
in the case of $e \mid k$, we have $r=0$. The expression of the tradeoff is obtained from \eref{tradeoff_e_not_divide_k} by setting $r=0$ and eliminating the last line.
We note that in this case, there are $\eta-1$ piece-wise linear portions on the trade-off curve.  
%%%%%%%%%%%%%%%%%%%%%%%%%%%%%%%%%%%%%%%%%%%%%%%%%%%%%%%%%%%%%%%%%%%%%%%%%%%%%%%%%%%%%%%%%%%%%%%%%%%%%%%%%%%%%%%%%%%%%%%%%%%%%%%%%%%%%%%%%%%%%%%%%%%%%%%%%%%%%%%%%%%%%%%%%%%%%%%%%%%%%%%%%%%%%%%%%%%%%%%%%%%%%%%%%%%%%%%%%%%%%%%%%%%%%%%%%%%%%%%%%%%%%%%%%%%%%%%%%%%%%%%%%%%%%%%%%%%%%%%%%%%%%%%%%%%%%%%%%%%%%%%%%%%%%%%%%%%%%%%%%%%%%%%%%%%%%%%%%%%%%%%%%%%%%%%%%%%%%%%%%%%%%%%%%%%%%%%%%%%%%%%%
%\subsection{Proof of \lref{prop_5} }
%\label{proof_prop_5}
%%%%%%%%%%%%%%%%%%%%%%%%%%%%%%%%%%%%%%%%%%%%%%%%%%%%%%%%%%%%%%%%%%%%%%%%%
\subsection{Derivations of \eref{coupling_expression_systematic_to_systematic}, \eref{coupling_expression_parity_to_systematic} and \eref{coupling_expression_parity_to_parity} for IA codes}
\label{Derivations_IA}
\noindent Consider two systematic nodes $l_1,l_2 \in [k], l_1 \neq l_2$, starting from  \eref{repair_systematic_expression} and noting that $ V^{'t} U^{'}= \kappa  P^{'}$, we obtain after simplification
\begin{align}
r_{l_1,l_2} = \mathbf{v}_{l_2}^{'t} \mathbf{w}_{l_1} 
&= \mathbf{v}_{l_2}^{'t}  ( U^{'} -  \frac{\kappa^2}{1+\kappa } V \mathbf{e}_{l_1} \mathbf{e}_{l_1}^t P^{'}  ) \begin{bmatrix}
\bar{s}_{1,{l_1}} -\sum\limits_{j \neq l_1}  p_{j,1}  r_{j,l_1}\\ 
\vdots\\ 
\bar{s}_{k,{l_1}} -\sum\limits_{j \neq l_1}  p_{j,k} r_{j,l_1}\\ 
\end{bmatrix} \nonumber\\
&=  \kappa \mathbf{e}_{l_2}^t P^{'} \begin{bmatrix}
\bar{s}_{1,{l_1}} -\sum\limits_{j \neq l_1}  p_{j,1} r_{j,l_1}\\ 
\vdots\\ 
\bar{s}_{k,{l_1}} -\sum\limits_{j \neq l_1}  p_{j,k} r_{j,l_1}\\ 
\end{bmatrix}\nonumber\\
 &= \sum\limits_{j \in [k] } ( \kappa   P^{'}_{l_2,j} )  
 \bar{s}_{j,l_1} - \kappa r_{l_2,l_1}.
%\label{coupling_expression_systematic_to_systematic}
\end{align}
\noindent Proceeding in a similar way, for a systematic node $l \in [k]$ and a parity node $m \in [k]$, starting from \eref{repair_parity_expression}, we obtain
\begin{align}
\bar{s}_{m,l} = \mathbf{v}_l^{'t} \bar{\mathbf{w}}_m
&=\mathbf{v}_l^{'t}   ( (1-\kappa^2)V+ (1+\kappa) U^{'} \mathbf{e}_m \mathbf{e}_m^{^t} P^t ) 
\begin{bmatrix}
s_{1,m} +\frac{ \kappa^2}{1-\kappa^2} \sum\limits_{j\neq m}   P_{1,j}^{'}  \bar{r}_{j,m}\\
\vdots\\
s_{k,m} +\frac{ \kappa^2}{1-\kappa^2} \sum\limits_{j\neq m}    P_{k,j}^{'} \bar{r}_{j,m} 
\end{bmatrix}
\nonumber\\
&= ( (1-\kappa^2)\mathbf{e}_l^t+ (1+\kappa) \kappa P^{'}_{l,m} \mathbf{e}_m^{^t} P^t ) 
\begin{bmatrix}
s_{1,m} +\frac{ \kappa^2}{1-\kappa^2} \sum\limits_{j\neq m}   P_{1,j}^{'}  \bar{r}_{j,m}\\
\vdots\\
s_{k,m} +\frac{ \kappa^2}{1-\kappa^2} \sum\limits_{j\neq m}    P_{k,j}^{'} \bar{r}_{j,m}
\end{bmatrix}\nonumber\\
&=(1-\kappa^2 + \kappa(1+\kappa) P_{l,m}^{'} P_{l,m}) s_{l,m}  
+ \sum\limits_{j \in [k] \backslash \{l \}}( \kappa(1+\kappa) P_{l,m}^{'}  P_{j,m} ) s_{j,m} 
+ \sum\limits_{j \in [k]} ( \kappa^2 P^{'}_{l,j}) \bar{r}_{j,m}.
%\label{coupling_expression_parity_to_systematic} 
\end{align}
%\begin{align}
%\bar{s}_{m,l} = \mathbf{v}_l^{'t} \bar{\mathbf{w}}_m=(1-\kappa^2 + \kappa(1+\kappa) P_{l,m}^{'} P_{l,m}) s_{l,m}  
%+ \sum\limits_{j \in [k] \backslash \{l \}}( \kappa(1+\kappa) P_{l,m}^{'}  P_{j,m} ) s_{j,m} 
%+ \sum\limits_{j \in [k]} ( \kappa^2 P^{'}_{l,j}) \bar{r}_{j,m}.
%\label{coupling_expression_parity_to_systematic} 
%\end{align}
\noindent Finally, consider two parity nodes $m_1,m_2 \in [k], m_1 \neq m_2$, starting from \eref{repair_parity_expression}, we obtain
\begin{align}
\bar{r}_{m_1,m_2}= \mathbf{u}_{m_2}^{t} \bar{\mathbf{w}}_{m_1} 
&=\mathbf{u}_{m_2}^{t}  ( (1-\kappa^2)V+ (1+\kappa) U^{'} \mathbf{e}_{m_1} \mathbf{e}_{m_1}^{^t} P^t ) 
   \begin{bmatrix}
s_{1,m_1} +\frac{ \kappa^2}{1-\kappa^2} \sum\limits_{j\neq m_1}   P_{1,j}^{'}  \bar{r}_{j,m_1}\\
\vdots\\
s_{k,m_1} +\frac{ \kappa^2}{1-\kappa^2} \sum\limits_{j\neq m_1}    P_{k,j}^{'} \bar{r}_{j,m_1}
\end{bmatrix}
\nonumber\\
&=  \frac{1-\kappa^2}{\kappa} \mathbf{e}_{m_2}^t P^t  \begin{bmatrix}
s_{1,m_1} +\frac{ \kappa^2}{1-\kappa^2} \sum\limits_{j\neq m_1}   P_{1,j}^{'}  \bar{r}_{j,m_1}\\
\vdots\\
s_{k,m_1} +\frac{ \kappa^2}{1-\kappa^2} \sum\limits_{j\neq m_1}    P_{k,j}^{'} \bar{r}_{j,m_1}
\end{bmatrix}\nonumber\\
&= \sum\limits_{ j\in [k]}   
( \frac{1-\kappa^2}{\kappa} P_{j,m_2}) s_{j,m_1}
+  \kappa \bar{r}_{m_2,m_1}.
 %\label{coupling_expression_parity_to_parity}
\end{align}
%\begin{align}
%\bar{r}_{m_1,m_2}&= \mathbf{u}_{m_2}^{t} \bar{\mathbf{w}}_{m_1}= \sum\limits_{ j\in [k]}   
%( \frac{1-\kappa^2}{\kappa} P_{j,m_2}) s_{j,m_1}
%+  \kappa \bar{r}_{m_2,m_1}.
% \label{coupling_expression_parity_to_parity}
%\end{align}
%%%%%%%%%%%%%%%%%%%%%%%%%%%%%%%%%%%%%%%%%%%%%%%%%%%%%%%%%%%%%%%%%%%%%%%%%
%%%%%%%%%%%%%%%%%%%%%%%%%%%%%%%%%%%%%%%%%%%%%%%%%%%%%%%%%%%%%%%%%%%%%%%%%
\subsection{Proof of Lemma \ref{prop_1}}
\label{prop_1_proof}
\begin{IEEEproof}
First, we note that when $j \geq \eta$, $I(W_L,W_A)=H(W_L)-H(W_L|A)=H(W_L)= e \alpha$. In the following, we assume $j <\eta$. We write
\begin{align}
I(W_L, W_A)&=H(W_L)-H(W_L|A) \nonumber\\
& =e \alpha - \min (e \alpha, (d-j e) \beta) \label{eq124} \\
&=e (\alpha-\min(\alpha,(d'-j)\beta))\nonumber\\
&= e (\alpha- (d'-j)\beta)^+ \nonumber\\
&= e ((j-p)\beta-\theta)^+,\nonumber
\end{align} 
\noindent where we use the notation $(x)^+ \triangleq \max(x,0)$. 
Here \eqref{eq124} follows from Lemma \ref{entropy_lemma}.
\end{IEEEproof}
%%%%%%%%%%%%%%%%%%%%%%%%%%%%%%%%%%%%%%%%%%%%%%%%%%%%%%%%%%%%%%%%%%%%%%%%%
\subsection{Proof of Lemma \ref{lem:contribution_of_nodes}}
\label{lem:contribution_of_nodes_proof}
\begin{IEEEproof}
%Similar to the proof of \lref{general_intersection_dimension}.
%
Partition the set of $d$ helpers into $A$ and $B$ such that $|A|=k-e$ and $|B|=d-k+e$, such that $m \in B$. We have $H(W_L|S_A^L)=\min(e \alpha, (d-k+e) \beta)=(d-k+e) \beta$, as $e \alpha \geq (d-k+e) \beta$	for all points on the tradeoff. Moreover, exact repair requires $H(W_L|S_A^L, S_B^L)=0$. Thus, $H(S_B^L) \geq (d-k+e) \beta$. This implies $H(S_B^L) = (d-k+e) \beta$. Moreover, it must hold that $H(S_m^L) = \beta$ in addition to $S_{m}^L$ and $S_{m'}^L$ being independent if $m \neq m'$. Moreover, by choosing $M \subseteq B$, one obtains $H(S_M^L)=e \beta$.
\end{IEEEproof}

%%%%%%%%%%%%%%%%%%%%%%%%%%%%%%%%%%%%%%%%%%%%%%%%%%%%%%%%%%%%%%%%%%%%%%%%%
\subsection{Proof of Lemma \ref{proof_helper_node_pooling}}
\label{proof_helper_node_pooling_proof}
\begin{IEEEproof} 
If the statement holds true for some $f,r'$, then it also holds true for all $f' \geq f$ and $r'' \le r'$. Thus, for the proof, we only need to consider $F= R\cup   M , |F|=f=e(p+3), |R|=r'e = (p+2)e, |M| = e$.
 
Consider repair of an arbitrary set of $e$ nodes $L \subseteq R$, where the set of helpers include $M$ and the $e(p+1)$ remaining nodes in $R$. Then, we write
\begin{align}
I(S_M^L;W_R)&= I(S_M^L;W_L, W_{R-{L} })\nonumber\\
&= I(S_M^L; W_{R-{L} })+ I(S_M^L;W_L| W_{R-{L} })\nonumber \\
&\geq I(S_M^L;W_L| W_{R-{L} }) \nonumber\\
& = H(W_L| W_{R-{L} })- H(W_L| W_{R-{L} }, S_M^L)\nonumber\\
& \geq H(W_L| W_{R-{L} })- H(W_L| S_{R-{L} }^L, S_M^L)\nonumber\\
&= \min(e \alpha, (d-e(p+1))\beta)-  \min(e \alpha,( d-e(p+2))\beta) \label{eq137}\\
&=   (d-e(p+1))\beta -  ( d-e(p+2))\beta = e \beta.\nonumber
\end{align}
Here \eqref{eq137} follows from Lemma \ref{entropy_lemma} and Corollary \ref{cor:H_W_given_S}.
Then, we obtain
\begin{align}
H(S_M^L|W_R)= H(S_M^L)-I(S_M^L;W_R) \le e \beta - e \beta =0.
\end{align}
Hence, $H(S_M^L| W_R)=0.$ Since $L$ is arbitrary, it follows that $H(S_M^{R} | W_R)=0$.
It follows from \lref{prop_1} that 
\begin{align*}
H(S_M^R)= I(S_M^R; W_R) \le I(W_M;W_R)=e (2 \beta- \theta).
\end{align*}
Hence the proof is completed.
\end{IEEEproof}
%%%%%%%%%%%%%%%%%%%%%%%%%%%%%%%%%%%%%%%%%%%%%%%%%%%%%%%%%%%%%%%%%%%%%%%%%
\subsection{Proof of Lemma \ref{prop_5} }
\label{prop_5_proof}
\begin{IEEEproof}
The set is $R$ assumed to consist of $|R|=e r'=e(p+1)$ nodes, and the set $F$ is such that $F=R\cup \{ M \}$, $|M|=e$. Similar to Lemma \ref{proof_helper_node_pooling},
\begin{align}
I(S_M^L;W_R)%&= I(S_M^L;W_{R-L} ,W_L) \nonumber\\
%&= I(S_M^L;W_{R-L}  )+  I(S_M^L;W_L|W_{R-L}  ) \nonumber\\
%& \geq I(S_M^L;W_L|W_{R-L}  ) \nonumber\\
%&= H(W_L|W_{R-L})- H(W_L|W_{R-L},S_M^L) \nonumber\\
& \geq H(W_L|W_{R-L})- H(W_L|S_{R-L}^L,S_M^L) \nonumber\\
& =\min(e \alpha, (d-(r'-1)e)\beta)   -\min(e \alpha, (d-r'e)\beta)\nonumber\\
&  =( d-pe)\beta - e \theta -( d-(p+1)e)\beta\nonumber\\
& =e (\beta-\theta).
\end{align}
Then, it must be that
\begin{align}\label{eq:etheta}
H(S_M^L|W_R)= H(S_M^L)- I(S_M^L;W_R) \le e \beta - e (\beta-\theta)= e \theta. 
\end{align}
Note that the last inequality holds for any set $L \subseteq R$. Next, consider $L_1,L_2 \subseteq	R$. For this, consider
\begin{align}
H(S_M^{L_1},S_M^{L_2}) 
&= I(W_R;S_M^{L_1},S_M^{L_2})+ 
H( S_M^{L_1} , S_M^{L_2} | W_R)\nonumber\\
& \le I(W_R; W_M)+ H( S_M^{L_1} , S_M^{L_2} | W_R) \nonumber\\
& = I(W_R; W_M)+ H( S_M^{L_1}   | W_R)+H(  S_M^{L_2} | W_R,
S_M^{L_1}  ) \nonumber\\
&\le e (\beta-\theta)+ e \theta +e \theta= e (\beta + \theta),
\end{align}
where the last inequality follows from \lref{prop_1} and \eqref{eq:etheta}. Then, we have
\begin{align}
H(S_M^{L_1} |  S_M^{L_2})&= H(S_M^{L_1} ,   S_M^{L_2})- H(S_M^{L_2}  ) \nonumber\\&= H(S_M^{L_1} ,   S_M^{L_2})- e \beta \nonumber\\&\le e (\beta + \theta) - e \beta= e \theta,
\end{align}
where the first equality follows from Lemma \ref{lem:contribution_of_nodes}. 

Finally, partitioning the nodes in $R$ into sets $R_1,R_2,\ldots, R_{r'}$ of size $e$, it follows
\begin{align}
H(S_M^R) \le H(S_M^{R_1})+ \sum\limits_{i=2}^{r'} H(S_M^{R_i}|S_M^{R_{i-1}}) \le e \beta + e (r'-1) \theta.
\end{align}
Thus the proof is completed.
\end{IEEEproof}
%%%%%%%%%%%%%%%%%%%%%%%%%%%%%%%%%%%%%%%%%%%%%%%%%%%%%%%%%%%%%%%%%%%%%%%%%
\subsection{Proof of \thref{non_achivability_2}}
\label{non_achivability_2_app}
\begin{IEEEproof}
Take a subnetwork $F$ of $d+e$ nodes. Let $L, M \subseteq F$ be two disjoint groups of $e$ nodes. Partition the $d-e$ remaining nodes  into two sets, $A$ of cardinality $e p$ and $B$ of cardinality $d-ep-e$. Exact repair requires
\begin{align}
H(W_L|S_A^L,S_B^L,S_M^L)&=0, \nonumber\\
H(W_M|S_A^M,S_B^M,S_L^M)&=0.
\end{align}
It follows that 
\begin{align}
 H(W_L,W_M|W_A,S_B^L,S_B^M,S_M^L) 
= H(W_L |W_A,S_B^L,S_B^M,S_M^L)  +H(W_M|W_L,W_A,S_B^L,S_B^M,S_M^L)
=0.
\end{align}
Therefore,
we have 
\begin{align}
 ( S_B^L,S_B^M,S_M^L)  
&\geq H(W_L,W_M|W_A) \nonumber\\
& =H(W_L|W_A) + H(W_M|W_A, W_L) \nonumber\\
& =H(W_L)-I(W_L;W_A)+ H(W_M) -I(W_M;W_A, W_L) \nonumber\\
& =e \alpha -0+ e \alpha -e(\beta-\theta)  \label{eq186} \\
&=2 e \alpha- e \beta + e \theta \nonumber\\
&=2  ((d-ep)\beta-e \theta)- e \beta + e \theta \nonumber\\
&=(2d-2e p-e) \beta - e \theta.\nonumber
\label{inter_point_first_case}
\end{align}
Here \eqref{eq186} follows from Lemma \ref{prop_1}.
We note that the lower bound does not depend on whether $d$ is a multiple of $e$.
Next, we obtain an an upper bound on the same quantity.

Partition $B$ into sets of size $e$, denoted by $L_i$. We will use $R = L \cup M, r'=2,$ in the helper node pooling. 
\paragraph*{case: $p+2 <\eta$} In this case, the parameters satisfy the condition in Lemma \ref{proof_helper_node_pooling}.

\begin{small}
\begin{align}
H( S_B^L,S_B^M,S_M^L)
& \le \sum\limits_{L_i \in B} H(S_{L_i}^L, S_{L_i}^M) + H(S_M^L) \\
& \le \sum\limits_{L_i \in B}  e(2 \beta-\theta)+ e \beta \label{eq191} \\
&= (d-pe-e)(2 \beta-\theta)+ e \beta \nonumber\\
&= (2d-2 e p-e) \beta - (d-e p-e) \theta,
\label{inter_point_second_case}
\end{align}
\end{small}
\noindent where the inequality \eqref{eq191} is obtained using \lref{lem:contribution_of_nodes} and \lref{proof_helper_node_pooling}.
Equations \eref{inter_point_first_case} and \eref{inter_point_second_case} are in contradiction if $d-ep -e >e \iff d > e(p+2)$, which is true as $d\geq k =ae > (p+2)e$. % and $\theta \neq 0$.

\paragraph*{case: $p+2 =\eta$}
In this case, \lref{prop_5} is used to derive an upper bound on $H( S_B^L,S_B^M,S_M^L)$. \lref{prop_5} does not hold if $\eta=2$. It holds for $\eta > 2 \iff k > 2 e$. 
Thus, we consider $k > 2e$. We have
\begin{align}
H( S_B^L,S_B^M,S_M^L)
& \le \sum\limits_{L_i \in B} H(S_{L_i}^L, S_{L_i}^M) + H(S_M^L) \\
& \le \sum\limits_{L_i \in B} e(\beta+\theta)+ e\beta \\
&= (d- e p) \beta + (d-e p -e) \theta.
\label{inter_point_second_case_2}
\end{align}
Equations \eref{inter_point_first_case} and \eref{inter_point_second_case_2} are in contradiction when
\begin{align}
\theta < \frac{d-ep -e}{d- ep} \beta.
\end{align}
\end{IEEEproof}
 
\bibliographystyle{IEEEtran}
\bibliography{biblio}

\end{document}